\documentclass{elsart3p}

\usepackage{amsmath,amssymb,graphicx}
\usepackage{natbib}

\newcommand{\D}{{\mathrm{d}}}
\newcommand{\Proba}[1]{ {\Bbb P}[ #1 ]}

\begin{document}
\begin{frontmatter}

\title{Dynamic and static limitation in multiscale reaction networks, revisited}
\author{A.~N.~Gorban \corauthref{cor1}}
 \ead{ag153@le.ac.uk}
\address{University of Leicester, UK}
\corauth[cor1]{Corresponding author: Department of Mathematics,
University of Leicester, LE1 7RH, UK}
\author{O. Radulescu}
 \ead{ovidiu.radulescu@univ-rennes1.fr}
\address{IRMAR, UMR 6625, University of Rennes 1, Campus de Beaulieu, 35042 Rennes , France}

\maketitle

\begin{abstract}
The concept of the limiting step gives the limit simplification: the
whole network behaves as a single step. This is the most popular
approach for model simplification in chemical kinetics. However, in
its elementary form this idea is applicable only to the simplest
linear cycles in steady states. For such simple cycles the
nonstationary behaviour is also limited by a single step, but not
the same step that limits the stationary rate. In this paper, we
develop a general theory of static and dynamic limitation for all
linear multiscale networks. Our main mathematical tools are
auxiliary discrete dynamical systems on finite sets and specially
developed algorithms of ``cycles surgery" for reaction graphs. New
estimates of eigenvectors for diagonally dominant matrices are used.

Multiscale ensembles of reaction networks with well separated
constants are introduced and typical properties of such systems
are studied. For any given ordering of reaction rate constants the
explicit approximation of steady state, relaxation spectrum and
related eigenvectors (``modes") is presented. In particular, we
prove that for systems with well separated constants eigenvalues
are real (damped oscillations are improbable). For systems with
modular structure, we propose the selection of such modules that
it is possible to solve the kinetic equation for every module in
the explicit form. All such ``solvable" networks are described.
The obtained multiscale approximations, that we call ``dominant
systems" are computationally cheap and robust. These dominant
systems can be used for direct computation of steady states and
relaxation dynamics, especially when kinetic information is
incomplete, for design of experiments and mining of experimental
data, and could serve as a robust first approximation in
perturbation theory or for preconditioning.
\end{abstract}
\end{frontmatter}

\section{Introduction\label{sec1}}

Which approach to model reduction is the most important?
Population is not the ultimate judge, and popularity is not a
scientific criterion, but ``Vox populi, vox Dei," especially in
the epoch of citation indexes, impact factors and bibliometrics.
Let us ask Google. It gave on 31st December 2006:
\begin{itemize}
\item{for ``quasi-equilibrium" -- 301000 links;}
\item{for ``quasi steady state" 347000 and for ``pseudo steady
state" 76200, 423000 together;}
\item{for our favorite ``slow manifold" (\citet{GorKar}, \citet{InChLANL}) 29800
links only, and for ``invariant manifold" slightly more, 98100;}
\item{for such a framework topic as ``singular perturbation"
Google gave 361000 links;}
\item{for ``model reduction" even more, as we did expect, 373000;}
\item{but for ``limiting step" almost two times more -- 714000!}
\end{itemize}

Our goal is the general theory of static and dynamic limitation
for multiscale networks. The concept of the limiting step gives,
in some sense, the limit simplification: the whole network behaves
as a single step. As the first result of our paper we introduce
further detail in this idea: the whole network behaves as a single
step in statics, and as {\it another} single step in dynamics:
even for simplest cycles the stationary rate and the relaxation
time to this stationary rate are limited by different reaction
steps, and we describe how to find these steps.

The concept of limitation is very attractive both for theorists
and experimentalists. It is very useful to find conditions when a
selected reaction step becomes the limiting step. We can change
conditions and study the network experimentally, step by step. It
is very convenient to model a system with limiting steps: the
model is extremely simple and can serve as a very elementary
building block for further study of more complex systems, a
typical situation both in industry and in systems biology.

In the IUPAC Compendium of Chemical Terminology (2007) one can
find two articles with a definition of limitation.
\begin{itemize}
\item{\cite{R-limit}: Rate-determining step (rate-limiting step):
``These terms are best regarded as synonymous with
rate-controlling step."} \item{\citet{R-cont}: ``A
rate-controlling (rate-determining or rate-limiting) step in a
reaction occurring by a composite reaction sequence is an
elementary reaction the rate constant for which exerts a strong
effect -- stronger than that of any other rate constant -- on the
overall rate."}
\end{itemize}
It is not wise to object to a definition and here we do not
object, but, rather, complement the definition by additional
comments. The main comment is that usually when people are talking
about limitation they expect significantly more: there exists  a
rate constant which exerts such a strong effect on the overall
rate that the effect of all other rate constants together is
significantly smaller. Of course, this is not yet a formal
definition, and should be complemented by a definition of
``effect", for example, by ``control function" identified by
derivatives  of the overall rate of reaction, or by other overall
rate ``sensitivity parameters" (\cite{R-cont}).

For the IUPAC Compendium definition a rate-controlling step always
exists, because among the control functions generically exists the
biggest one. On the contrary,  for the notion of limitation that
is used in practice, there exists a difference between systems
with limitation and systems without limitation.

An additional problem arises: are systems without limitation rare
or should they be treated equitably with limitation cases? The
arguments in favor of limitation typicality are as follows: the
real chemical networks are very multiscale with very different
constants and concentrations. For such systems it is improbable to
meet a situation with compatible effects of all different stages.
Of course, these arguments are statistical and apply to generic
systems from special ensembles.

During last century, the concept of the limiting step was revised
several times. First simple idea of a ``narrow place" (a least
conductive step) could be applied without adaptation only to a
simple cycle of irreversible steps that are of the first order
(see Chap. 16 of the book \cite{Johnston} or the paper of
\cite{Boyd}). When researchers try to apply this idea in more
general situations they meet various difficulties such as:
\begin{itemize}
 \item{Some reactions have to be ``pseudomonomolecular."
 Their constants depend on
 concentrations of outer components, and are constant only
 under condition that these outer components are present in
 constant concentrations, or change sufficiently slow. For example, the simplest Michaelis--Menten
 enzymatic reaction is $E+S \to ES \to E+P$ ($E$ here stands for enzyme,
 $S$ for substrate, and $P$ for product), and the linear catalytic cycle
 here is $S \to ES \to S$.   Hence, in  general we must consider nonlinear systems.}
 \item{Even under fixed outer components concentration,
 the simple ``narrow place" behaviour
 could be spoiled by branching or by reverse reactions. For
 such reaction systems definition of a limiting step simply as a step with
 the smallest constant does not work. The simplest example is given by the cycle:
 $A_1 \leftrightarrow A_2 \to A_3 \to A_1$. Even if the constant of the last step
 $A_3 \to A_1$ is the smallest one, the stationary rate may be much smaller than
 $k_3b$ (where $b$ is the overall balance of concentrations, $b=c_1+c_2+c_3$), if the
 constant of the reverse reaction $A_2 \to A_1$ is sufficiently big.}
\end{itemize}

In a series of papers, \cite{Northrop1,Northrop2} clearly
explained these difficulties with many examples based on the
isotope effect analysis and suggested that the concept of
rate--limiting step is ``outmoded". Nevertheless, the main idea of
limiting is so attractive that  Northrop's arguments stimulated
the search for modification and improvement of the main concept.

\cite{Ray} proposed the use of sensitivity analysis. He considered
cycles of reversible reactions and suggested a definition: {\it
The rate--limiting step in a reaction sequence is that forward
step for which a change of its rate constant produces the largest
effect on the overall rate}. In his formal definition of
sensitivity functions the reciprocal reaction rate ($1/W$) and
rate constants ($1/k_i$) were used and the connection between
forward and reverse step constants (the equilibrium constant) was
kept fixed.

Ray's approach was revised by \cite{BrownCo} from the system
control analysis point of view (see the book of \cite{CorBow}).
They stress again that there is no unique rate--limiting step
specific for an enzyme, and this step, even if it exists, depends
on substrate, product and effector concentrations. They
demonstrated also that the control coefficients
$$C^W_{k_i}=\left(\frac{k_i}{W} \frac{\partial W}{\partial
k_i}\right)_{[S],[P],...},$$ where $W$ is the stationary reaction
rate and $k_i$ are constants, are additive and obey the summation
theorems (as concentrations do). A simple relation between control
coefficients of rate constants and intermediate concentrations was
reported by \cite{Kholod}. This relation connects two type of
experiments: measurement of intermediate levels and steady--state
rate measurements.

For the analysis of nonlinear cycles the new concept of {\it
kinetic polynomial} was developed (\cite{YabLaz1,YabLaz2}). It was
proven that the stationary state of the single-route reaction
mechanism of catalytic reaction can be described by a single
polynomial equation for the reaction rate. The roots of the
kinetic polynomial are the values of the reaction rate in the
steady state. For a system with limiting step the kinetic
polynomial can be approximately solved and the reaction rate found
in the form of a series in powers of the limiting step constant
(\cite{YabLazLim}).

In our approach, we analyze not only the steady state reaction
rates, but also the relaxation dynamics of multiscale systems. We
focused mostly on the  case when all the elementary processes have
significantly different time scales. In this case, we obtain
``limit simplification" of the model: all stationary states and
relaxation processes could be analyzed ``to the very end", by
straightforward computations, mostly analytically.  Chemical
kinetics is an inexhaustible source of examples of multiscale
systems for analysis. It is not surprising that many ideas and
methods for such analysis were first invented for chemical
systems.

In Sec.~\ref{sec2} we  analyze a simple example and the source of
most generalizations, the catalytic cycle, and demonstrate the
main notions on this example. This analysis is quite elementary,
but includes many ideas elaborated in full in subsequent sections.

There exist several estimates for relaxation time in chemical
reactions (developed, for example, by \cite{CherYab}), but even
for the simplest cycle with limitation the main property of
relaxation time is not widely known. For a simple irreversible
catalytic cycle with limiting step the stationary rate is
controlled by the smallest constant, but the relaxation time is
determined by the second in order constant. Hence, if in the
stationary rate experiments for that cycle we mostly  extract the
smallest constant, in relaxation experiments another, the second
in order constant will be observed.

It is also proven that for cycles with well separated constants
damped oscillations are impossible, and spectrum of the matrix of
kinetic coefficients is real. For general reaction networks with
well separated constants this property is proven in
Sec.~\ref{sec4}.

Another general effect observed for a cycle is robustness of
stationary rate and relaxation time. For multiscale systems with
random constants, the standard deviation of constants that
determine stationary rate (the smallest constant for a cycle) or
relaxation time (the second in order constant) is approximately
$n$ times smaller than the standard deviation of the individual
constants (where $n$ is the cycle length). Here we deal with the
so-called ``order statistics". This decrease of the deviation as
$n^{-1}$ is much faster than for the standard error summation,
where it decreases with increasing $n$ as $n^{-1/2}$.

In more general settings, robustness of the relaxation time was
studied  by \cite{GorOvRobu} for chemical kinetics  models of
genetic and signalling networks. \cite{GorOvRobu} proved that for
large multiscale systems with hierarchical distribution of time
scales the variance of the inverse relaxation time (as well as the
variance of the stationary rate) is much lower than the variance of
the separate constants. Moreover, it can tend to 0 faster than
$1/n$, where $n$ is the number of reactions. It was demonstrated
that similar phenomena are valid in the nonlinear case as well. As a
numerical illustration we used a model of a signalling network that
can be applied to important transcription factors such as NFkB.

Each multiscale system is characterized by its structure (the
system of elementary processes) and by the rate constants of these
processes. To make any general statement about such systems when
the structure is given but the constants are unknown it is useful
to take the constant set as random and independent.  But it is not
obvious how to chose the random distribution. The usual idea to
take normal or uniform distribution meets obvious difficulties,
the time scales are not sufficiently well separated.

The statistical approach to chemical kinetics was developed by
\cite{Rabi1,Rabi2}, and high-dimensional model representations
(HDMR) were proposed as efficient tools to provide a fully global
statistical analysis of a model. The work of \cite{Rabi3} was
focused on how the network properties are affected by random rate
constant changes. The rate constants were transformed to a
logarithmic scale to ensure an even distribution over the large
space.

The log-uniform distribution on sufficiently wide interval helps us
to improve the situation, indeed, but a couple of extra parameters
appears: $\alpha = \min \log k$ and $\beta = \max \log k$. We have
to study the asymptotics $\alpha \to -\infty$, $\beta \to \infty$.
This approach could be formalized by means of the uniform invariant
distributions of $\log k$ on $\mathbb{R}^n$. These distributions are
finite--additive, but not countable--additive (not
$\sigma$-additive).

The probability and measure theory without countable additivity
has a long history. In Euclid's time only arguments based on
finite--additive properties of volume were legal. Euclid meant by
equal area the scissors congruent area. Two polyhedra are
scissors--congruent if one of them can be cut into finitely many
polyhedral pieces which can be re-assembled to yield the second.
But all proofs of the formula for the volume of a pyramid involve
some form of limiting process. Hilbert asked in his third problem:
are two Euclidean polyhedra of the same volume scissors congruent?
The answer is ``no" (a review of old and recent results is
presented by \cite{Neum}). There is another invariant of cutting
and gluing polyhedra.

Finite--additive invariant measures on non-compact groups were
studied by  \cite{Birkhoff} (see also the book of \cite{HewRoss},
Chap. 4). The frequency--based Mises approach to probability
theory foundations (\cite{Mises}), as well as logical foundations
of probability by \cite{Carnap} do not need $\sigma$-additivity.
Non-Kolmogorov probability theories are discussed now in the
context of quantum physics (\cite{Hren}), nonstandard analysis
(\cite{Loeb}) and many other problems (and we do not pretend
provide here a full review of related works).

We answer the question: What does it mean ``to pick a multiscale
system at random"?  We introduce and analyze a notion of
multiscale ensemble of reaction systems. These ensembles with well
separated variables are presented in Sec.~\ref{sec3}.

The best geometric example that helps us to understand this
problem is one of Lewis Carroll's Pillow Problems published in
1883 (\cite{Carroll}): ``Three points are taken at random on an
infinite plane. Find the chance of their being the vertices of an
obtuse-angled triangle." (In an acute-angled triangle all angles
are comparable, in an obtuse-angled triangle the obtuse angle is
bigger than others and could be much bigger.) The solution of this
problem depends significantly on the ensemble definition. What
does it mean ``points are taken at random on an infinite plane"?
Our intuition requires translation invariance, but the normalized
translation invariant measure on the plain could not be
$\sigma$-additive. Nevertheless, there exist finite--additive
invariant measures.

Lewis Carroll proposed a solution that did not satisfy some of
modern scientists. There exists a lot of attempts to improve the
problem statement (\cite{Guy,Portnoy,Eisenb,RumaSam}): reduction
from infinite plane to a bounded set, to a compact symmetric
space, etc. But the elimination of paradox destroys the essence of
Carroll's problem. If we follow the paradox  and try to give a
meaning to ``points are taken at random on an infinite plane" then
we replace $\sigma$-additivity of the probability measure by
finite--additivity and come to the applied probability theory for
finite--additive probabilities. Of course, this theory for
abstract probability spaces would be too poor, and some additional
geometric and algebraic structures are necessary to build rich
enough theory.

This is not just a beautiful geometrical problem, but rather an
applied question about the proper definition of multiscale
ensembles. We need such a definition to make any general statement
about multiscale systems, and briefly analyze lessons of Carroll's
problem in Sec.~\ref{sec3}.

In this section we use some mathematics to define the multiscale
ensembles with well separated constants. This is necessary
background for the analysis of systems with limitation, and
technical consequences are rather simple. We need only two
properties of a typical system from the multiscale ensemble with
well separated constants:
\begin{enumerate}
\item{Every two reaction rate constants $k$, $k'$, are connected by
the relation $k \gg k'$ or $k \ll k'$ (with probability close to
1);}
\item{The first property persists (with probability close to 1),
if we delete two constants $k$,
$k'$ from the list of constants, and add a number $kk'$ or a
number $k/k'$ to that list.}
\end{enumerate}
If the reader can use these properties (when it is necessary)
without additional clarification, it is possible to skip reading
Sec.~\ref{sec3} and go directly to more applied sections. In.
Sec.~\ref{sec4} we study static and dynamic properties of linear
multiscale reaction networks. An important instrument for that
study is a hierarchy of auxiliary discrete dynamical system. Let
$A_i$ be nodes of the network (``components"), $A_i \to A_j$ be
edges (reactions), and $k_{ji}$ be the constants of these
reactions (please pay attention to the inverse order of
subscripts). A discrete dynamical system $\phi$ is a map that maps
any node $A_i$ in a node $A_{\phi(i)}$. To construct a first
auxiliary dynamical system for a given network we find for each
$A_i$ the maximal constant of reactions $A_i \to A_j$: $
k_{\phi(i) i} \geq k_{ji}$ for all $j$, and $\phi(i)=i$ if there
are no reactions $A_i \to A_j$. Attractors in this discrete
dynamical system are cycles and fixed points.

The fast stage of relaxation of a complex reaction network could be
described as mass transfer from nodes to correspondent attractors of
auxiliary dynamical system and mass distribution in the attractors.
After that, a slower process of mass redistribution between
attractors should play a more important role. To study the next
stage of relaxation, we should glue cycles of the first auxiliary
system (each cycle transforms into a point), define constants of the
first derivative network on this new set of nodes, construct for
this new network a (first) auxiliary discrete dynamical system, etc.
The process terminates when we get a discrete dynamical system with
one attractor. Then the inverse process of cycle restoration and
cutting starts. As a result, we create an explicit description of
the relaxation process in the reaction network, find estimates of
eigenvalues and eigenvectors for the kinetic equation, and provide
full analysis of steady states for systems with well separated
constants.

The problem of multiscale asymptotics of eigenvalues of
non-selfadjoint matrices  was studied by \cite{ViLju} and
\cite{Lid}. Recently, some generalizations were obtained by
idempotent (min-plus) algebra methods (\cite{Akian}). These
methods provide a natural language for discussion of some
multiscale problems (\cite{LitMas}). In the
Vishik--Ljusternik--Lidskii theorem and its generalizations the
asymptotics of eigenvalues and eigenvectors for the family of
matrices
$\mathcal{A}_{ij}(\epsilon)=a_{ij}\epsilon^{A_{ij}}+o(\epsilon^{A_{ij}})$
is studied for  $\epsilon > 0$, $\epsilon \to 0$.

In the chemical reaction networks that we study, there is no small
parameter $\epsilon$ with a  given distribution of the orders
$\epsilon^{A_{ij}}$ of the matrix nodes. Instead of these powers of
$\epsilon$ we have orderings of rate constants. Furthermore, the
matrices of kinetic equations have some specific properties. The
possibility to operate with the graph of reactions (cycles surgery)
significantly helps in our constructions. Nevertheless, there exists
some similarity between these problems and, even for general
matrices, graphical representation is useful. The language of
idempotent algebra (\cite{LitMas}), as well as nonstandard analysis
with infinitisemals (\cite{NonSt}), can be used for description of
the multiscale reaction networks, but now we postpone this for later
use.

We summarize results of relaxation analysis and describe the
algorithm of approximation of steady state and relaxation in
Subsec.~\ref{subsecSurgery}. After that, several examples of
networks are analyzed. In Sec.~\ref{sec:triangle} we illustrate
the analysis of dominant systems on a simple example, the
reversible triangle of reactions: $ A_1 \leftrightarrow A_2
\leftrightarrow A_3 \leftrightarrow A_1$. This simplest example
became very popular for the lumping analysis case study after the
well known work of \cite{Wei62}.  The most important mathematical
proofs are presented in the appendices.

In multiscale asymptotic analysis of reaction network we found
several very attractive {\it zero-one laws}. First of all,
components eigenvectors are close to 0 or $\pm 1$. This law
together with two other zero-one laws are discussed in
Sec.~\ref{sec:01}: ``Three zero-one laws and nonequilibrium phase
transitions in multiscale systems."

A multiscale system where every two constants have very different
orders of magnitude is, of course, an idealization. In parametric
families of multiscale systems there could appear systems with
several constants of the same order. Hence, it is necessary to
study effects that appear due to a group of constants of the same
order in a multiscale network. The system can have modular
structure, with different time scales in different modules, but
without separation of times inside modules. We discuss systems
with modular structure in Sec.~\ref{sec5}. The full theory of such
systems is a challenge for future work, and here we study
structure of one module. The elementary modules have to be
solvable. That means that the kinetic equations could be solved in
explicit analytical form. We give the necessary and sufficient
conditions for solvability of reaction networks. These conditions
are presented constructively, by algorithm of analysis of the
reaction graph.

It is necessary to repeat our study for nonlinear networks. We
discuss this problem and perspective of its solution in the
concluding Sec.~\ref{sec:conclud}. Here we again use the
experience summarized in the IUPAC Compendium (\cite{R-cont})
where the notion of controlling step is generalized onto nonlinear
elementary reaction by inclusion of some concentration into
``pseudo-first order rate constant".

\section{Static and dynamic limitation in a linear chain and a
simple catalytic cycle \label{sec2}}

\subsection{Linear chain}

A linear chain of reactions, $A_1 \to A_2 \to ... A_n$, with
reaction rate constants $k_i$ (for $A_i \to A_{i+1}$), gives the
first example of limitation. Let the reaction rate constant $k_q$
be the smallest one. Then we expect the following behaviour of the
reaction chain in time scale $\sim 1/k_q$: all the components
$A_1,... A_{q-1}$ transform fast into $A_q$, and all the
components $A_{q+1}, ... A_{n-1}$ transform fast into $A_n$, only
two components, $A_q$ and $A_n$ are present (concentrations of
other components are small) , and the whole dynamics in this time
scale can be represented by a single reaction $A_q \to A_n$ with
reaction rate constant $k_q$. This picture becomes more exact when
$k_q$ becomes smaller with respect to other constants.

The kinetic equation for the linear chain is
\begin{equation}\label{chainkin}
\dot{c_i}=k_{i-1} c_{i-1}-k_i c_i,
\end{equation}
where $c_i$ is concentration of $A_i$ and $k_{i-1}=0$ for $i=1$.
The coefficient matrix $K$ of this equations is very simple. It
has nonzero elements only on the main diagonal, and one position
below. The eigenvalues of $K$ are $-k_i$ ($i=1,...n-1$) and 0. The
left and right eigenvectors for 0 eigenvalue, $l^0$ and $r^0$,
are:
\begin{equation}\label{chain0eigen}
l^0=(1,1,...1), \;\; r^0=(0,0,...0,1),
\end{equation}
all coordinates of $l^0$ are equal to 1, the only nonzero
coordinate of $r^0$ is $r^0_n$ and we represent vector--column
$r^0$ in row.

Below we use explicit form of $K$ left and right eigenvectors. Let
vector--column $r^i$ and vector--row $l^i$ be right and left
eigenvectors of $K$ for eigenvalue $-k_i$. For coordinates of
these eigenvectors we use notation $r^i_j$ and $l^i_j$. Let us
choose a normalization condition $r^i_i=l^i_i=1$. It is
straightforward to check that $r^i_{j}=0$ $(j<i)$ and $l^i_{j}=0$
$(j>i)$, $r^i_{j+1} =k_j r_j /(k_{j+1}-k_i)$ $(j\geq i)$ and
$l^i_{j-1}=k_{j-1} l_j/(k_{j-1}-k_j)$ $(j \leq i)$, and
\begin{equation}\label{ChainEigen}
 r^i_{i+m}=\prod_{j=1}^m \frac{k_{i+j-1}}{k_{i+j}-k_i};
 \; l^i_{i-m}=\prod_{j=1}^m \frac{k_{i-j}}{k_{i-j}-k_i}.
\end{equation}
It is convenient to introduce formally $k_0=0$. Under selected
normalization condition, the inner product of eigenvectors is:
$l^i r^j = \delta_{ij}$, where $\delta_{ij}$ is the Kronecker
delta.

If the rate constants are well separated (i.e., any two constants,
$k_i$, $k_j$ are connected by relation $k_i \gg k_j$ or $k_i \ll
k_j$,
\begin{equation}\label{firstMultiL}
 \frac{k_{i-j}}{k_{i-j}-k_i}
 \approx \left\{
    \begin{aligned}
        & 1, \; &\mbox{if} \; k_i \ll k_{i-j};
        \\
        & 0, \;&  \mbox{if} \; k_i \gg k_{i-j},
    \end{aligned}\right.
\end{equation}
Hence, $|l^i_{i-m}| \approx 1$ or $|l^i_{i-m}| \approx 0$. To
demonstrate that also $|r^i_{i+m}| \approx 1$ or $|r^i_{i+m}|
\approx 0$, we shift nominators in the product (\ref{ChainEigen}) on
such a way: $$ r^i_{i+m}= \frac{k_i}{k_{i+m}-k_i} \prod_{j=1}^{m-1}
\frac{k_{i+j}}{k_{i+j}-k_i}.$$ Exactly as in (\ref{firstMultiL}),
each multiplier $${k_{i+j}}/({k_{i+j}-k_i})$$ here is either almost
1 or almost 0, and $${k_i}/({k_{i+m}-k_i})$$ is either almost 0 or
almost $-1$. In this zero-one asymptotics
\begin{equation}\label{01cyrcleAsy}
\begin{split}
l^i_i= &1, \; l^i_{i-m}\approx 1 \; \\ &\mbox{if} \; k_{i-j}>k_i \;
\mbox{for all} \; j=1, \ldots m, \; \mbox{else} \; l^i_{i-m}\approx
0;
\\
r^i_i=&1, \; r^i_{i+m}\approx -1  \; \\& \mbox{if} \; k_{i+j} >k_i
\; \mbox{for all} \; j=1, \ldots m-1 \; \\ &\mbox{and} \; k_{i+m} <
k_i, \;  \mbox{else} \; r^i_{i+m}\approx 0 .
\end{split}
\end{equation}
In this asymptotic, only two coordinates of right eigenvector
$r^i$ can have nonzero values, $r^i_i = 1$ and $r^i_{i+m}\approx
-1$ where $m$ is the first such positive integer that $i+m < n$
and $k_{i+m} < k_i$. Such $m$ always exists because $k_n=0$. For
left eigenvector $l^i$, $l^i_i \approx \ldots l^i_{i-q}\approx 1$
and $l^i_{i-q-j}\approx 0$ where $j>0$ and $q$ is the first such
positive integer that $i-q-1>0$ and $k_{i-q-1}<k_i$. It is
possible that such $q$ does not exist. In that case, all
$l^i_{i-j}\approx 1$ for $j\geq 0$. It is straightforward to check
that in this asymptotic $l^i r^j = \delta_{ij}$.

The simplest example gives the order $k_1 \gg k_{2} \gg ... \gg
k_{n-1}$: $l^i_{i-j}\approx 1$ for $j\geq 0$, $r^i_i = 1$,
$r^i_{i+1}\approx -1$ and all other coordinates of eigenvectors
are close to zero. For the inverse order, $k_1 \ll k_{2} \ll ...
\ll k_{n-1}$, $l^i_{i} = 1$, $r^i_i = 1$, $r^i_{n}\approx -1$  and
all other coordinates of eigenvectors are close to zero.

For less trivial example, let us find the asymptotic of left and
right eigenvectors for a chain of reactions:
 $$A_1 {\rightarrow^{\!\!\!\!\!\!5}}\,\,
A_2{\rightarrow^{\!\!\!\!\!\!3}}\,\,
A_3{\rightarrow^{\!\!\!\!\!\!4}}\,\,
A_4{\rightarrow^{\!\!\!\!\!\!1}}\,\,
A_5{\rightarrow^{\!\!\!\!\!\!2}}\,\, A_6 ,$$
 where  the upper index marks the order of rate constants:
 $k_4 \gg k_5 \gg k_2\gg k_3\gg k_1$ ($k_i$ is the
 rate constant of reaction $A_i \to ...$).

 For left eigenvectors, rows $l^i$, we have the
following asymptotics:
\begin{equation}\label{LeftApproxLin}
\begin{split}
&l^1\approx (1,0,0,0,0,0), \;  l^2\approx (0,1,0,0,0,0), \; \\
&l^3\approx (0,1,1,0,0,0),  l^4\approx (0,0,0,1,0,0), \; \\
&l^5\approx (0,0,0,1,1,0).
\end{split}
\end{equation}
For right eigenvectors, columns $r^i$, we have the following
asymptotics (we write  vector-columns in rows):
\begin{equation}\label{RightApproxLin}
\begin{split}
&r^1\approx
 (1, 0, 0, 0, 0, -1), \;
r^2\approx
 (0, 1, -1, 0, 0, 0), \; \\
&r^3\approx
 (0, 0, 1, 0, 0, -1),
r^4\approx
 (0, 0, 0, 1, -1, 0), \; \\
&r^5\approx
 (0, 0, 0, 0, 1, -1).
\end{split}
\end{equation}
The correspondent approximation to the general solution of the
kinetic equations is:
\begin{equation}\label{genSol}
c(t)=(l^0 c(0)) r^0 + \sum_{i=1}^{n-1}  (l^i c(0)) r^i \exp(-k_i
t),
\end{equation}
where $c(0)$ is the initial concentration vector, and for left and
right eigenvectors $l^i$ and $r^i$ we use their zero-one
asymptotic.

Asymptotic formulas allow us to transform kinetic matrix $K$ to a
matrix with value of diagonal element could not be smaller than
the value of any element from the correspondent column and row.

Let us represent the kinetic matrix $K$ in the basis of
approximations to eigenvectors (\ref{RightApproxLin}). The
transformed matrix is $\widetilde{K}_{ij}=l^i K r^j$
($i,j=0,1,...5$):
\begin{equation}
\begin{split}
& K=\left[\begin{array}{cccccc}
 -k_1 &  0 & 0 & 0 & 0 & 0 \\
  k_1 &-k_2& 0 & 0 & 0 & 0\\
  0  & k_2&-k_3& 0 & 0 & 0\\
  0  & 0& k_3&-k_4& 0 & 0\\
  0  & 0& 0 &  k_4 & -k_5 & 0 \\
  0  & 0& 0 &  0 & k_5 & 0 \\
\end{array}\right], \;\; \\
&\widetilde{K}=\left[\begin{array}{cccccc}
 0 &  0 &   0 &  0  & 0 & 0 \\
  0 &-k_1&  0 &  0  & 0 & 0\\
  0  & k_1&-k_2&  0 & 0 & 0\\
  0  & k_1& k_3&-k_3& 0 & 0\\
  0  & 0&  -k_3& k_3 & -k_4 & 0 \\
  0  & 0&  -k_3 &k_3 & -k_5 &-k_5
   \\
\end{array}\right].
\end{split}
\end{equation}
The transformed matrix has an important property
 $$|\widetilde{K}_{ij}| \leq \min\{|\widetilde{K}_{ii}|,
 |\widetilde{K}_{jj}| \}.$$
The initial matrix $K$ is diagonally dominant in  columns, but its
rows can include elements that are much bigger than the
correspondent diagonal elements.

We mention that a naive expectation $\widetilde{K}_{ij} \approx
\delta_{ij}$ is not realistic: some of the nondiagonal matrix
elements $\widetilde{K}_{ij}$ are of the same order than
$\min\{\widetilde{K}_{ii},\widetilde{K}_{jj}\}$. This example
demonstrates that a good approximation to an eigenvector could be
not an approximate eigenvector. If $K e = \lambda e$ and $\|e-f\|$
is small then $f$ is an approximation to eigenvector $e$. If $K f
\approx \lambda f$ (i.e. $\| K f - \lambda f \|$ is small), then
$f$ is an approximate eigenvector for eigenvalue $\lambda$. Our
kinetic matrix $K$ is very ill--conditioned. Hence, nobody can
guarantee that an approximation to eigenvector is an approximate
eigenvector, or, inverse, an approximate eigenvector (a
``quasimode") is an approximation to an eigenvector.

The question is, what do we need for approximation of the
relaxation process (\ref{genSol}). The answer is obvious: for
approximation of general solution (\ref{genSol}) with guaranteed
accuracy we need approximation to the genuine eigenvectors
(``modes") with the same accuracy. The zero-one asymptotic
(\ref{01cyrcleAsy}) gives this approximation. Below we always find
the modes approximations and not quasimodes.

\subsection{General properties of a cycle} The catalytic cycle is
one of the most important substructures that we study in reaction
networks. In the reduced form the catalytic cycle is a set of
linear reactions: $$A_1 \to A_2 \to \ldots A_n \to A_1.$$ Reduced
form means that in reality some of these reaction are not
monomolecular and include some other components (not from the list
$A_1, \ldots A_n$). But in the study of the isolated  cycle
dynamics, concentrations of these components are taken as constant
and are included into kinetic constants of the cycle linear
reactions.

For the constant of  elementary reaction $A_i \to $ we use the
simplified notation $k_i$ because the product of this elementary
reaction  is known, it is $A_{i+1}$ for $i<n$ and $A_1$ for $i=n$.
The elementary reaction rate is $w_i = k_i c_i$, where $c_i$ is
the concentration of $A_i$. The kinetic equation is:
\begin{equation}\label{kinCyc}
\dot{c}_i = w_{i-1} - w_i,
\end{equation}
where by definition $w_0=w_n$. In the stationary state ($\dot{c}_i
= 0$), all the $w_i$ are equal: $w_i=w$. This common rate $w$ we
call the cycle stationary rate, and
\begin{equation}\label{CycleRate}
w = \frac{b}{\frac{1}{k_1}+\ldots \frac{1}{k_n}}; \; \; c_i
=\frac{w}{k_i},
\end{equation}
where $b=\sum_i c_i$ is the conserved quantity for reactions in
constant volume (for general case of chemical kinetic equations
see elsewhere, for example, the book by \cite{Yab}). The
stationary rate $w$ (\ref{CycleRate}) is a product of the
arithmetic mean of concentrations, $b/n$, and the harmonic mean of
constants (inverse mean of inverse $k_i$).

\subsection{Static limitation in a cycle} If one of the constants,
$k_{\min}$,  is much smaller than others (let it be $k_{\min} =
k_n$), then
\begin{equation}\label{CycleLimRate}
\begin{split}
&c_n = b\left(1 -  \sum_{i<n}\frac{k_n}{k_i} +   o\left(
\sum_{i<n}\frac{k_n}{k_i}\right)\right), \\ &c_i = b
\left(\frac{k_n}{k_i} + o\left(
\sum_{i<n}\frac{k_n}{k_i}\right)\right),  \\  &w= k_n b \left(1+
O\left( \sum_{i<n}\frac{k_n}{k_i}\right)\right),
\end{split}
\end{equation}
or simply in linear approximation
\begin{equation}\label{CycleLimRateLin}
c_n = b\left(1 -  \sum_{i<n}\frac{k_n}{k_i}\right),  \; c_i = b
\frac{k_n}{k_i},\;  w= k_n b,
\end{equation}
where we should keep the first--order terms in $c_n$ in order not
to violate the conservation law.

The simplest zero order approximation for the steady state gives
\begin{equation}\label{CycleLimZero}
c_n = b,  \; c_i = 0\; (i \neq n).
\end{equation}
This is trivial: all the concentration is collected at the starting
point of the ``narrow place", but may be useful as an origin point
for various approximation procedures.

So, the stationary rate of a cycle is determined by the smallest
constant, $k_{\min}$, if $k_{\min}$ is sufficiently small:
\begin{equation}\label{limitation}
w=k_{\min} b \; \; \; \mbox {if} \;\; \sum_{ k_i \neq
k_{\min}}\frac{k_{\min}}{k_i} \ll 1.
\end{equation}
In that case we say that the cycle has a limiting step with
constant $k_{\min}$.

\subsection{Dynamical limitation in a cycle}
If ${k_n}/{k_i}$ is  small for all $i<n$, then the kinetic behaviour
of the cycle is extremely simple: the coefficients matrix on the
right hand side of kinetic equation (\ref{kinCyc}) has one simple
zero eigenvalue that corresponds to the conservation law $\sum c_i =
b$ and $n-1$ nonzero eigenvalues
\begin{equation}\label{cycle spectra}
\lambda_i = - {k_i} + \delta_i \; (i<n).
\end{equation}
where $\delta_i \to 0$ when $\sum_{i<n}\frac{k_n}{k_i} \to 0$.

It is easy to demonstrate (\ref{cycle spectra}): let us exclude the
conservation law (the zero eigenvalue) $\sum c_i = b$ and use
independent coordinates $c_i$ ($i= 1,\ldots n-1$); $c_n = b -
\sum_{i<n} c_i$. In these coordinates the kinetic equation
(\ref{kinCyc}) has the form
\begin{equation}\label{kinCycLim}
\dot{c} = K c - k_n A c + k_n b e^1
\end{equation}
where $c$ is the vector--column with components $c_i$ $(i<n)$, $K$
is the lower triangle matrix with nonzero elements only in two
diagonals: $(K)_{ii} = - k_i$ $(i=1,\ldots n-1)$, $(K)_{i+1,\, i}=
k_i$ $(i=1,\ldots n-2)$ (this is the kinetic matrix for the linear
chain of $n-1$ reactions $A_1 \to A_2 \to ... A_n$); $A$ is the
matrix with nonzero elements only in the first row:
$(A)_{1i}\equiv 1$, $e^1$ is the first basis vector ($e^1_1=1$,
$e^1_i=0$ for $1<i<n$). After that, eq. (\ref{cycle spectra})
follows simply from continuous dependence of spectra on matrix.

The relaxation time of a stable linear system (\ref{kinCycLim}) is,
by definition, $$\tau = [\min \{Re (-\lambda_i)\, | \, i=1,\ldots
n-1\}]^{-1}.$$ For small $k_n$,
\begin{equation}
\tau \approx 1/k_{\tau}, \;\; k_{\tau}=\min \{k_i  \, | \,
i=1,\ldots n-1\}.
\end{equation}
In other words, $k_{\tau}$ is the second slowest rate constant:
$k_{\min} \leq k_{\tau}\leq ... $\ .

\subsection{Relaxation equation for a cycle rate}

A definition of the cycle rate is clear for steady states because
stationary rates of all elementary reactions in cycle coincide.
There is no common definition of the cycle rate for nonstationary
regimes. In practice, one of steps is the step of product release
(the ``final" step of the catalytic transformation), and we can
consider its rate as the rate of the cycle. Formally, we can take
any step and study relaxation of its rate to the common stationary
rate. The single relaxation time approximation gives for rate
$w_i$ of any step:
\begin{equation}\label{relappr}
\begin{split}
&\dot{w}_i = k_{\tau}(k_{\min} b-w_i); \\ & w_i(t) = k_{\min} b +
{\rm e}^ {-k_{\tau} t} (w_i(0) -k_{\min} b),
\end{split}
\end{equation}
where $k_{\min}$ is the limiting (the minimal) rate constant of the
cycle, $k_{\tau}$ is the second in order rate constant of the cycle.

So, for catalytic cycles with the limiting constant $k_{\min}$,
the relaxation time is also determined by one constant, but
another one. This is $k_{\tau}$, the second in order rate
constant. It should be stressed that the only smallness condition
is required, $k_{\min}$ should be much smaller than other
constants. The second constant, $k_{\tau}$  should be just smaller
than others (and bigger than $k_{\min}$), but there is no $\ll$
condition for $k_{\tau}$ required.

One of the methods for measurement of chemical reaction constants
is the relaxation spectroscopy (\cite{Eigen}). Relaxation of a
system after an impact gives us a relaxation time or even a
spectrum of relaxation times. For catalytic cycle with limitation,
the relaxation experiment gives us the second constant $k_{\tau}$,
while the measurement of stationary rate gives the smallest
constant, $k_{\min}$. This simple remark may be important for
relaxation spectroscopy of open system.

\subsection{Ensembles of cycles and robustness of stationary rate
and relaxation time}

Let us consider a catalytic cycle with random rate constants. For
a given sample constants $k_1, \ldots k_n$ the  $i$th order
statistics is equal its $i$th-smallest value.  We are interested
in the first order (the minimal) and the second order statistics.

For independent identically distributed constants the variance of
$k_{\min}= \min \{k_1, \ldots k_n\}$ is significantly smaller then
the variance of each $k_i$, ${\rm Var} (k)$. The same is true for
statistic of every order. For many important distributions (for
example, for uniform distribution), the variance of $i$th order
statistic is of order $\sim {\rm Var} (k) / n^2$. For big $n$ it
goes to zero faster than variance of the mean that is of order
$\sim {\rm Var} (k) / n$. To illustrate this, let us consider $n$
constants  distributed in  interval $[a,b]$. For each set of
constants, $k_1, \ldots k_n$ we introduce ``symmetric coordinates"
$s_i$: first, we order the constants, $a \leq k_{i_1}\leq
k_{i_2}\leq \ldots k_{i_n}\leq b$, then calculate $s_0=k_{i_1}-a$,
$s_j =k_{i_{j+1}}-k_{i_{j}}$   $(j=1,\ldots n-1)$,
$s_n=b-k_{i_n}$. Transformation $(k_1, \ldots k_n)\mapsto (s_0,
\ldots s_{n})$ maps a cube $[a,b]^n$ onto n-dimensional simplex
$\Delta_n = \{ (s_0, \ldots s_{n})\, | \,\sum_i s_i = b-a\}$ and
uniform distribution on a cube transforms into uniform
distribution on a simplex.

For large $n$, almost all volume of the simplex is concentrated in a
small neighborhood of its center and this effect is an example of
measure concentration effects that play important role in modern
geometry and analysis (\cite{Gromov99}). All $s_i$ are identically
distributed, and for normalized variable $s=s_i/(b-a)$ the first
moments are: $\mathbf{E}(s)=1/(n+1)=1/n +o(1/n)$,
$\mathbf{E}(s^2)=2/[(n+1)(n+2)]=2/n^2 +o(1/n^2)$,
\begin{equation*}
\begin{split}
\mbox{Var}(s)= & \mathbf{E}(s^2)-(\mathbf{E}(s))^2 \\
= & \frac{n}{(n+1)^2(n+2)}=
\frac{1}{n^2}+o\left(\frac{1}{n^2}\right).
\end{split}
\end{equation*}
 Hence, for example,
$\mbox{Var}(k_{\min})=(b-a)^2/n^2 + o({1}/{n^2})$. The standard
deviation of $k_{\min}$ goes to zero as $1/n$ when $n$ increases.
This is much faster than $1/\sqrt{n}$ prescribed to the deviation of
the mean value of independent observation (the ``law of errors").
The same asymptotic $\sim 1/n$ is true for the standard deviation of
the second constant also. These parameters fluctuate much less than
individual constants, and even less than mean constant (for more
examples with applications to statistical physics we address to the
paper by \cite{Order}).

It is impossible to use this observation for cycles  with
limitation directly, because the inequality of limitation
(\ref{limitation}) is not true for uniform distribution. According
to this inequality, ratios  $k_i / k_{\min}$ should be
sufficiently small (if $k_i \neq k_{\min}$). To provide this
inequality we need to use at least the log-uniform distribution:
$k_i=\exp \Delta_i$ and $\Delta_i$ are independent variables
uniformly distributed in interval $[\alpha,\beta]$ with
sufficiently big $(\beta-\alpha)/n$.

One can interpret  the log-uniform distribution through the
Arrhenius law: $k=A \exp(- \Delta G /kT)$, where $\Delta G$ is the
change of the Gibbs free energy inreaction (it includes both
energetic and entropic terms: $\Delta G=\Delta H - T\Delta S$,
where $\Delta H$ is enthalpy change and $\Delta S$ is entropy
change in reaction, $T$ is temperature). The log-uniform
distribution of $k$ corresponds to the uniform distribution of
$\Delta G$.

For log-uniform distribution of constants $k_1, \ldots$ $k_n$, if
the interval of distribution is sufficiently big (i.e.
$(\beta-\alpha)/n \gg 1$), then the cycle with these constants has
the limiting step with probability close to one. More precisely we
can show that for any two constants $k_i,k_j$ the probability
$\Proba{k_i/k_j>r \; \text{or}\; k_j/k_i>r} = (1- \log(r)/(\beta -
\alpha))^2$ approaches one for any fixed $r>1$ when $\beta - \alpha
\to \infty$. Relaxation time of this cycle is determined by the
second constant $k_{\tau}$ (also with probability close to one).
Standard deviations of $k_{\min}$ and $k_{\tau}$ are much smaller
than standard deviation of single constant $k_i$ and even smaller
than standard deviation of mean constant $\sum_i k_i /n$. This
effect of  stationary rate and relaxation time robustness seems to
be important for understanding robustness in biochemical networks:
behaviour of the entire system is much more stable than the
parameters of its parts; even for large fluctuations of parameters,
the system does not change significantly the stationary rate
(statics) and the relaxation time (dynamics).

\subsection{Systems with well separated constants and monotone
relaxation}

The log-uniform identical distribution of independent constants
$k_1, \ldots k_n$ with sufficiently big interval of distribution
($(\beta-\alpha)/n \gg 1$) gives us the first example of ensembles
with well separated constants: any two constants are connected by
relation $\gg$ or $\ll$ with probability close to one. Such
systems (not only cycles, but much more complex networks too)
could be studied analytically ``up to the end".

Some of their properties are simpler than for general networks.
For example, the damping oscillations are impossible, i.e. the
eigenvalues of kinetic matrix are real (with probability close to
one). If constants are not separated, damped oscillations could
exist, for example, if all constants of the cycle are equal,
$k_1=k_2=\ldots =k_n=k$, then $(1+\lambda/k)^{n}=1$ and $\lambda_m
= k (\exp(2\pi i m/n)-1)$ $(m=1,\ldots n-1)$, the case $m=0$
corresponds to the linear conservation law. Relaxation time of
this cycle may be relatively big: $\tau =
\frac{1}{k}(1-\cos(2\pi/n))^{-1} \sim n^2/(2 \pi k)$ (for big
$n$).

The catalytic cycle without limitation can have relaxation time
much bigger then $1/k_{\min}$, where $k_{\min}$ is the minimal
reaction rate constant. For example, if all $k$ are equal, then
for $n=11$ we get $\tau \approx 20/k$. In more detail the possible
relations between $\tau$ and the slowest constant were discussed
by \cite{YabCher}. In that paper, a variety of cases with
different relationships between the steady-state reaction rate and
relaxation was presented.

For catalytic cycle, if a matrix $K - k_n A$ (\ref{kinCycLim}) has
a pair of complex eigenvalues with nonzero imaginary part, then
for some $g \in [0,1]$ the matrix $K - g k_n A$ has a degenerate
eigenvalue (we use a simple continuity argument). With probability
close to one, $k_{\min} \ll |k_i-k_j|$ for any two $k_i, \, k_j$
that are not minimal. Hence, the $k_{\min}$-small perturbation
cannot transform matrix $K$ with eigenvalues $k_i$ (\ref{cycle
spectra}) and given structure  into a matrix with a degenerate
eigenvalue. For proof of this statement it is sufficient to refer
to diagonal dominance of $K$ (the absolute value of each diagonal
element is greater than the sum of the absolute values of the
other elements in its column) and classical inequalities.

The matrix elements of $A$ in the eigenbasis of $K$ are $(A)_{ij}
= l^i  A r^j$. From obtained estimates for eigenvectors we get
$|(A)_{ij}| \lesssim 1$ (with probability close to one). This
estimate does not depend on values of kinetic constants. Now, we
can apply the Gershgorin theorem (see, for example, the review of
\cite{MM} and for more details the book of \cite{Varga}) to the
matrix $K- k_n A$ in the eigenbasis of $K$: the characteristic
roots of $K- k_n A$ belong to discs $|z+k_i|\leq k_n R_i(A)$,
where $R_i(A)=\sum_j |(A)_{ij}|$. If the discs do not intersect,
then each of them contains one and only one characteristic number.
For ensembles with well separated constants these discs do not
intersect (with probability close to one). Complex conjugate
eigenvalues could not belong to different discs. In this case, the
eigenvalues are real -- there exist no damped oscillations.

\subsection{Limitation by two steps with comparable constants}

If we consider  one-parametric families of systems, then
appearance of systems with two comparable constants may be
unavoidable. Let us imagine a continuous path $k_i(s)$ ($s \in
[0,1]$, $s$ is a parameter along the path) in the space of
systems, which goes from one system with well separated constants
($s=0$) to another such system ($s=1$). On this path $k_i(s)$ such
a point $s$ that $k_i(s)=k_j(s)$ may exist, and this existence may
be stable, that is, such a point persists under continuous
perturbations. This means that on a path there may be points where
not all the constants are well separated, and trajectories of some
constants may intersect.

For catalytic cycle, we are interested in the following
intersection only: $k_{\min}$ and the second constant are of the
same order, and are much smaller than other constants. Let these
constants be $k_j$ and $k_l$, $j \neq l$. The limitation condition
is
\begin{equation}\label{limit2}
 \frac{1}{k_j} + \frac{1}{k_l} \gg \sum_{i \neq j,l}
 \frac{1}{k_i}.
\end{equation}
 The steady state reaction rate and relaxation time are
determined by these two constants. In that case their effects  are
coupled. For the steady state we get in first order approximation
instead of (\ref{CycleLimRateLin}):
\begin{equation}\label{CycleLimRateLin2}
\begin{split}
&w= \frac{k_j k_l}{k_j + k_l} b; \; \; c_i = \frac{w}{k_i}=
\frac{b}{k_i} \frac{k_j k_l}{k_j + k_l}\; (i\neq j,l);\\ &c_j
=\frac{bk_l}{k_j + k_l}\left(1-\sum_{i \neq j,l}\frac{1}{k_i}
\frac{k_j k_l}{k_j + k_l}\right); \\ &c_l =\frac{bk_j}{k_j +
k_l}\left(1-\sum_{i \neq j,l}\frac{1}{k_i} \frac{k_j k_l}{k_j +
k_l}\right).
\end{split}
\end{equation}
Elementary analysis shows that under the limitation condition
(\ref{limit2}) the relaxation time is
\begin{equation}\label{RelaxCyc2}
\tau= \frac{1}{k_j+k_l}.
\end{equation}

The single relaxation time approximation for all elementary
reaction rates in a cycle with two limiting reactions is
\begin{equation}\label{relappr2}
\begin{split}
&\dot{w}_i = {k_j k_l}b - ({k_j+k_l})w_i; \\& w_i(t) = \frac{k_j
k_l}{k_j + k_l}  b + {\rm e}^ {-({k_j+k_l}) t} \left(w_i(0)
-\frac{k_j k_l}{k_j + k_l} b\right).
\end{split}
\end{equation}

The catalytic cycle with two limiting reactions has the same
stationary rate $w$ (\ref{CycleLimRateLin2}) and relaxation time
(\ref{RelaxCyc2}) as a reversible reaction $A\leftrightarrow B$
with $k^+=k_j$, $k^-=k_l$.

In two-parametric families three constants can meet. If three
smallest constants $k_j,k_l,k_m$ have comparable values and are much
smaller than others, then static and dynamic properties would be
determined by these three constants. Stationary rate $w$ and dynamic
of relaxation for the whole cycle would be the same as for
3-reaction cycle $A \rightarrow B \rightarrow C \rightarrow A$ with
constants $k_j,k_l,k_m$. The damped oscillation here are possible,
for example, if $k_j=k_l=k_m=k$, then there are complex eigenvalues
$\lambda=k (-\frac{3}{2}\pm i \frac{\sqrt{3}}{2})$. Therefore, if a
cycle manifests damped oscillation, then at least three slowest
constants are of the same order. The same is true, of course, for
more general reaction networks.

In $N$-parametric families of systems $N+1$ smallest constants can
meet, and near such a ``meeting point" a slow auxiliary cycle of
$N+1$ reactions determines behaviour of the entire cycle.

\subsection{Irreversible cycle with one inverse reaction
\label{subsecCycle+1}}

In this subsection, we represent a simple example that gives the
key to most of subsequent constructions of ``cycles surgery". Let
us add an inverse reaction to the irreversible cycle: $A_1 \to ...
\to A_i \leftrightarrow A_{i+1} \to... \to A_n \to A_1$. We use
the previous notation $k_1,... k_n$ for the cycle reactions, and
$k_i^-$ for the inverse reaction $A_i \leftarrow A_{i+1}$. For
well-separated constants, influence of $k_i^-$ on the whole
reaction is determined by relations of three constants: $k_i$,
$k_i^- $ and $k_{i+1}$. First of all, if $k_i^- \ll k_{i+1}$ then
in the main order there is no such influence, and dynamic of the
cycle is the same as for completely irreversible cycle.

If the opposite inequality is true,  $k_i^- \gg k_{i+1}$, then
equilibration between $A_i$ and $A_{i+1}$ gives $k_i c_i x\approx
k_i^- c_{i+1}$. If we introduce a lumped component $A^1_i$ with
concentration $c_i^1=c_i+c_{i+1}$, then $c_i\approx k_i^-
c_i^1/(k_i + k_i^-)$ and $c_{i+1}\approx k_i c_i^1/(k_i + k_i^-)$.
Using  this component instead of the pair $A_i, A_{i+1}$ we can
consider an irreversible cycle with $n-1$ components and $n$
reactions $A_1 \to ... \to A_{i-1} \to A_i^1 \to A_{i+2} \to...
\to A_n \to A_1$. To estimate the reaction rate constant $k^1_i$
for a new reaction, $A_i^1 \to A_{i+2}$, let us mention that the
correspondent reaction rate should be $k_{i+1}c_{i+1} \approx
k_{i+1}k_i c_i^1/(k_i + k_i^-)$. Hence, $$k^1_i \approx
k_{i+1}k_i/(k_i + k_i^-).$$ For systems with well separated
constants this expression can be simplified: if $k_i \gg k_i^-$
then $k^1_i \approx k_{i+1}$ and if $k_i \ll k_i^-$ then $k^1_i
\approx k_{i+1}k_i/ k_i^-$.  The first case, $k_i \gg k_i^-$ is
limitation in the small cycle (of length two) $A_i \leftrightarrow
A_{i+1}$ by the inverse reaction $A_i \leftarrow A_{i+1}$. The
second case, $k_i \ll k_i^-$, means the the direct reaction is the
limiting step in this small cycle.

In order to estimate eigenvectors, we can, after identification of
the limiting step in the small cycle, delete this step and
reattach the outgoing reaction to the beginning of this step. For
the first case, $k_i \gg k_i^-$, we get the irreversible cycle,
$A_1 \to ... \to A_i \to A_{i+1} \to... \to A_n \to A_1$, with the
same reaction rate constants. For the second case, $k_i \ll k_i^-$
we get a new system of reactions: a shortened cycle $A_1 \to ...
\to A_i \to A_{i+2} \to... \to A_n \to A_1$ and an ``appendix"
$A_{i+1} \to A_i$. For the new elementary reaction $A_{i} \to
A_{i+2}$ the reaction  rate constant is $k^1_i \approx k_{i+1}k_i/
k_i^-$. All other elementary reactions have the same rate
constants, as they have in the initial system. After deletion of
the limiting step from the ``big cycle" $A_1 \to ... \to A_i \to
A_{i+2} \to... \to A_n \to A_1$, we get an acyclic system that
approximate relaxation of the initial system.

So, influence of a single inverse reaction on the irreversible
catalytic cycle with well-separated constants is determined by
relations of three constants: $k_i$, $k_i^- $ and $k_{i+1}$. If
$k_i^- $ is much smaller than at least one of $k_i$, $k_{i+1}$,
then there is no influence in the main order. If $k_i^- \gg k_i$
and $k_i^- \gg k_{i+1}$ then the relaxation of the initial cycle
can be approximated by relaxation of the auxiliary acyclic system.

Asymptotic equivalence (for $k_i^- \gg k_i, \,
k_{i+1}$) of the reaction network $A_i \leftrightarrow A_{i+1} \to
A_{i+2}$ with rate constants $k_i$, $k_i^-$ and $k_{i+1}$ to the
reaction network $A_{i+1} \to A_i \to A_{i+2}$ with rate constants
$k_i^-$ (for the reaction $A_{i+1} \to A_i$) and $k_{i+1} k_i
/k_i^-$ (for the reaction $A_i \to A_{i+2}$) is simple, but
slightly surprising fact. The kinetic matrix for the first network
in coordinates $c_i, \, c_{i+1}, \, c_{i+2}$ is $$
K=\left[\begin{array}{ccc} -k_i & k_i^-& 0 \\ k_i &
-(k_i^-+k_{i+1}) & 0
\\  0   & k_{i+1} & 0 \\  \end{array}\right]
$$ The eigenvalues are 0 and
\begin{equation*}
\begin{split}
& \lambda_{1,2}=\frac{1}{2}\bigg[-(k_i+k_i^-+k_{i+1}) \\ & \pm
\sqrt{(k_i+k_i^-+k_{i+1})^2-4k_ik_{i+1}}\; \bigg],
\end{split}
\end{equation*}
 $\lambda_1=-k_{i+1}
k_i(1+o(1)) /k_i^-$, $\lambda_2=-k_i^-(1+o(1))$, where $o(1) \ll 1$.
Right eigenvector $r^0$ for zero eigenvalue is $(0,0,1)$ (we write
vector columns in rows). For $\lambda_1$ the eigenvector is
$r^1=(1,0,-1)+o(1)$, and for $\lambda_2$ it is $r^2=(1,-1,0)+o(1)$.
For the linear chain of reactions, $A_{i+1} \to A_i \to A_{i+2}$,
with rate constants $k_i^-$ and $k_{i+1} k_i /k_i^-$ eigenvalues are
$-k_i^-$ and $-k_{i+1} k_i /k_i^-$. These values approximate
eigenvalues of the initial system with small relative error. The
linear chain has the same zero-one asymptotic of the correspondent
eigenvectors.

This construction, a small cycle inside a big system, a quasi
steady state in the small cycle, and deletion of the limiting step
with reattaching of reactions (see Fig.~\ref{CycleSurg} below)
appears in this paper many times in much general settings. The
uniform estimates that we need for approximation of eigenvalues
and eigenvectors by these procedures are proven in Appendices.

\section{Multiscale ensembles and finite--additive distributions \label{sec3}}

\subsection{Ensembles with well separated constants, formal
approach}

In previous section, ensembles with well separated constants
appear. We represented them by a log-uniform distribution in a
sufficiently big interval $\log k \in [\alpha, \beta]$, but we
were not interested in most of probability distribution
properties, and  did not use them. The only property we really
used is: if $k_i > k_j$, then $k_i / k_j \gg 1$ (with probability
close to one). It means that we can assume that $k_i / k_j \gg a$
for any preassigned value of $a$ that does not depend on $k$
values. One can interpret this property as an asymptotic one for
$\alpha \to -\infty$, $\beta \to \infty$.

That property allows us to simplify algebraic formulas. For
example, $k_i+k_j$ could be substituted by $\max\{k_i,k_j\}$ (with
small relative error), or $$\frac{a k_i + b k_j}{ck_i+dk_j}
  \approx \left\{
    \begin{aligned}
        & a/c, \; &\mbox{if} \; k_i \gg k_{j};
        \\
        & b/d, \;&  \mbox{if} \; k_i \ll k_{j},
    \end{aligned}\right.$$
for nonzero $a,b,c,d$ (see, for example, (\ref{firstMultiL})).

Of course, some ambiguity can be introduced, for example, what is
it, $(k_1+k_2)-k_1$, if $k_1 \gg k_2$? If we first simplify the
expression in brackets, it is zero, but if we open brackets without
simplification, it is $k_2$. This is a standard difficulty in use of
relative errors for round-off. If we estimate the error in the final
answer, and then simplify, we shall avoid this difficulty. Use of
$o$ and $\mathcal{O}$ symbols also helps to control the error
qualitatively: if $k_1 \gg k_2$, then we can write
$(k_1+k_2)=k_1(1+o(1))$, and $k_1(1+o(1))-k_1=k_1o(1)$. The last
expression is neither zero, nor absolutely small -- it is just
relatively small with respect to $k_1$.

The formal approach is: for any ordering of rate constants, we use
relations $\gg$ and $\ll$, and assume that $k_i / k_j \gg a$ for
any preassigned value of $a$ that does not depend on $k$ values.
This approach allows us to perform asymptotic analysis of reaction
networks. A special version of this approach consists of group
ordering: constants are separated on several groups, inside groups
they are comparable, and between groups there are relations $\gg$
or $\ll$. An example of such group ordering was discussed at the
end of previous section (several limiting constants in a cycle).

\subsection{Probability approach: finite additive measures}

The asymptotic analysis of multiscale systems for log-uniform
distribution of independent constants on an interval $\log k \in
[\alpha,\beta]$ ($-\alpha,\beta \to \infty$) is possible, but
parameters $\alpha,\beta$ do not present in any answer, they just
should be sufficiently big. A natural question arises, what is the
limit? It is a log-uniform distribution on a line, or, for $n$
independent identically distributed constants, a log-uniform
distribution on $\mathbb{R}^n$).

It is well known that the uniform distribution on $\mathbb{R}^n$
is impossible: if a cube has positive probability $\epsilon>0$
(i.e. the distribution has positive density) then the union of $N
> 1/\epsilon$ such disjoint  cubes has probability bigger than 1
(here we use the finite--additivity of probability). This is
impossible. But if that cube has probability zero, then the whole
space has also zero probability, because it can be covered by
countable family of the cube translation. Hence, translation
invariance and $\sigma$-additivity (countable additivity) are in
contradiction (if we have no doubt about probability
normalization).

Nevertheless, there exists  finite--additive probability which is
invariant with respect to Euclidean group $E(n)$ (generated by
rotations and translations). Its values are densities of sets.

Let $\lambda$ be the Lebesgue measure in $\mathbb{R}^n$ and $D
\subset \mathbb{R}^n$ be a Lebesgue measurable subset. Density of
$D$ is the limit (if it exists):
\begin{equation}\label{density}
\rho(D) = \lim_{r \to \infty} \frac{\lambda (D \cap
\mathbb{B}^n_r)}{\lambda (\mathbb{B}^n_r)},
\end{equation}
where $\mathbb{B}^n_r$ is a ball with radius $r$ and centre at
origin. Density of $\mathbb{R}^n$ is 1, density of every
half--space is 1/2, density of bounded set is zero, density of a
cone is its solid angle (measured as a sphere surface fractional
area). Density (\ref{density})  and translation and rotational
invariant. It is finite-additive: if densities $\rho(D)$ and
$\rho(H)$ (\ref{density}) exist and $D \cap H =\varnothing$ then
$\rho(D \cup H)$ exists and $\rho(D \cup H) = \rho(D) + \rho(H)$.

Every polyhedron has a density. A polyhedron could be defined as
the union of a finite number of convex polyhedra. A convex
polyhedron is  the intersection of a finite number of half-spaces.
It may be bounded or unbounded. The family of polyhedra is closed
with respect to union, intersection and subtraction of sets. For
our goals, polyhedra form sufficiently rich class. It is important
that in definition of polyhedron {\it finite} intersections and
unions are used. If one uses countable unions, he gets too many
sets including all open sets, because open convex polyhedra (or
just cubes with rational vertices) form a basis of standard
topology.

Of course, not every  measurable set has density. If it is
necessary, we can use the Hahn--Banach theorem \cite{Rudin} and
study extensions $\rho_{\rm Ex}$ of $\rho$ with the following
property:
$$\underline{\rho}(D) \leq \rho_{\rm Ex}(D) \leq
\overline{\rho}(D),$$ where
\begin{equation*}
\begin{split}
&\underline{\rho}(D) = \lim_{r  \to \infty}\! \! \inf \frac{\lambda
(D \cap \mathbb{B}^n_r)}{\lambda (\mathbb{B}^n_r)}, \\
 &\overline{\rho}(D) = \lim_{r \to \infty} \! \! \sup
\frac{\lambda (D \cap \mathbb{B}^n_r)}{\lambda (\mathbb{B}^n_r)}.
\end{split}
\end{equation*}

 Functionals $\underline{\rho}(D)$ and $\overline{\rho}(D)$ are
 defined for all measurable $D$. We should stress that such
 extensions are not unique. Extension of density
(\ref{density}) using the Hahn--Banach theorem for picking up a
random integer was used in a very recent work by \cite{Adam}.

One of the most important concepts of any probability theory is
the conditional probability. In the density--based approach we can
introduce the conditional density. If densities $\rho(D)$ and
$\rho(H)$ (\ref{density}) exist, $\rho(H) \neq 0$ and the
following limit $\rho(D|H)$ exists, then we call it conditional
density:
\begin{equation}\label{conditden}
\rho(D|H)= \lim_{r \to \infty} \frac{\lambda (D \cap H \cap
\mathbb{B}^n_r)}{\lambda (H \cap \mathbb{B}^n_r)}.
\end{equation}
For polyhedra the situation is similar to usual probability
theory: densities $\rho(D)$ and $\rho(H)$ always exist  and if
$\rho(H) \neq 0$ then conditional density exists too. For general
measurable sets the situation is not so simple, and existence of
$\rho(D)$ and $\rho(H)\neq 0$ does not guarantee existence of
$\rho(D|H)$.

On a line, convex polyhedra are just intervals, finite or
infinite.  The probability defined on polyhedra is: for finite
intervals and their finite unions it is zero, for half--lines $x>
\alpha$ or $x < \alpha$ it is  1/2, and for the whole line
$\mathbb{R}$ the probability is 1. If one takes a set of positive
probability and adds or subtracts a zero--probability set, the
probability does not change.

If independent random variables $x$ and $y$ are  uniformly
distributed on a line, then their linear combination  $z = \alpha
x + \beta y$ is also uniformly distributed on a line. (Indeed,
vector $(x,y)$ is uniformly distributed on a plane (by
definition), a set $z > \gamma$ is  a half-plane, the
correspondent probability is 1/2.) This is a simple, but useful
stability property. We shall use this result in the following
form. If independent random variables $k_1, \ldots k_n$ are
log-uniformly distributed on a line, then the monomial
$\prod_{i=1}^n k_i^{\alpha_i}$ for real ${\alpha_i}$ is also
log-uniformly distributed on a line, if some of ${\alpha_i}\neq
0$.

\subsection{Carroll's obtuse problem and paradoxes of conditioning}

Lewis Carroll's Pillow Problem \#58 (\cite{Carroll}): ``Three
points are taken at random on an infinite plane. Find the chance
of their being the vertices of an obtuse--angled triangle."

A random triangle on an infinite plane is presented by a point
equidistributed in $\mathbb{R}^6$.  Due to the density -- based
definition, we should take and calculate the density of the set of
obtuse--angled triangles in  $\mathbb{R}^6$. This is equivalent to
the problem: find a fraction of the sphere $\mathbb{S}^5 \subset
\mathbb{R}^6$ that corresponds to obtuse--angled triangles. Just
integrate... \ . But there remains a problem. Vertices of triangle
are independent. Let us use the standard logic for discussion of
independent trials: we take the first point $A$ at random, then
the second point $B$, and then  the third point $C$. Let us draw
the first side $AB$. Immediately we find that for almost all
positions of the the third point $C$ the triangle is
obtuse--angled (\cite{Guy}). L. Carroll proposed to take another
condition: let $AB$ be the longest side and let $C$ be uniformly
distributed in the allowed area. The answer then is easy -- just a
ratio of areas of two simple figures. But there are absolutely no
reasons for uniformity of $C$ distribution. And it is more
important that the absolutely standard reasoning for independently
chosen points gives another answer than could be found on the base
of joint distribution. Why these approaches are in disagreement
now? Because there is no classical Fubini theorem for our
finite--additive probabilities, and we cannot easily transfer from
a multiple integral to a repeated one.

There exists a much simpler example. Let $x$ and $y$ be independent
positive real number. This means that vector $(x,y)$ is uniformly
distributed in the first quadrant. What is probability that $x\geq
y$? Following the definition of probability based on the density of
sets, we take the correspondent angle and find immediately that this
probability is 1/2. This meets our intuition well. But let us take
the first number $x$ and look for possible values of $y$. The
result: for given $x$ the second number $y$ is uniformly distributed
on $[0, \infty)$, and only a finite interval $[0,x]$ corresponds to
$x\geq y$. For the infinite rest we have $x< y$. Hence, $x< y$ with
probability 1. This is nonsense because of symmetry. So, for our
finite--additive measure we cannot use repeated integrals (or, may
be, should use them in a very peculiar manner).

\subsection{Law of total probability and orderings}

For polyhedra, there appear no conditioning problems. The law of
total probabilities holds: if $\mathbb{R}^n=\cup_{i=1}^m H_i$,
$H_i$ are polyhedra, $\rho(H_i)>0$, $\rho(H_i \cap H_j)=0$ for
$i\neq j$, and $D \subset \mathbb{R}^n$ is a polyhedron, then
\begin{equation}\label{LTOTPROB}
\rho(D)=\sum_{i=1}^m \rho(D \cap H_i) = \sum_{i=1}^m \rho(D | H_i)
\rho(H_i).
\end{equation}
Our basic example of multiscale ensemble is log-uniform
distribution of reaction constants in $\mathbb{R}^n_+$ ($\log k_i$
are independent and uniformly distributed on the line). For every
ordering $k_{j_1}>k_{j_2}> \ldots > k_{j_n}$ a polyhedral cone
$H_{{j_1}{j_2}\ldots {j_n}}$ in $\mathbb{R}^n$ is defined. These
cones have equal probabilities $\rho(H_{{j_1}{j_2}\ldots {j_n}})=
1/n!$ and probability of intersection of cones for different
orderings is zero. Hence, we can apply the law of total
probability (\ref{LTOTPROB}). This means that we can study every
event $D$ conditionally, for different orderings, and then combine
the results of these studies in the final answer (\ref{LTOTPROB}).

For example, if we study a simple cycle then formula
(\ref{CycleLimRateLin}) for steady state is valid with any given
accuracy with probability one for any ordering with the given
minimal element $k_n$.

For cycle with given ordering of constants we can find zero-one
approximation of left and right eigenvectors (\ref{01cyrcleAsy}).
This approximation is valid with any given accuracy for this
ordering with probability one.

If we consider sufficiently wide log-uniform distribution of
constants on a bounded interval instead of the infinite axis then
these statements are true with probability close to 1.

For general system that we study below the situation is slightly
more complicated: new terms, auxiliary reactions with  monomial
rate constants $k_{\varsigma}=\prod_i k_i^{\varsigma_i}$ could
appear with integer (but not necessary positive) $\varsigma_i$,
and we should include these $k_{\varsigma}$ in ordering. It
follows from stability property that these monomials are
log-uniform distributed on infinite interval, if $k_i$ are.
Therefore the situation seems to be similar to ordering of
constants, but there is a significant difference: monomials are
not independent, they depend on $k_i$ with $\varsigma_i \neq 0$.

Happily, in the forthcoming analysis when we include auxiliary
reactions with constant $k_{\varsigma}$, we always exclude at
least one of reactions with rate constant $k_i$ and $\varsigma_i
\neq 0$. Hence, for we always can use the following statement (for
the new list of constants, or for the old one): if
$k_{j_1}>k_{j_2}> \ldots > k_{j_n}$ then $k_{j_1}\gg k_{j_2}\gg
\ldots \gg  k_{j_n}$, where $a \gg b$ for positive $a,b$ means:
for any given $\varepsilon >0$ the inequality $\varepsilon a > b$
holds with unite probability.

If we use sufficiently wide but finite log-uniform distribution
then $\varepsilon$ could not be arbitrarily small (this depends on
the interval with), and probability is not unite but close to one.
For given $\varepsilon >0$ probability tends to one when the
interval width goes to infinity. It is important that we use only
finite number of auxiliary reactions with monomial constants, and
this number is bounded from above for given number of elementary
reactions. For completeness, we should mention here general
algebraic theory of orderings that is necessary in more
sophisticated cases (\cite{orderRobb,orderPfi}).

Finally, what is a most general form of multiscale ensemble? Our
basic example of is the log-uniform distribution of reaction
constants: $\log k_i$  ($i=1,…n$) are independent and uniformly
distributed on the line. Of course, the main candidate to the
general definition of multiscale ensemble is {\it a distribution
which is absolutely continuous with respect to that log-uniform
distribution}. If $\mu$ and $\nu$ are measures on the same algebra
of sets, then $\mu$ is said to be {\it absolutely continuous} with
respect to $\nu$ if $\mu(A) = 0$ for every set $A$ for which
$\nu(A) = 0$. It is written as ``$\mu \ll \nu$": $$\mu \ll \nu
\iff \left( \nu(A) = 0 \implies \mu (A) = 0 \right).$$

For example, let us take an event (an infinite polyhedron in
logarithmic coordinates) $U\in R^n$ with nonzero probability in
this distribution. The conditional probability distribution under
condition $U$ is an  multiscale ensemble. We have such an ensemble
for each polyhedral cone $U$ with non-empty interior in the space
of logarithms of constants.

\section{Relaxation of multiscale networks and hierarchy of
auxiliary discrete dynamical systems \label{sec4}}

\subsection{Definitions, notations and auxiliary results}

\subsubsection{Notations}

In this Sec., we consider a general network of linear
(monomolecular) reactions. This network is represented as a
directed graph (digraph): vertices correspond to components $A_i$,
edges correspond to reactions $A_i \to A_j$ with kinetic constants
$k_{ji} > 0$. For each vertex, $A_i$, a positive real variable
$c_i$ (concentration) is defined. A basis vector $e^i$ corresponds
to $A_i$ with components $e^i_j=\delta_{ij}$, where $\delta_{ij}$
is the Kronecker delta.  The kinetic equation for the system is
\begin{equation}\label{kinur}
\frac{\D c_i}{\D t}=\sum_j (k_{ij} c_j - k_{ji}c_i),
\end{equation}
or in vector form: $\dot{c} = Kc$.

To write another form of (\ref{kinur}) we use stoichiometric
vectors: for a reaction $A_i \to A_j$ the stoichiometric vector
$\gamma_{ji}$ is a vector in concentration space with $i$th
coordinate $-1$, $j$th coordinate 1, and zero other coordinates.
The reaction rate $w_{ji}=k_{ji}c_i$. The kinetic equation has the
form
\begin{equation}\label{kinur1}
\frac{\D c}{\D t}=\sum_{i,j} w_{ji} \gamma_{ji},
\end{equation}
where $c$ is the concentration vector. One more form of
(\ref{kinur}) describes  directly dynamics of reaction rates:
\begin{equation}\label{kinur2}
\frac{\D w_{ji}}{\D t}\left(= k_{ji}\frac{\D c_i}{\D t}\right) =
k_{ji} \sum_{l}( w_{il} - w_{li}).
\end{equation}
It is necessary to mention that, in general, system (\ref{kinur2})
is not equivalent to (\ref{kinur1}), because there are additional
connections between variables $w_{ji}$. If there exists at least
one $A_i$ with two different outgoing reactions, $A_i \to A_j$ and
$A_i \to A_l$ ($j \neq l$), then $w_{ji}/w_{li}\equiv
k_{ji}/k_{li}$. If the reaction network generates a discrete
dynamical system $A_i \to A_j$ on the set of $A_i$ (see below),
then the variables $w_{ji}$ are independent, and (\ref{kinur2})
gives equivalent representation of kinetics.

For analysis of kinetic systems, linear conservation laws and
positively invariant polyhedra are important. A linear
conservation law is a linear function defined on the
concentrations $b(c)= \sum_{i} b_i c_i$, whose value is preserved
by the dynamics \eqref{kinur}. The  conservation laws coefficient
vectors $b_i$ are left eigenvectors of the matrix $K$
corresponding to the zero eigenvalue. The set of all the
conservation laws forms the left kernel of the matrix $K$.
Equation \eqref{kinur} always has a linear conservation law:
$b^0(c)=\sum_i c_i = {\rm const}$. If there is no other
independent linear conservation law, then the system is {\it
weakly ergodic}.

A set $E$ is {\it positively invariant} with respect to kinetic
equations (\ref{kinur}), if any solution $c(t)$ that starts in $E$
at time $t_0$ ($c(t_0) \in E$) belongs to $E$ for $t>t_0$ ($c(t)
\in E$ if $t>t_0$). It is straightforward to check that the
standard simplex $\Sigma = \{c\,| \, c_i \geq 0, \, \sum_i c_i
=1\}$ is positively invariant set for kinetic equation
(\ref{kinur}): just to check that if  $c_i=0$ for some $i$, and
all $c_j \geq 0$ then $\dot{c}_i \geq 0$. This simple fact
immediately implies the following properties of ${K}$:
\begin{itemize}
\item{All eigenvalues $\lambda$ of ${K}$ have non-positive real
parts, $Re \lambda \leq 0$, because solutions cannot leave
$\Sigma$ in positive time;}
\item{If $Re \lambda = 0$ then $\lambda = 0$, because intersection of
$\Sigma$ with any plain is a polygon, and a polygon cannot be
invariant with respect of rotations to sufficiently small angles;}
\item{The Jordan cell of ${K}$ that corresponds to zero eigenvalue
is diagonal -- because all solutions should be bounded in $\Sigma$
for positive time.} \item{The shift in time operator $\exp({K} t)$
is a contraction in the $l_1$ norm for $t>0$: for positive $t$ and
any two solutions of (\ref{kinur}) $c(t), c'(t) \in \Sigma$
$$\sum_i |c_i(t) -c'_i(t)| \leq \sum_i |c_i(0) -c'_i(0)|.$$}
\end{itemize}

Two vertices are called adjacent if they share a common edge. A
path is a sequence of adjacent vertices. A graph is connected if
any two of its vertices are linked by a path. A maximal connected
subgraph of graph $G$ is called a connected component of $G$.
Every graph can be decomposed into connected components.

A directed path is a sequence of adjacent edges where each step
goes in direction of an edge. A vertex $A$ is {\it reachable} by a
vertex $B$, if there exists an oriented path from $B$ to $A$.

A nonempty set $V$ of graph vertexes forms a {\it sink}, if there
are no oriented edges from $A_i \in V$ to any $A_j \notin V$. For
example, in the reaction graph $A_1\leftarrow A_2 \rightarrow A_3$
the one-vertex sets $\{A_1\}$ and $\{A_3\}$ are sinks. A sink is
minimal if it does not contain a strictly smaller sink. In the
previous example, $\{A_1\}$, $\{A_3\}$  are minimal sinks. Minimal
sinks are also called ergodic components.

A digraph is strongly connected, if every  vertex $A$ is reachable
by any other vertex $B$. Ergodic components are maximal strongly
connected subgraphs of the graph, but inverse is not true: there
may exist maximal strongly connected subgraphs that have outgoing
edges and, therefore, are not sinks.

We study ensembles of systems with a given graph and independent
and well separated  kinetic constants $k_{ij}$. This means that we
study asymptotic behaviour of ensembles with  independent
identically distributed constants, log-uniform distributed in
sufficiently big interval $\log k \in [\alpha, \beta]$, for
$\alpha \to -\infty$, $\beta \to \infty$, or just a log-uniform
distribution on infinite axis, $\log k \in \mathbb{R}$.

\subsubsection{Sinks and ergodicity}

If there is no other independent linear conservation law, then the
system is weakly ergodic. The weak ergodicity of the network
follows from its topological properties.

The following properties are equivalent and each one of them can
be used as an alternative definition of weak ergodicity:
\begin{enumerate}
\item{There exist the only independent linear conservation law for
kinetic equations (\ref{kinur}) (this is $b^0(c)=\sum_i c_i = {\rm
const}$).}
 \item{For any normalized initial state $c(0)$ ($b^0(c)=1$)
there exists a limit state $$c^*= \lim_{t\rightarrow \infty }
\exp(Kt) \, c(0)$$ that is the same for all normalized initial
conditions: For all $c$, $$\lim_{t\rightarrow \infty } \exp(Kt) \, c
= b^0(c)  c^*.$$}
\item{For each two vertices  $A_i, \: A_j \: (i
\neq j)$ we can find such a vertex $A_k$ that is reachable both by
$A_i$ and by $A_j$. This means that the following structure
exists:
 $$A_i \to \ldots \to A_k \leftarrow \ldots \leftarrow A_j.$$
One of the paths can be degenerated: it may be $i=k$ or $j=k$.}
\item{The network has only one minimal sink (one ergodic
component).}
\end{enumerate}

For every monomolecular kinetic system, the Jordan cell for zero
eigenvalue of matrix $K$ is diagonal and the maximal number of
independent linear conservation laws (i.e. the geometric
multiplicity of the zero eigenvalue of the matrix $K$) is equal to
the maximal number of disjoint ergodic components (minimal sinks).

Let $G= \{A_{i_1}, \ldots A_{i_l} \}$ be an ergodic component. Then
there exists a unique vector (normalized invariant distribution)
$c^G$ with the following properties: $c^G_i= 0$ for $i \notin \{i_1,
\ldots i_l\}$, $c^G_i > 0$ for all $i \in \{i_1, \ldots i_l\}$,
$b^0(c^G)=1$, $Kc^G =0$.

If $G_1, \ldots G_m$ are all ergodic components of the system,
then there exist $m$ independent positive linear functionals
$b^1(c)$, ...  $b^m(c)$ such that $\sum_{i=1}^m b^i = b^0$ and for
each $c$
\begin{equation}\label{TimeEv}
\lim_{t \to \infty} \exp(Kt) c = \sum_{i=1}^m b^i(c) c^{G_i}.
\end{equation}
 So, for any solution of kinetic equations (\ref{kinur}), $c(t)$, the
limit at $t \to \infty$ is a linear combination of normalized
invariant distributions $c^{G_i}$ with coefficients $b^i(c(0))$.
In the simplest example, $A_1\leftarrow A_2 \to A_3$, $G_1=
\{A_1\}$, $G_2= \{A_3\}$, components of vectors $c^{G_1}$,
$c^{G_2}$ are $(1,0,0)$ and $(0,0,1)$, correspondingly. For
functionals $b^{1,2}$ we get:
\begin{equation}\label{massDistri}
b^1(c)=c_1+\frac{k_1}{k_1 + k_2}c_2; \; \; b^2(c)=\frac{k_2}{k_1 +
k_2}c_2 + c_3,
\end{equation}
 where $k_1, k_2$ are rate constants for reaction $A_2 \to A_1$, and
 $A_2 \to A_3$, correspondingly. We can mention that for
 well separated constants either $k_1 \gg k_2$ or $k_1 \ll k_2$.
 Hence, one of the coefficients ${k_1}/({k_1 + k_2})$, ${k_2}/({k_1 +
k_2})$ is close to 0, another is close to 1. This is an example of
the general zero--one law for multiscale systems: for any $l, i$,
the value of functional $b^l$ (\ref{TimeEv}) on basis vector
$e^i$, $b^l (e^i)$, is either close to one or close to zero (with
probability close to 1).

We can understand better this asymptotics by using the Markov chain
language. For non-separated constants a particle in $A_2$ has
nonzero probability to reach $A_1$ and nonzero probability to reach
$A_3$. The zero--one law in this simplest case means that the
dynamics of the particle becomes deterministic: with probability one
it chooses to go to one of vertices $A_2, \, A_3$ and to avoid
another. Instead of branching, $A_2 \to A_1$ and
 $A_2 \to A_3$, we select only one way: either $A_2 \to A_1$ or
$A_2 \to A_3$. Graphs without branching represent discrete dynamical
systems.

\subsubsection{Decomposition of discrete dynamical systems}

Discrete dynamical system on a finite set $V= \{A_1,A_2, \ldots
A_n \}$ is a semigroup $1, \phi, \phi^2, ...$, where $\phi$ is a
map $\phi: V \to V$.  $A_i \in V$ is a periodic point, if $\phi^l
(A_i)=A_i$ for some $l>0$; else $A_i$ is a transient point. A
cycle of period $l$ is a sequence of $l$ distinct periodic points
$A, \phi(A), \phi^2(A), \ldots \phi^{l-1}(A)$ with $\phi^{l}(A) =
A$. A cycle of period one consists of one fixed point,
$\phi(A)=A$. Two cycles, $C, C'$ either coincide or have empty
intersection.

The set of periodic points, $V^{\rm p}$, is always nonempty. It is
a union of cycles: $V^{\rm p}= \cup_j C_j$. For each point $A \in
V$ there exist such a positive integer $\tau(A)$ and a cycle
$C(A)=C_j$ that $\phi^q(A) \in C_j$ for $q \geq \tau(A)$. In that
case we say that $A$ belongs to basin of attraction of cycle $C_j$
and use notation $Att(C_j)= \{A\ | \ C(A) = C_j \}$. Of course,
$C_j \subset Att(C_j)$. For different cycles, $Att(C_j) \cap
Att(C_l) = \varnothing$. If $A$ is periodic point then
$\tau(A)=0$. For transient points $\tau(A)>0$.

So, the phase space $V$ is divided onto subsets $Att(C_j)$. Each
of these subsets includes one cycle (or a fixed point, that is a
cycle of length 1). Sets $Att(C_j)$ are $\phi$-invariant:
$\phi(Att(C_j)) \subset Att(C_j) $. The set $Att(C_j) \setminus
C_j$ consist of transient points and there exists such positive
integer $\tau$ that $\phi^q(Att(C_j))=C_j$ if $q \geq \tau$.

\subsection{Auxiliary discrete dynamical systems and relaxation
analysis}

\subsubsection{Auxiliary discrete dynamical system}

For each $A_i$, we define $\kappa_i$ as the maximal kinetic
constant for reactions $A_i \to A_j$: $\kappa_i =\max_j
\{k_{ji}\}$. For correspondent $j$ we use notation $\phi(i)$:
$\phi(i)={\rm arg \,max}_j \{k_{ji}\}$. The function $\phi(i)$ is
defined under condition that for $A_i$ outgoing reactions $A_i \to
A_j$ exist. Let us extend the definition: $\phi(i)=i$ if there
exist no such outgoing reactions.

The map $\phi$ determines discrete dynamical system on a set of
components $V=\{A_i\}$. We call it the auxiliary discrete
dynamical system for a given network of monomolecular reactions.
Let us decompose this system and find the cycles $C_j$ and their
basins of attraction, $Att(C_j)$.

Notice that for the graph that represents a discrete dynamic
system,  attractors are ergodic components, while basins are
connected components.

An auxiliary reaction network is associated with the auxiliary
discrete dynamical system. This is the set of reactions $A_i \to
A_{\phi(i)}$ with kinetic constants $\kappa_i$. The correspondent
kinetic equation is
\begin{equation}\label{auxkinet}
\dot{c}_i =  -\kappa_i c_i + \sum_{\phi(j)=i} \kappa_j c_j,
\end{equation}
or in vector notations (\ref{kinur1})
\begin{equation}\label{kinurAux}
\frac{\D c}{\D t} = \tilde{K} c =\sum_{i} \kappa_i c_i
\gamma_{\phi(i)\, i}; \;\; \tilde{K}_{ij}= - \kappa_j \delta_{ij}
+ \kappa_j \delta_{i \, \phi (j)}.
\end{equation}

For deriving of the auxiliary discrete dynamical system we do not
need the values of rate constants. Only the ordering is important.
Below we consider multiscale  ensembles of kinetic systems with
given ordering and with well separated kinetic constants
($k_{\sigma(1)}\gg  k_{\sigma(2)} \gg ...$ for some permutation
$\sigma$).

In the following, we analyze first the situation when the system
is connected and has only one attractor. This can be a point or a
cycle. Then, we discuss the general situation with any number of
attractors.

\subsubsection{Eigenvectors for acyclic auxiliary kinetics}

Let us study kinetics (\ref{auxkinet}) for acyclic discrete
dynamical system (each vertex has one or zero outgoing reactions,
and there are no cycles). Such acyclic reaction networks have many
simple properties. For example, the nonzero eigenvalues are
exactly minus reaction rate constants, and it is easy to find all
left and right eigenvectors in explicit form. Let us find left and
right eigenvectors of matrix $\tilde{K}$ of auxiliary kinetic
system (\ref{auxkinet}) for acyclic auxiliary dynamics. In this
case, for any vertex $A_i$ there exists is an eigenvector. If
$A_i$ is a fixed point of the discrete dynamical system (i.e.
$\phi(i)=i$) then this eigenvalue is zero. If $A_i$ is not a fixed
point (i.e. $\phi(i) \neq i$ and reaction $A_i \to A_{\phi(i)}$
has nonzero rate constant $\kappa_i$) then this eigenvector
corresponds to eigenvalue $-\kappa_i$. For left and right
eigenvectors of $\tilde{K}$ that correspond to $A_i$ we use
notations $l^i$ (vector-raw) and $r^i$ (vector-column),
correspondingly, and apply normalization condition
$r^i_i=l^i_i=1$.

First, let us find the eigenvectors for zero eigenvalue. Dimension
of zero eigenspace is equal to the number of fixed points in the
discrete dynamical system. If $A_i$ is a fixed point then the
correspondent eigenvalue is zero, and the right eigenvector $r^i$
has only one nonzero coordinate, concentration of $A_i$: $r^i_j =
\delta_{ij}$.

To construct the correspondent left eigenvectors $l^i$ for zero
eigenvalue (for fixed point $A_i$), let us mention that $l^i_j$
could have nonzero value only if there exists such $q\geq 0$ that
$\phi^q(j)=i$ (this $q$ is unique because absence of cycles). In
that case (for $q>0$), $$(l^i \tilde{K})_j=-\kappa_j l^i_j +
\kappa_j l^i_{\phi(j)}=0.$$ Hence, $l^i_j = l^i_{\phi(j)}$, and
$l^i_j=1$ if $\phi^q(j)=i$ for some $q>0$.

For nonzero eigenvalues, right eigenvectors will be constructed by
recurrence starting from the vertex $A_i$ and moving in the
direction of the flow. The construction is in opposite direction
for left eigenvectors. Nonzero eigenvalues of $\tilde{K}$
(\ref{auxkinet}) are $-\kappa_i$.

For given $i$, $\tau_i$ is the minimal integer such that
$\phi^{\tau_i}(i)=\phi^{\tau_i+1}(i)$ (this is a {\it relaxation
time} i.e. the number of steps to reach a fixed point). All indices
$\{\phi^k(i) \ | \ k=0,1,\ldots \tau_i \}$ are different. For right
eigenvector $r^i$ only coordinates $r^i_{\phi^k(i)}$ ($k=0,1,\ldots
\tau_i$) could have nonzero values, and
\begin{equation*}
\begin{split}
(\tilde{K}r^i)_{\phi^{k+1}(i)}&=-\kappa_{\phi^{k+1}(i)}r^i_{\phi^{k+1}(i)}+
\kappa_{\phi^{k}(i)}r^i_{\phi^k(i)}\\ & =-\kappa_i
r^i_{\phi^{k+1}(i)}.
\end{split}
\end{equation*}
Hence,
\begin{equation}\label{rightAcyc}
\begin{split}
r^i_{\phi^{k+1}(i)}=
\frac{\kappa_{\phi^{k}(i)}}{\kappa_{\phi^{k+1}(i)}-\kappa_i}
r^i_{\phi^{k}(i)}= \prod_{j=0}^k
\frac{\kappa_{\phi^{j}(i)}}{\kappa_{\phi^{j+1}(i)}-\kappa_i} \\
 = \frac{\kappa_i}{\kappa_{\phi^{k+1}(i)}-\kappa_i}
 \prod_{j=0}^{k-1}\frac{\kappa_{\phi^{j+1}(i)}}{\kappa_{\phi^{j+1}(i)}-\kappa_i}.
\end{split}
\end{equation}
The last transformation is convenient for estimation of the
product for well separated constants (compare to
(\ref{firstMultiL})):
\begin{equation}\label{right01}
\begin{split}
&\frac{\kappa_{\phi^{j+1}(i)}}{\kappa_{\phi^{j+1}(i)}-\kappa_i}
 \approx \left\{\begin{aligned} 1, \; &\mbox{if} \;
 \kappa_{\phi^{j+1}(i)}\gg
 \kappa_i,\\ 0, \; &\mbox{if} \; \kappa_{\phi^{j+1}(i)}\ll
 \kappa_i; \end{aligned} \right. \\
&\frac{\kappa_{i}}{\kappa_{\phi^{k+1}(i)}-\kappa_i}
 \approx \left\{\begin{aligned} -1, \; &\mbox{if} \;
 \kappa_{i}\gg
 \kappa_{\phi^{k+1}(i)},\\ 0, \; &\mbox{if} \; \kappa_{i}\ll
 \kappa_{\phi^{k+1}(i)}. \end{aligned} \right.
\end{split}
\end{equation}

For left eigenvector $l^i$ coordinate $l^i_j$  could have nonzero
value only if there exists such $q\geq 0$ that $\phi^q(j)=i$ (this
$q$ is unique because the auxiliary dynamical system has no
cycles). In that case (for $q>0$), $$(l^i \tilde{K})_j=-\kappa_j
l^i_j + \kappa_j l^i_{\phi(j)}=-\kappa_i l^i_j.$$ Hence,
\begin{equation}\label{leftAcyc}
l^i_j=\frac{\kappa_j}{\kappa_j - \kappa_i} l^i_{\phi(j)}=
\prod_{k=0}^{q-1} \frac{\kappa_{\phi^k(j)}}{\kappa_{\phi^k(j)} -
\kappa_i}.
\end{equation}
For every fraction in (\ref{leftAcyc}) the following estimate
holds:
\begin{equation}\label{left01}
\frac{\kappa_{\phi^k(j)}}{\kappa_{\phi^k(j)} - \kappa_i}
 \approx \left\{\begin{aligned} 1, \; &\mbox{if} \; \kappa_{\phi^k(j)}\gg
 \kappa_i, \\  0, \; &\mbox{if} \; \kappa_{\phi^k(j)}\ll
 \kappa_i. \end{aligned} \right.
\end{equation}

As we can see, every coordinate of left and right eigenvectors of
$\tilde{K}$ (\ref{rightAcyc}), (\ref{leftAcyc}) is either 0 or $\pm
1$, or close to 0 or to $\pm 1$ (with probability close to 1). We
can write this asymptotic representation explicitly (analogously to
(\ref{01cyrcleAsy})). For left eigenvectors, $l^i_i=1$ and $l^i_j=1$
(for $i\neq j$) if there exists such $q$ that $\phi^q(j)=i$, and
$\kappa_{\phi^d(j)}>\kappa_i$ for all $d= 0, \ldots q-1$, else
$l^i_j=0$. For right eigenvectors, $r^i_i=1$ and
$r^i_{\phi^k(i)}=-1$ if $\kappa_{\phi^k(i)}< \kappa_i$ and for all
positive $m<k$ inequality  $\kappa_{\phi^m(i)}> \kappa_i$ holds,
i.e. $k$ is first such positive integer that $\kappa_{\phi^k(i)}<
\kappa_i$ (for fixed point $A_p$ we use $\kappa_p=0$). Vector $r^i$
has not more than two nonzero coordinates. It is straightforward to
check that in this asymptotic $l^i r^j = \delta_{ij}$.

In general, coordinates of eigenvectors $l^i_j$, $r^i_j$ are
simultaneously nonzero only for one value $j=i$  because the
auxiliary system is acyclic. On the other hand, $l^i r^j= 0$ if $i
\neq j$, just because that are eigenvectors for different
eigenvalues, $\kappa_i$ and $\kappa_j$. Hence, $l^i r^j=
\delta_{ij}$.

For example, let us find the asymptotic of left and right
eigenvectors for a branched acyclic system of reactions:
\begin{equation*}
%\begin{split}
A_1 {\rightarrow^{\!\!\!\!\!\!7}}\,\,
A_2{\rightarrow^{\!\!\!\!\!\!5}}\,\,
A_3{\rightarrow^{\!\!\!\!\!\!6}}\,\,
A_4{\rightarrow^{\!\!\!\!\!\!2}}\,\,
A_5{\rightarrow^{\!\!\!\!\!\!4}}\,\, A_8, \;\;
A_6{\rightarrow^{\!\!\!\!\!\!1}}\,\,
A_7{\rightarrow^{\!\!\!\!\!\!3}}\,\,A_4
%\end{split}
\end{equation*}
 where  the upper index marks the order of rate constants:
 $\kappa_6>\kappa_4>\kappa_7>\kappa_5>\kappa_2>\kappa_3>\kappa_1$ ($\kappa_i$ is the
 rate constant of reaction $A_i \to ...$).

For zero eigenvalue, the left and right eigenvectors are $$l^8=
(1,1,1,1,1,1,1,1,1), \; r^8= (0,0,0,0,0,0,0,1).$$ For left
eigenvectors, rows $l^i$, that correspond to nonzero eigenvalues we
have the following asymptotics:
\begin{equation}\label{leftAcycBran}
\begin{split}
&l^1\approx (1,0,0,0,0,0,0,0), \; l^2\approx (0,1,0,0,0,0,0,0), \;
\\
&l^3\approx (0,1,1,0,0,0,0,0), l^4\approx (0,0,0,1,0,0,0,0), \; \\
&l^5\approx (0,0,0,1,1,1,1,0), \; l^6\approx (0,0,0,0,0,1,0,0). \\
&l^7\approx (0,0,0,0,0,1,1,0)
\end{split}
\end{equation}
For the correspondent right eigenvectors, columns $r^i$, we have
the following asymptotics (we write vector-columns in rows):
\begin{equation}
\begin{split}
&r^1 \!\approx \!
 (1, 0, 0, 0, 0, 0, 0, -1),  \,
r^2 \!\approx  \!
 (0, 1, -1, 0, 0, 0, 0, 0 ), \\
&r^3 \!\approx  \!
 (0, 0, 1,  0, 0, 0, 0, -1), \,
r^4 \!\approx  \!
 (0, 0, 0, 1,  -1,0 ,0 ,0 ), \\
&r^5 \!\approx  \!
 (0, 0, 0, 0, 1 ,0 ,0 , -1), \,
r^6 \!\approx  \!
 (0, 0, 0,  0, 0, 1, -1, 0),  \\
&r^7 \!\approx  \!
 (0, 0, 0, 0,  -1, 0 ,1, 0  ).
\end{split}
\end{equation}

\subsubsection{The first case: auxiliary dynamical system is
acyclic and has one attractor}

In the simplest case, the  auxiliary discrete dynamical system for
the reaction network $\mathcal{W}$ is acyclic and has only one
attractor, a fixed point. Let this point be $A_n$ ($n$ is the
number of vertices). The correspondent eigenvectors for zero
eigenvalue are $r^n_j=\delta_{nj}$, $l^n_j=1$. For such a system,
it is easy to find explicit analytic solution of kinetic equation
(\ref{auxkinet}).

Acyclic auxiliary dynamical system with one attractor have a
characteristic property among all auxiliary dynamical systems: the
stoichiometric vectors of reactions $A_i \to A_{\phi(i)}$ form a
basis in the subspace of concentration space with $\sum_i c_i =0$.
Indeed, for such a system there exist $n-1$ reactions, and their
stoichiometric vectors are independent. On the other hand, existence
of cycles implies linear connections between stoichiometric vectors,
and existence of two attractors in acyclic system implies that the
number of reactions is less $n-1$, and their stoichiometric vectors
could not form a basis in $n-1$-dimensional space.

Let us assume that the auxiliary dynamical system is acyclic and
has only one attractor, a fixed point. This means that
stoichiometric vectors $\gamma_{\phi(i)\, i}$ form a basis in a
subspace of concentration space with $\sum_i c_i =0$. For every
reaction $A_{i} \to A_l$ the following linear operators $Q_{i l}$
can be defined:
\begin{equation}\label{OprQ}
Q_{i l} (\gamma_{\phi(i)\, i}) = \gamma_{li}, \; Q_{i l}(
\gamma_{\phi(p)\, p}) = 0 \; \mbox{for} \; p \neq i.
\end{equation}
The kinetic equation for the whole reaction network (\ref{kinur1})
could be transformed in the form
\begin{equation}\label{kinurWholeAcy}
\begin{split}
\frac{\D c}{\D t}&=\sum_i \left(1+ \sum_{l, \, l \neq \phi(i)}
\frac{k_{li}}{\kappa_i} Q_{il}\right) \gamma_{\phi(i)\, i} \kappa_i
c_i\\  &= \left(1+ \ \sum_{j,l \, (l \neq \phi(j))}
\frac{k_{lj}}{\kappa_j} Q_{jl}\right) \sum_i  \gamma_{\phi(i)\, i}
\kappa_i c_i \\ &=\left(1+ \ \sum_{j,l \, (l \neq \phi(j))}
\frac{k_{lj}}{\kappa_j} Q_{jl}\right) \tilde{K}c,
\end{split}
\end{equation}
where $\tilde{K}$ is kinetic matrix of auxiliary kinetic equation
(\ref{kinurAux}). By construction of auxiliary dynamical system,
${k_{li}}\ll{\kappa_i}$ if ${l \neq \phi(i)}$. Notice also that
$|Q_{jl}|$ does not depend on rate constants.

Let us represent system (\ref{kinurWholeAcy}) in eigenbasis of
$\tilde{K}$ obtained in previous subsection. Any matrix $B$ in
this eigenbasis has the form $B=(\tilde{b}_{ij})$,
$\tilde{b}_{ij}= l^i B r^j= \sum_{qs}l^i_q b_{qs} r^j_s $, where
$(b_{qs})$ is matrix $B$ in the initial basis, $l^i$ and $r^j$ are
left and right eigenvectors of $\tilde{K}$ (\ref{rightAcyc}),
(\ref{leftAcyc}). In eigenbasis of $\tilde{K}$ the Gershgorin
estimates of eigenvalues and estimates of eigenvectors are much
more efficient than in original coordinates: the system is
stronger diagonally dominant. Transformation to this basis is an
effective preconditioning for perturbation theory that uses
auxiliary kinetics as a first approximation to the kinetics of the
whole system.

First of all, we can exclude the conservation law. Any solution of
(\ref{kinurWholeAcy}) has the form $c(t)=br^n+\tilde{c}(t)$, where
$b=l^n c(t)= l^n c(0)$ and  $\sum_i \tilde{c}_i(t)=0$. On  the
subspace of concentration space with $\sum_i c_i =0$ we get
\begin{equation}
\frac{\D c}{\D t}= (1+ \mathcal{E}){\rm diag}\{-\kappa_1, \ldots
-\kappa_{n-1}\} c,
\end{equation}
where $\mathcal{E}=(\varepsilon_{ij})$, $|\varepsilon_{ij}| \ll
1$, and ${\rm diag} \{-\kappa_1, \ldots -\kappa_{n-1}\}$ is
diagonal matrix with $-\kappa_1, \ldots -\kappa_{n-1}$ on the main
diagonal. If $|\varepsilon_{ij}| \ll 1$ then we can use the
Gershgorin theorem and state that eigenvalues of matrix $(1+
\mathcal{E}){\rm diag}\{-\kappa_1, \ldots -\kappa_{n-1}\}$ are
real and have the form $\lambda_i = -\kappa_i+\theta_i$ with
$|\theta_i| \ll \kappa_i$.

To prove inequality $|\varepsilon_{ij}| \ll 1$ (for ensembles with
well separated constants, with probability close to 1) we use that
the left and right eigenvectors of $\tilde{K}$ (\ref{rightAcyc}),
(\ref{leftAcyc}) are uniformly bounded under some non-degeneracy
conditions and those conditions are true for well separated
constants. For ensembles with well separated constants, for any
given positive $g <1$ and all $i,j$ ($i \neq j$) the following
inequality is true with probability close to 1: $|\kappa_i -
\kappa_j| > g \kappa_i$. Let us select a value of $g$ and assume
that this {\it diagonal gap condition} is always true. In this case,
for every fraction in (\ref{rightAcyc}), (\ref{leftAcyc}) we have
estimate $$\frac{\kappa_i}{|\kappa_j-\kappa_i|} < \frac{1}{g}.$$
Therefore, for coordinates of right and left eigenvectors of
$\tilde{K}$ (\ref{rightAcyc}), (\ref{leftAcyc}) we get
\begin{equation}
|r^i_{\phi^{k+1}(i)}| < \frac{1}{g^k}< \frac{1}{g^n}, \; |l^i_j| <
\frac{1}{g^q} < \frac{1}{g^n}.
\end{equation}

We can estimate $|\varepsilon_{ij}|$ and $|\theta_i|/\kappa_i$ from
above as $const \times \max_{l \neq \phi(s)}\{k_{ls}/\kappa_s\}$.
So, the eigenvalues for kinetic matrix of the whole system
(\ref{kinurWholeAcy}) are real and close to eigenvalues of auxiliary
kinetic matrix $\tilde{K}$ (\ref{kinurAux}). For eigenvectors, the
Gershgorin theorem gives no result, and additionally to diagonal
dominance we must assume the diagonal gap condition.  Based on this
assumption, we proved the Gershgorin type estimate of eigenvectors
in Appendix 1. In particular, according to this estimate,
eigenvectors for the whole reaction network are arbitrarily close to
eigenvectors of $\tilde{K}$ (with probability close to 1).

So, if the auxiliary discrete dynamical system is acyclic and has
only one attractor (a fixed point), then the relaxation of the
whole reaction network could be approximated by the auxiliary
kinetics (\ref{auxkinet}):
\begin{equation}\label{ApproxRelAux}
c(t)=(l^nc(0))r^n + \sum_{i=1}^{n-1} (l^i c(0))r^i \exp(-\kappa_i
t),
\end{equation}
For $l^i$ and $r^i$ one can use exact formulas (\ref{rightAcyc})
and (\ref{leftAcyc}) or zero-one asymptotic representations based
on (\ref{left01}) and (\ref{right01}) for multiscale systems.
This approximation (\ref{ApproxRelAux}) could be improved by
iterative methods, if necessary.

\subsubsection{The second case: auxiliary system has one cyclic attractor}

The second simple particular case on the way to general case is a
reaction network with components $A_1,\ldots A_n$  whose auxiliary
discrete dynamical system has one attractor, a cycle with period
$\tau >1$: $A_{n-\tau+1} \to A_{n-\tau+2} \to \ldots A_n \to
A_{n-\tau+1}$ (after some change of enumeration). We assume that
the limiting step in this cycle (reaction with minimal constant)
is $A_n \to A_{n-\tau+1}$. If auxiliary discrete dynamical system
has only one attractor then the whole network is weakly ergodic.
But the attractor of the auxiliary system may not coincide with a
sink of the reaction network.

There are two possibilities:
\begin{enumerate}
\item{In the whole network, all the outgoing reactions from the cycle
have the form $A_{n-\tau+i}\to A_{n-\tau+j}$ ($i,j>0$). This means
that the cycle vertices $A_{n-\tau+1}, A_{n-\tau+2}, \ldots A_n$
form a sink for the whole network.}
\item{There exists a reaction from a cycle vertex $A_{n-\tau+i}$ to
$A_m$, $m \leq n-\tau$. This means that the set $\{A_{n-\tau+1},
A_{n-\tau+2}, \ldots A_n\}$ is not a sink for the whole network.}
\end{enumerate}
In the first case, the limit (for $t \to \infty$) distribution for
the auxiliary kinetics is the well-studied  stationary
distribution of the cycle $A_{n-\tau+1}, A_{n-\tau+2}, \ldots A_n$
described in Sec.~\ref{sec2} (\ref{CycleRate}),
(\ref{CycleLimRate}) (\ref{CycleLimRateLin}), (\ref{limitation}).
The set $\{A_{n-\tau+1}, A_{n-\tau+2}, \ldots A_n\}$ is the only
ergodic component for the whole network too, and the limit
distribution for that system is nonzero on vertices only. The
stationary distribution for the cycle $A_{n-\tau+1} \to
A_{n-\tau+2} \to \ldots A_n \to A_{n-\tau+1}$ approximates the
stationary distribution for the whole system. To approximate the
relaxation process, let us delete the limiting step $A_n \to
A_{n-\tau+1}$  from this cycle. By this deletion we produce an
acyclic system with one fixed point, $A_n$, and auxiliary kinetic
equation (\ref{kinurAux}) transforms into
\begin{equation}\label{AckinurAux}
\frac{\D c}{\D t} = \tilde{K}_0 c =\sum_{i=1}^{n-1} \kappa_i c_i
\gamma_{\phi(i)\, i}.
\end{equation}
As it is demonstrated, dynamics of this system approximates
relaxation of the whole network in subspace $\sum_i c_i = 0$.
Eigenvalues for (\ref{AckinurAux}) are $-\kappa_i$ ($i < n$), the
corresponded eigenvectors are represented by (\ref{rightAcyc}),
(\ref{leftAcyc}) and zero-one multiscale asymptotic representation
is based on (\ref{left01}) and (\ref{right01}).

In the second case, the set $$\{A_{n-\tau+1}, A_{n-\tau+2}, \ldots
A_n\}$$ is not a sink for the whole network. This means that there
exist outgoing reactions from the cycle, $A_{n-\tau+i}\to A_j$ with
$A_j \notin \{A_{n-\tau+1}, A_{n-\tau+2}, \ldots A_n\}$. For every
cycle vertex $A_{n-\tau+i}$ the rate constant $\kappa_{n-\tau+i}$
that corresponds to the cycle reaction $A_{n-\tau+i} \to
A_{n-\tau+i+1}$ is much bigger than any other constant $k_{j, \,
n-\tau+i}$ that corresponds to a ``side" reaction $A_{n-\tau+i} \to
A_j$ ($j \neq {n-\tau+i+1}$): $\kappa_{n-\tau+i} \gg k_{j, \,
n-\tau+i}$. This is because definition of auxiliary discrete
dynamical system and assumption of ensemble with well separated
constants (multiscale asymptotics).  This inequality allows us to
separate motion and to use for computation of the rates of outgoing
reaction $A_{n-\tau+i}\to A_j$ the quasi steady state distribution
in the cycle.  This means that we can glue the cycle into one vertex
$A^1_{n-\tau+1}$ with the correspondent concentration
$c^1_{n-\tau+1}=\sum_{1 \leq i \leq \tau }c_{n-\tau+i}$ and
substitute the reaction $A_{n-\tau+i} \to A_j$ by $A^1_{n-\tau+1}
\to A_j$ with the rate constant renormalization: $k^{1}_{j, \,
{n-\tau+1}}=  k_{j, \, n-\tau+i} c_{n-\tau+i}^{\rm
QS}/c^1_{n-\tau+1}$. By the superscript QS we mark here the
quasistationary concentrations  for given total cycle concentration
$c^1_{n-\tau+1}$. Another possibility is to recharge the link
$A_{n-\tau+i} \to A_j$ to another vertex of the cycle (usually to
$A_n$): we can substitute the reaction $A_{n-\tau+i} \to A_j$ by the
reaction  $A_{n-\tau+q} \to A_j$ with the rate constant
renormalization:
\begin{equation}\label{renormrecharge}
k_{j, \,{n-\tau+q}}=k_{j, \, n-\tau+i}c_{n-\tau+i}^{\rm
QS}/c_{n-\tau+q}^{\rm QS} .
\end{equation}
The new rate constant is smaller than the initial one: $k_{j,
\,{n-\tau+q}}\leq k_{j, \, n-\tau+i}$, because $c_{n-\tau+i}^{\rm
QS}\leq c_{n-\tau+q}^{\rm QS}$ due to definition.

We apply this approach now and demonstrate its applicability in
more details later in this section. For the quasi-stationary
distribution on the cycle we get $c_{n-\tau+i} = c_n
\kappa_n/\kappa_{n-\tau+i}$ ($1 \leq i < \tau$). The original
reaction network is transformed by gluing the cycle
$\{A_{n-\tau+1}$, $A_{n-\tau+2}, \ldots$ $A_n\}$ into a point
$A^1_{n-\tau+1}$. We say that components $A_{n-\tau+1}$,
$A_{n-\tau+2}, \ldots$ $A_n$ of the original system belong to the
component $A^1_{n-\tau+1}$ of the new system. All the reactions
$A_i \to A_j$ with $i,j \leq n-\tau$ remain the same with rate
constant $k_{ji}$. Reactions of the form $A_i \to A_j$ with $i
\leq n-\tau$, $j > n-\tau$ (incoming reactions  of the cycle
$\{A_{n-\tau+1}, A_{n-\tau+2}, \ldots A_n\}$) transform into $A_i
\to A^1_{n-\tau+1}$ with the same rate constant $k_{ji}$.
Reactions of the form $A_i \to A_j$ with $i > n-\tau $, $j\leq
n-\tau $ (outgoing reactions  of the cycle $\{A_{n-\tau+1},
A_{n-\tau+2}, \ldots A_n\}$) transform into reactions
$A^1_{n-\tau+1} \to A_j$ with the ``quasistationary" rate constant
$k^{\rm QS}_{j i} = k_{ji} \kappa_n/\kappa_{n-\tau+i}$. After
that, we select the maximal $k^{\rm QS}_{j i}$ for given $j$:
$k^{(1)}_{j, \,n-\tau+1} = \max_{i > n-\tau}k^{\rm QS}_{j i}$.
This $k^{(1)}_{j, \,n-\tau+1}$ is the rate constant for reaction
$A^1_{n-\tau+1} \to A_j$ in the new system. Reactions $A_i \to
A_j$ with $i,j > n-\tau $ (internal reactions of the site) vanish.

Among rate constants for reactions of the form $A_{n-\tau+i} \to
A_m$ ($m \geq n-\tau$) we find
\begin{equation}\label{kappamax}
\kappa^{(1)}_{n-\tau+i}=\max_{i,m}\{k_{m, \,
n-\tau+i}\kappa_n/\kappa_{n-\tau+i}\}.
\end{equation}
Let the correspondent $i,m$ be $i_1,\,m_1$.

After that, we create a new auxiliary discrete dynamical system
for the new reaction network on the set $\{A_1, \ldots A_{n-\tau},
A^1_{n-\tau+1} \}$. We can describe this new auxiliary system as a
result of transformation of the first auxiliary discrete dynamical
system of initial reaction network. All the reactions from this
first auxiliary system of the form $A_i \to A_j$ with $i,j \leq
n-\tau$ remain the same with rate constant $\kappa_i$. Reactions
of the form $A_i \to A_j$ with $i \leq n-\tau$, $j > n-\tau$
transform into $A_i \to A^1_{n-\tau+1}$ with the same rate
constant $\kappa_i$. One more reaction is to be added:
$A^1_{n-\tau+1} \to A_{m_1}$ with rate constant
$\kappa^{(1)}_{n-\tau+i}$. We ``glued" the cycle into one vertex,
$A^1_{n-\tau+1}$, and added new reaction from this vertex to
$A_{m_1}$ with maximal possible constant (\ref{kappamax}). Without
this reaction the new auxiliary dynamical system has only one
attractor, the fixed point $A^1_{n-\tau+1}$. With this additional
reaction that point is not fixed, and a new cycle appears:
$A_{m_1} \to\ldots A^1_{n-\tau+1} \to A_{m_1}$.

Again we should analyze, whether this new cycle is a sink in the
new reaction network, etc. Finally, after a chain of
transformations, we should  come to an auxiliary discrete
dynamical system with one attractor, a cycle, that is the sink of
the transformed whole reaction network. After that, we can find
stationary distribution by restoring of glued cycles in auxiliary
kinetic system and applying formulas (\ref{CycleRate}),
(\ref{CycleLimRate}) (\ref{CycleLimRateLin}), (\ref{limitation})
from Sec.~\ref{sec2}. First, we find the stationary state of the
cycle constructed on the last iteration, after that for each
vertex $A^k_j$ that is a glued cycle we know its concentration
(the sum of all concentrations)  and can find the stationary
distribution, then if there remain some vertices that are glued
cycles we find distribution of concentrations in these cycles,
etc. At the end of this process we find all stationary
concentrations with high accuracy, with probability close to one.

\begin{figure}[t]
\centering{ \includegraphics[width=70mm]{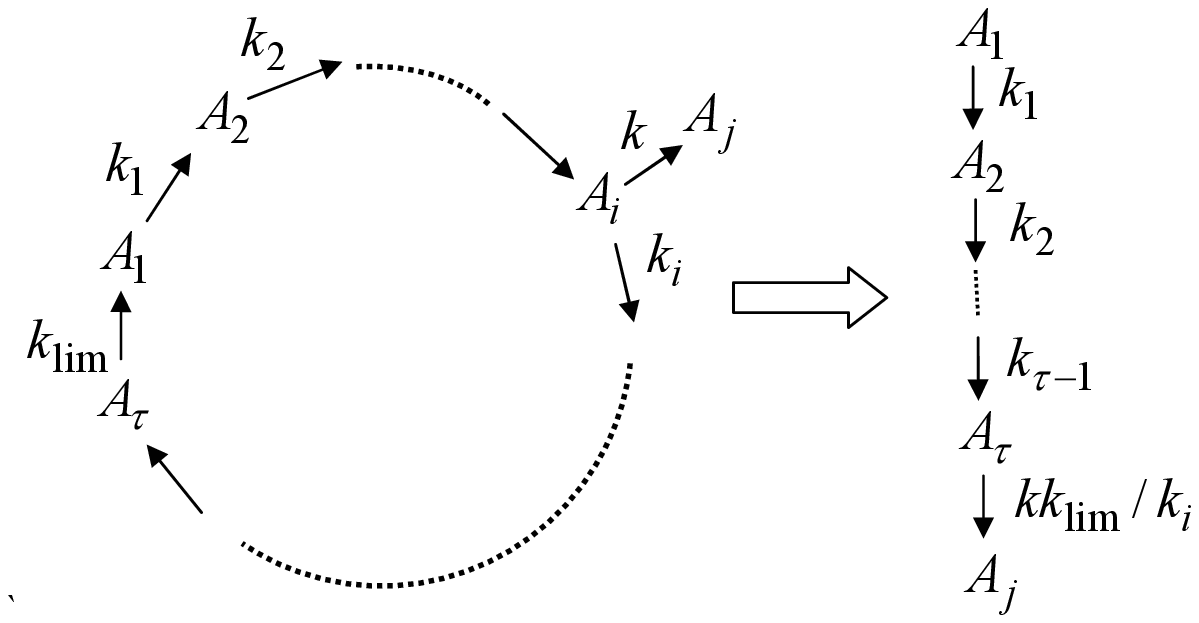} }
\caption{\label{CycleSurg} The main operation of the cycle surgery:
on a step back we get a cycle $A_{1} \to... \to A_{\tau} \to A_{1}$
with the limiting step $A_{\tau} \to A_{1}$ and one outgoing
reaction $A_{i} \to A_j$. We should delete the limiting step,
reattach (``recharge") the outgoing reaction $A_{i} \to A_j$ from
$A_i$ to $A_{\tau}$ and change its rate constant $k$ to the rate
constant $k k_{\lim}/k_{i}$. The new value of reaction rate constant
is always smaller than the initial one: $k k_{\lim}/k_{i} < k$ if
$k_{\lim}\neq k_{i}$. For this operation only one condition $k \ll
k_i$ is necessary ($k$ should be small with respect to reaction
$A_{i}\to A_{i+1}$ rate constant, and can exceed any other reaction
rate constant).}
\end{figure}

As a simple example we use the following system, a chain
supplemented by three reactions:
\begin{equation}\label{chain+}
\begin{split}
 A_1 {\rightarrow^{\!\!\!\!\!\!1}}\,\,
A_2{\rightarrow^{\!\!\!\!\!\!2}}\,\,
A_3{\rightarrow^{\!\!\!\!\!\!3}}\,\,
A_4{\rightarrow^{\!\!\!\!\!\!4}}\,\,
A_5{\rightarrow^{\!\!\!\!\!\!5}}\,\, A_6, \\
A_6{\rightarrow^{\!\!\!\!\!\!6}}\,\, A_4, \;\;
A_5{\rightarrow^{\!\!\!\!\!\!7}}\,\, A_2, \;\;
A_3{\rightarrow^{\!\!\!\!\!\!8}}\,\, A_1,
\end{split}
\end{equation}
where  the upper index marks the order of rate constants.

Auxiliary discrete dynamical system for the network (\ref{chain+})
includes the chain and one reaction: $$A_1
{\rightarrow^{\!\!\!\!\!\!1}}\,\,
A_2{\rightarrow^{\!\!\!\!\!\!2}}\,\,
A_3{\rightarrow^{\!\!\!\!\!\!3}}\,\,
A_4{\rightarrow^{\!\!\!\!\!\!4}}\,\,
A_5{\rightarrow^{\!\!\!\!\!\!5}}\,\,
A_6{\rightarrow^{\!\!\!\!\!\!6}}\,\, A_4.$$ It has one attractor,
the cycle $A_4{\rightarrow^{\!\!\!\!\!\!4}}\,\,
A_5{\rightarrow^{\!\!\!\!\!\!5}}\,\,
A_6{\rightarrow^{\!\!\!\!\!\!6}}\,\, A_4$. This cycle is not a
sink for the whole system, because there exists an outgoing
reaction $A_5{\rightarrow^{\!\!\!\!\!\!7}}\,\, A_2$.

By gluing the cycle $A_4{\rightarrow^{\!\!\!\!\!\!4}}\,\,
A_5{\rightarrow^{\!\!\!\!\!\!5}}\,\,
A_6{\rightarrow^{\!\!\!\!\!\!6}}\,\, A_4$ into a vertex $A^1_4$ we
get new network with a chain supplemented by two reactions:
\begin{equation}\label{transnet}
A_1 {\rightarrow^{\!\!\!\!\!\!1}}\,\,
A_2{\rightarrow^{\!\!\!\!\!\!2}}\,\,
A_3{\rightarrow^{\!\!\!\!\!\!3}}\,\, A_4^1, \;\; A^1_4
{\rightarrow^{\!\!\!\!\!\!?}}\,\, A_2, \;\; A_3
{\rightarrow^{\!\!\!\!\!\!?}}\,\, A_1.
\end{equation}
Here the new rate constant is $k^{(1)}_{2 4}=k_{2 5}
\kappa_6/\kappa_5$ ($\kappa_6=k_{46}$ is the limiting step of the
cycle $A_4{\rightarrow^{\!\!\!\!\!\!4}}\,\,
A_5{\rightarrow^{\!\!\!\!\!\!5}}\,\,
A_6{\rightarrow^{\!\!\!\!\!\!6}}\,\, A_4$, $\kappa_5=k_{65}$).

Here we can make a simple but important observation: the new
constant $k^1_{2 4}=k_{2 5} \kappa_6/\kappa_5$ has the same
log-uniform distribution on the whole axis as  constants $k_{2
5}$, $\kappa_6$ and $\kappa_5$ have. The new constant $k^1_{2 4}$
depends on $k_{2 5}$ and the internal cycle constants $\kappa_6$
and $\kappa_5$, and  is independent from other constants.

Of course, $k^{(1)}_{2 4} < \kappa_5$, but relations between
$k^{(1)}_{2 4}$ and $k_{13}$ are a priori unknown. Both orderings,
$k^{(1)}_{2 4}
> k_{13}$ and $k^{(1)}_{2 4} < k_{13}$, are possible,  and should be
considered separately, if necessary. But for both orderings the
auxiliary dynamical system for network (\ref{transnet}) is
 $$A_1 {\rightarrow^{\!\!\!\!\!\!1}}\,\,
A_2{\rightarrow^{\!\!\!\!\!\!2}}\,\,
A_3{\rightarrow^{\!\!\!\!\!\!3}}\,\, A_4^1
{\rightarrow^{\!\!\!\!\!\!?}}\,\, A_2$$ (of course,
$\kappa^{(1)}_4 < \kappa_3<\kappa_2<\kappa_1$). It has one
attractor, the cycle $A_2{\rightarrow^{\!\!\!\!\!\!2}}\,\,
A_3{\rightarrow^{\!\!\!\!\!\!3}}\,\,
A^1_4{\rightarrow^{\!\!\!\!\!\!?}}\,\, A_2$. This cycle is not a
sink for the whole system, because there exists an outgoing
reaction $A_3 {\rightarrow^{\!\!\!\!\!\!?}}\,\, A_1$. The limiting
constant for this cycle is $\kappa^{(1)}_4=k^{(1)}_{24}=k_{2 5}
k_{46}/k_{65}$. We glue this cycle into one point, $A^2_2$. The
new transformed system is very simple, it is just a two step
cycle: $A_1 {\rightarrow^{\!\!\!\!\!\!1}}\,\,
A_2^2{\rightarrow^{\!\!\!\!\!\!?}}\,\, A_1$. The new reaction
constant  is $k^{(2)}_{12}=k_{13}\kappa^{(1)}_4/ \kappa_3=k_{13}
k_{2 5}k_{46}/(k_{65}k_{43})$. The auxiliary discrete dynamical
system is the same graph $A_1 {\rightarrow^{\!\!\!\!\!\!1}}\,\,
A_2^2{\rightarrow^{\!\!\!\!\!\!?}}\,\, A_1$, this is a cycle, and
we do not need further transformations.

Let us find the steady state on the way back, from this final
auxiliary system to the original one. For steady state of each
cycle we use formula (\ref{CycleLimRateLin}).

The steady state for the final system is
$c_1=bk^{(2)}_{12}/k_{21}$, $c^2_2=b(1-k^{(2)}_{12}/k_{21})$. The
component $A^2_2$ includes the cycle
$A_2{\rightarrow^{\!\!\!\!\!\!2}}\,\,
A_3{\rightarrow^{\!\!\!\!\!\!3}}\,\,
A^1_4{\rightarrow^{\!\!\!\!\!\!?}}\,\, A_2$. The steady state of
this cycle is $c_2=c^{(2)}_2 k^{(1)}_{24}/k_{32}$, $c_3=c^{(2)}_2
k^{(1)}_{24}/ k_{43}$,
$c^{(1)}_4=c^{(2)}_2(1-k^{(1)}_{24}/k_{32}-k^{(1)}_{24}/ k_{43})$.
The component $A^1_4$ includes the cycle
$A_4{\rightarrow^{\!\!\!\!\!\!4}}\,\,
A_5{\rightarrow^{\!\!\!\!\!\!5}}\,\,
A_6{\rightarrow^{\!\!\!\!\!\!6}}\,\, A_4$. The steady state of
this cycle is $c_4=c^{(1)}_4 k_{46}/k_{54}$, $c_5=c^{(1)}_4
k_{46}/k_{65}$, $c_6=c^{(1)}_4(1-k_{46}/k_{54}-k_{46}/k_{65})$.

 For one catalytic cycle, relaxation in the subspace $\sum_i c_i
=0$ is approximated by relaxation of a chain that is produced from
the cycle by cutting the limiting step (Sec.~\ref{sec2}). For
reaction networks under consideration (with one cyclic attractor
in auxiliary discrete dynamical system) the direct generalization
works: for approximation of relaxation in the subspace $\sum_i c_i
=0$ it is sufficient to perform the following procedures:
\begin{itemize}
\item{To glue iteratively attractors (cycles) of the auxiliary system
that are not sinks of the whole system;}
\item{To restore these cycles from the end of the first procedure to its
beginning. For each of cycles (including the last one that is a
sink) the limited step should be deleted, and the outgoing
reaction should be reattached to the head of the limiting steps
(with the proper normalization), if it was not deleted before as a
limiting step of one of the cycles.}
\end{itemize}

The heads of outgoing reactions of that cycles should be reattached
to the heads of the limiting steps. Let for a cycle this limiting
step be $A_m \to A_q$. If for a glued cycle $A^k$ there exists an
outgoing reaction $A^k \to A_j$ with the constant $\kappa$
(\ref{kappamax}), then after restoration we add the outgoing
reaction $A_m \to A_j$ with the rate constant $\kappa$. Kinetic of
the resulting acyclic system approximates relaxation of the initial
network (under assumption of well separated constants, for given
ordering, with probability close to 1).

Let us construct this acyclic network for  the same example
(\ref{chain+}). The final cycle is $A_1
{\rightarrow^{\!\!\!\!\!\!1}}\,\,
A_2^2{\rightarrow^{\!\!\!\!\!\!?}}\,\, A_1$. The limiting step in
this cycle is $A_2^2{\rightarrow^{\!\!\!\!\!\!?}}\,\, A_1$. After
cutting we get $A_1 {\rightarrow^{\!\!\!\!\!\!1}}\,\, A_2^2$. The
component $A_2^2$ is glued cycle
$A_2{\rightarrow^{\!\!\!\!\!\!2}}\,\,
A_3{\rightarrow^{\!\!\!\!\!\!3}}\,\,
A^1_4{\rightarrow^{\!\!\!\!\!\!?}}\,\, A_2$. The reaction $A_1
{\rightarrow^{\!\!\!\!\!\!1}}\,\, A_2^2$ corresponds to the
reaction $A_1 {\rightarrow^{\!\!\!\!\!\!1}}\,\, A_2$ (in this
case, this is the only reaction from $A_1$ to cycle; in other case
one should take the reaction from $A_1$ to cycle with maximal
constant). The limiting step in the cycle is
$A^1_4{\rightarrow^{\!\!\!\!\!\!?}}\,\, A_2$. After cutting, we
get a system $A_1 {\rightarrow^{\!\!\!\!\!\!1}}\,\,
A_2{\rightarrow^{\!\!\!\!\!\!2}}\,\,
A_3{\rightarrow^{\!\!\!\!\!\!3}}\,\, A^1_4$. The component $A^1_4$
is the glued cycle $A_4{\rightarrow^{\!\!\!\!\!\!4}}\,\,
A_5{\rightarrow^{\!\!\!\!\!\!5}}\,\,
A_6{\rightarrow^{\!\!\!\!\!\!6}}\,\, A_4$ from the previous step.
The limiting step in this cycle is
$A_6{\rightarrow^{\!\!\!\!\!\!6}}\,\, A_4$. After restoring this
cycle and cutting the limiting step, we get an acyclic system $A_1
{\rightarrow^{\!\!\!\!\!\!1}}\,\,
A_2{\rightarrow^{\!\!\!\!\!\!2}}\,\,
A_3{\rightarrow^{\!\!\!\!\!\!3}}\,\, A_4
{\rightarrow^{\!\!\!\!\!\!4}}\,\,
A_5{\rightarrow^{\!\!\!\!\!\!5}}\,\, A_6$ (as one can guess from
the beginning: this coincidence is provided by the simple constant
ordering selected in (\ref{chain+})). Relaxation of this system
approximates relaxation of the whole initial network.

To demonstrate possible branching of described algorithm for
cycles surgery (gluing, restoring and cutting) with necessity of
additional orderings, let us consider the following system:

\begin{equation}\label{chain+-}
A_1 {\rightarrow^{\!\!\!\!\!\!1}}\,\,
A_2{\rightarrow^{\!\!\!\!\!\!6}}\,\,
A_3{\rightarrow^{\!\!\!\!\!\!2}}\,\,
A_4{\rightarrow^{\!\!\!\!\!\!3}}\,\,
A_5{\rightarrow^{\!\!\!\!\!\!4}}\,\, A_3, \;\;
A_4{\rightarrow^{\!\!\!\!\!\!5}}\,\, A_2, \;\;
\end{equation}

The auxiliary discrete dynamical system for reaction network
(\ref{chain+-}) is $$A_1 {\rightarrow^{\!\!\!\!\!\!1}}\,\,
A_2{\rightarrow^{\!\!\!\!\!\!6}}\,\,
A_3{\rightarrow^{\!\!\!\!\!\!2}}\,\,
A_4{\rightarrow^{\!\!\!\!\!\!3}}\,\,
A_5{\rightarrow^{\!\!\!\!\!\!4}}\,\, A_3.$$
 It has only one attractor, a cycle
$A_3{\rightarrow^{\!\!\!\!\!\!2}}\,\,
A_4{\rightarrow^{\!\!\!\!\!\!3}}\,\,
A_5{\rightarrow^{\!\!\!\!\!\!4}}\,\, A_3$. This cycle is not a
sink for the whole network (\ref{chain+-}) because reaction
$A_4{\rightarrow^{\!\!\!\!\!\!5}}\,\, A_2$ leads from that cycle.
After gluing the cycle into a vertex $A^1_3$ we get the new
network $A_1 {\rightarrow^{\!\!\!\!\!\!1}}\,\,
A_2{\rightarrow^{\!\!\!\!\!\!6}}\,\,
A^1_3{\rightarrow^{\!\!\!\!\!\!?}}\,\,A_2$. The rate constant for
the reaction $A^1_3{\rightarrow}A_2$ is $k^1_{23}=
k_{24}k_{35}/k_{54}$, where $k_{ij}$ is the rate constant for the
reaction $A_j \to A_i$ in the initial network ($k_{35}$ is the
cycle limiting reaction). The new network coincides with its
auxiliary system and has one cycle,
$A_2{\rightarrow^{\!\!\!\!\!\!6}}\,\,
A^1_3{\rightarrow^{\!\!\!\!\!\!?}}\,\,A_2$. This cycle is a sink,
hence, we can start the back process of cycles restoring and
cutting. One question arises immediately: which constant is
smaller, $k_{32}$ or $k^1_{23}$. The smallest of them is the
limiting constant, and the answer depends on this choice. Let us
consider two possibilities separately: (1) $k_{32} > k^1_{23}$ and
(2) $k_{32} < k^1_{23}$. Of course, for any choice the stationary
concentration of the source component $A_1$ vanishes: $c_1=0$.

(1) Let as assume that  $k_{32} > k^1_{23}$. In this case, the
steady state of the cycle $A_2{\rightarrow^{\!\!\!\!\!\!6}}\,\,
A^1_3{\rightarrow^{\!\!\!\!\!\!?}}\,\,A_2$ is (according to
(\ref{CycleLimRateLin})) $c_2=b k^1_{23}/k_{32}$,
$c^1_3=b(1-k^1_{23}/k_{32})$, where $b=\sum c_i$. The component
$A^1_3$ is a glued cycle $A_3{\rightarrow^{\!\!\!\!\!\!2}}\,\,
A_4{\rightarrow^{\!\!\!\!\!\!3}}\,\,
A_5{\rightarrow^{\!\!\!\!\!\!4}}\,\, A_3$. Its steady state is
$c_3=c^1_3 k_{35}/k_{43}$, $c_4=c^1_3 k_{35}/k_{54}$, $c_5=c^1_3
(1-k_{35}/k_{43}-k_{35}/k_{54})$.

Let us construct an acyclic system that approximates relaxation of
(\ref{chain+-}) under the same assumption (1) $k_{32} > k^1_{23}$.
The final auxiliary system after gluing cycles is $A_1
{\rightarrow^{\!\!\!\!\!\!1}}\,\,
A_2{\rightarrow^{\!\!\!\!\!\!6}}\,\,
A^1_3{\rightarrow^{\!\!\!\!\!\!?}}\,\,A_2$. Let us delete  the
limiting reaction $A^1_3{\rightarrow^{\!\!\!\!\!\!?}}\,\,A_2$ from
the cycle. We get an acyclic system $A_1
{\rightarrow^{\!\!\!\!\!\!1}}\,\,
A_2{\rightarrow^{\!\!\!\!\!\!6}}\,\, A^1_3$. The component $A^1_3$
is the glued cycle $A_3{\rightarrow^{\!\!\!\!\!\!2}}\,\,
A_4{\rightarrow^{\!\!\!\!\!\!3}}\,\,
A_5{\rightarrow^{\!\!\!\!\!\!4}}\,\, A_3$. Let us restore this
cycle and delete the limiting reaction
$A_5{\rightarrow^{\!\!\!\!\!\!4}}\,\, A_3$. We get an acyclic
system $A_1 {\rightarrow^{\!\!\!\!\!\!1}}\,\,
A_2{\rightarrow^{\!\!\!\!\!\!6}}\,\,A_3{\rightarrow^{\!\!\!\!\!\!2}}\,\,
A_4{\rightarrow^{\!\!\!\!\!\!3}}\,\, A_5$. Relaxation of this
system approximates relaxation of the initial network
(\ref{chain+-}) under additional condition $k_{32} > k^1_{23}$.

(2) Let as assume now that  $k_{32} < k^1_{23}$. In this case, the
steady state of the cycle $A_2{\rightarrow^{\!\!\!\!\!\!6}}\,\,
A^1_3{\rightarrow^{\!\!\!\!\!\!?}}\,\,A_2$ is (according to
(\ref{CycleLimRateLin})) $c_2=b(1-k_{32}/k^1_{23})$,
$c^1_3=bk_{32}/k^1_{23}$. The further analysis is the same as it
was above: $c_3=c^1_3 k_{35}/k_{43}$, $c_4=c^1_3 k_{35}/k_{54}$,
$c_5=c^1_3 (1-k_{35}/k_{43}-k_{35}/k_{54})$ (with another
$c^1_3$).

Let us construct an acyclic system that approximates relaxation of
(\ref{chain+-}) under assumption (2) $k_{32} < k^1_{23}$. The
final auxiliary system after gluing cycles is the same, $A_1
{\rightarrow^{\!\!\!\!\!\!1}}\,\,
A_2{\rightarrow^{\!\!\!\!\!\!6}}\,\,
A^1_3{\rightarrow^{\!\!\!\!\!\!?}}\,\,A_2$, but the limiting step
in the cycle is different, $A_2{\rightarrow^{\!\!\!\!\!\!6}}\,\,
A^1_3$. After cutting this step, we get acyclic system $A_1
{\rightarrow^{\!\!\!\!\!\!1}}\,\, A_2{\leftarrow^{\!\!\!\!?}}\,
A^1_3$, where the last reaction has rate constant $k^1_{23}$.

The component $A^1_3$ is the glued cycle
$$A_3{\rightarrow^{\!\!\!\!\!\!2}}\,\,
A_4{\rightarrow^{\!\!\!\!\!\!3}}\,\,
A_5{\rightarrow^{\!\!\!\!\!\!4}}\,\, A_3\, .$$ Let us restore this
cycle and delete the limiting reaction
$A_5{\rightarrow^{\!\!\!\!\!\!4}}\,\, A_3$. The connection from
glued cycle $A^1_3{\rightarrow^{\!\!\!\!\!\!?}}\,\,A_2$ with
constant $k^1_{23}$ transforms into connection
$A_5{\rightarrow^{\!\!\!\!\!\!?}}\,\,A_2$ with the same constant
$k^1_{23}$.

We get the acyclic system: $$A_1 {\rightarrow^{\!\!\!\!\!\!1}}\,\,
A_2\, ,  \; A_3{\rightarrow^{\!\!\!\!\!\!2}}\,\,
A_4{\rightarrow^{\!\!\!\!\!\!3}}\,\, A_5
{\rightarrow^{\!\!\!\!\!\!?}}\,\,A_2\, .$$ The order of constants is
now known: $k_{21}>k_{43}>k_{54}>k^1_{23}$, and we can substitute
the sign ``?" by ``4": $A_3{\rightarrow^{\!\!\!\!\!\!2}}\,\,
A_4{\rightarrow^{\!\!\!\!\!\!3}}\,\, A_5
{\rightarrow^{\!\!\!\!\!\!4}}\,\,A_2$.

For both cases, $k_{32} > k^1_{23}$ ($k^1_{23}=
k_{24}k_{35}/k_{54}$) and $k_{32} < k^1_{23}$ it is easy to find
the eigenvectors explicitly  and to write the solution to the
kinetic equations in explicit form.

\subsection{The general case: cycles surgery
for auxiliary discrete dynamical system
with arbitrary family of attractors \label{subsecSurgery}}

In this subsection, we summarize results of relaxation analysis
and describe the algorithm of approximation of steady state and
relaxation process for arbitrary reaction network with well
separated constants.

\subsubsection{Hierarchy of cycles gluing}

Let us consider a reaction network $\mathcal{W}$ with a given
structure and fixed ordering of constants. The set of vertices of
$\mathcal{W}$ is $\mathcal{A}$ and the set of elementary reactions
is $\mathcal{R}$. Each reaction from $\mathcal{R}$ has the form
$A_i \to A_j$, $A_i, A_j \in \mathcal{A}$. The correspondent
constant is $k_{ji}$. For each $A_i \in \mathcal{A}$ we define
$\kappa_i= \max_j \{k_{ji} \}$ and $\phi (i) = {\rm arg \, max}_j
\{k_{ji} \}$. In addition, $\phi (i)=i$ if $k_{ji}=0$ for all $j$.

The auxiliary discrete dynamical system for the reaction network
$\mathcal{W}$ is the dynamical system $\Phi=\Phi_{\mathcal{W}}$
defined by the map $\phi$ on the set $\mathcal{A}$. Auxiliary
reaction network $\mathcal{V}=\mathcal{V}_{\mathcal{W}}$ has the
same set of vertices $\mathcal{A}$ and the set of reactions $A_i
\to A_{\phi(i)}$ with reaction constants $\kappa_i$. Auxiliary
kinetics is described by $\dot c = \tilde K c$, where $\tilde
{K}_{ij}= - \kappa_j \delta_{ij} +  \kappa_j \delta_{i \, \phi
(j)} $.

Every fixed point of $\Phi_{\mathcal{W}}$ is also a sink for the
reaction network $\mathcal{W}$. If all attractors of the system
$\Phi_{\mathcal{W}}$ are fixed points $A_{f1}, A_{f2}, ... \in
\mathcal{A}$ then  the set of stationary distributions for the
initial kinetics as well as for the auxiliary kinetics is the set
of distributions concentrated the set of fixed points $\{A_{f1},
A_{f2}, ... \}$. In this case, the auxiliary reaction network is
acyclic, and the auxiliary kinetics approximates relaxation of the
whole network $\mathcal{W}$.

In general case, let the system $\Phi_{\mathcal{W}}$ have several
attractors that are not fixed points, but cycles $C_1,C_2, ...$
with periods $\tau_1, \tau_2,...>1$. By gluing these cycles in
points, we transform the reaction network $\mathcal{W}$ into
$\mathcal{W}^1$. The dynamical system $\Phi_{\mathcal{W}}$ is
transformed into $\Phi^1$.  For these new system and network, the
connection $\Phi^1=\Phi_{\mathcal{W}^1}$ persists: $\Phi^1$ is the
auxiliary discrete dynamical system for $\mathcal{W}^1$.

For each cycle, $C_i$, we introduce a new vertex $A^i$. The new
set of vertices, $\mathcal{A}^1=\mathcal{A} \cup\{A^1,A^2,...\}
\setminus (\cup_i C_i) $ (we delete cycles $C_i$ and add vertices
$A^i$).

All the reaction between $A \to B$ ($A, B \in \mathcal{A}$) can be
separated into 5 groups:
\begin{enumerate}
\item{both $A,B \notin \cup_i C_i$;}
\item{$A \notin \cup_i C_i$, but $B \in C_i$;}
\item{$A\in C_i$, but $B \notin \cup_i C_i$;}
\item{$A \in C_i$, $B \in C_j$, $i \neq j$;}
\item{$A,B \in C_i$.}
\end{enumerate}
Reactions from the first group do not change. Reaction from the
second group transforms into $A\to A^i$ (to the whole glued cycle)
with the same constant. Reaction of the third type changes into
$A^i \to B$ with the rate constant renormalization
(\ref{renormrecharge}): let the cycle $C^i$ be the following
sequence of reactions $A_{1} \to A_{2} \to ... A_{\tau_i} \to
A_1$, and the reaction rate constant for $A_i \to A_{i+1}$ is
$k_i$ ($k_{\tau_i}$ for $A_{\tau_i} \to A_1$). For the limiting
reaction of the cycle $C_i$ we use notation $k_{\lim \, i}$. If
$A=A_j$ and $k$ is the rate reaction for $A \to B$, then the new
reaction $A^i \to B$ has the rate constant $k k_{\lim \, i}/ k_j$.
This corresponds to a quasistationary distribution on the cycle
(\ref{CycleLimRateLin}). It is obvious that the new rate constant
is smaller than the initial one: $k k_{\lim \, i}/ k_j < k$,
because  $k_{\lim \, i} < k_j $ due to definition of limiting
constant. The same constant renormalization is necessary for
reactions of the fourth type. These reactions transform into $A^i
\to A^j$. Finally, reactions of the fifth type vanish.

After we glue all the cycles of auxiliary dynamical system in the
reaction network $\mathcal{W}$, we get $\mathcal{W}^1$. Strictly
speaking, the whole network $\mathcal{W}^1$ is not necessary, and
in efficient realization  of the algorithm for large networks the
computation could be significantly reduced. What we need, is the
correspondent auxiliary dynamical system
$\Phi^1=\Phi_{\mathcal{W}^1}$ with auxiliary kinetics.

To find the auxiliary kinetic system, we should glue all cycles in
the first auxiliary system, and then add several reactions: for
each $A^i$ it is necessary to find in $\mathcal{W}^1$ the reaction
of the form $A^i \to B$ with maximal constant and add this
reaction to the auxiliary network. If there is no reaction of the
form $A^i \to B$ for given $i$ then the point $A^i$ is the fixed
point for $\mathcal{W}^1$ and  vertices of the  cycle $C_i$ form a
sink for the initial network.

After that, we decompose the new auxiliary dynamical system, find
cycles and repeat gluing. Terminate when all attractors of the
auxiliary dynamical system $\Phi^m$ become fixed points.

\subsubsection{Reconstruction of steady states}

After this termination, we can find all steady state distributions
by restoring cycles in the auxiliary reaction network
$\mathcal{V}^m$. Let $A^m_{f1}, A^m_{f2}, ...$ be fixed points of
$\Phi^m$. The set of steady states for $\mathcal{V}^m$ is the set
of all distributions on the set of fixed points $\{A^m_{f1},
A^m_{f2}, ...\}$. Let us take one of these distributions,
$c=(c^m_{f1}, c^m_{f2}, ...)$ (we mark the concentrations by the
same indexes as the vertex has; other $c_i=0$).

To make a step of cycle restoration we select those vertexes
$A^m_{fi}$ that are glued cycles and substitute them in the list
$A^m_{f1}, A^m_{f2}, ...$ by all the vertices of these cycles. For
each of those cycles we find the limiting rate constant and
redistribute the concentration $c^m_{fi}$ between the vertices of
the correspondent cycle by the rule (\ref{CycleLimRateLin}) (with
$b=c^m_{fi}$). As a result, we get a set of vertices  and a
distribution on this set of vertices. If among these vertices
there are glued cycles, then we repeat the procedure of cycle
restoration. Terminate when there is no glued cycles in the
support of the distribution. The resulting distribution is the
approximation to a steady state of $\mathcal{W}$, and all steady
states for $\mathcal{W}$ can be approximated by this method.

In order to construct the approximation to the basis of stationary
distributions of $\mathcal{W}$, it is sufficient to apply the
described algorithm to distributions concentrated on a single fixed
point $A^m_{fi}$, $c^m_{fj}=\delta_{ij}$, for every $j$.

The steady state approximation on the base of the rule
(\ref{CycleLimRateLin}) is a linear function of the
restored-and-cut cycles rate limiting constants. It is the first
order approximation.

The zero order approximation also makes sense. For one cycle is
gives (\ref{CycleLimZero}): all the concentration is collected at
the start of the limiting step. The algorithm for the zero order
approximation is even simpler than for the first order. Let us start
from the distributions concentrated on a single fixed point
$A^m_{fi}$, $c^m_{fj}=\delta_{ij}$ for some $i$. If this point is a
glued cycle then restore that cycle, and find the limiting step. The
new distribution is concentrated at the starting vertex of that
step. If this vertex is a glued cycle, then repeat. If it is not
then terminate. As a result we get a distribution concentrated in
one vertex of $\mathcal{A}$.

\subsubsection{Dominant kinetic system for approximation of
relaxation}

To construct an approximation to the relaxation process in the
reaction network $\mathcal{W}$, we also need to restore cycles, but
for this purpose we should start from the whole glued network
$\mathcal{V}^m$ on $\mathcal{A}^m$ (not only from fixed points as we
did for the steady state approximation). On a step back, from the
set $\mathcal{A}^m$ to $\mathcal{A}^{m-1}$ and so on some of glued
cycles should be restored and cut. On each step we build an acyclic
reaction network, the final network is defined on the initial vertex
set and approximates relaxation of $\mathcal{W}$.

To make one step back from $\mathcal{V}^m$ let us select the
vertices of $\mathcal{A}^m$ that are glued cycles from
$\mathcal{V}^{m-1}$. Let these vertices be $A^m_{1}, A^m_{2},
...$. Each $A^m_i$ corresponds to a glued cycle from
$\mathcal{V}^{m-1}$, $A^{m-1}_{i1} \to A^{m-1}_{i2} \to ...
A^{m-1}_{i \tau_i} \to A^{m-1}_{i1}$, of the length $\tau_i$. We
assume that the limiting steps in these cycles are $A^{m-1}_{i
\tau_i} \to A^{m-1}_{i1}$. Let us substitute each vertex $A^m_i$
in $\mathcal{V}^m$ by $\tau_i$ vertices $A^{m-1}_{i1} ,
A^{m-1}_{i2}, ... A^{m-1}_{i\tau_i}$ and add to $\mathcal{V}^m$
reactions $A^{m-1}_{i1} \to A^{m-1}_{i2} \to ... A^{m-1}_{i
\tau_i}$ (that are the cycle reactions without the limiting step)
with correspondent constants from $\mathcal{V}^{m-1}$.

If there exists an outgoing reaction $A^m_i \to B$ in
$\mathcal{V}^m$ then we substitute it by the reaction $A^{m-1}_{i
\tau_i} \to B$ with the same constant, i.e. outgoing reactions
$A^m_i \to...$ are reattached to the heads of the limiting steps.
Let us rearrange reactions from $\mathcal{V}^m$ of the form $B \to
A^m_i$. These reactions have prototypes in $\mathcal{V}^{m-1}$
(before the last gluing). We simply restore these reactions. If
there exists a reaction $A^m_i \to A^m_j$ then we find the
prototype in $\mathcal{V}^{m-1}$, $A \to B$, and substitute the
reaction by $A^{m-1}_{i \tau_i} \to B$ with the same constant, as
for $A^m_i \to A^m_j$.

After that step is performed, the vertices set is
$\mathcal{A}^{m-1}$, but the reaction set differs from the
reactions of the network $\mathcal{V}^{m-1}$: the limiting steps
of cycles are excluded and the outgoing reactions of glued cycles
are included (reattached to the heads of the limiting steps). To
make the next step, we select vertices of $\mathcal{A}^{m-1}$ that
are glued cycles from $\mathcal{V}^{m-2}$, substitute these
vertices by vertices of cycles, delete the limiting steps, attach
outgoing reactions to the heads of the limiting steps, and for
incoming reactions restore their prototypes from
$\mathcal{V}^{m-2}$, and so on.

After all, we restore all the glued cycles, and construct an
acyclic reaction network on the set $\mathcal{A}$. This acyclic
network approximates relaxation of the network $\mathcal{W}$. We
call this system the dominant system of $\mathcal{W}$ and use
notation ${\rm dom \, mod}(\mathcal{W})$.

\subsection{Example: a prism of reactions}

\begin{figure}[t]
\centering{ a)\includegraphics[height=23mm]{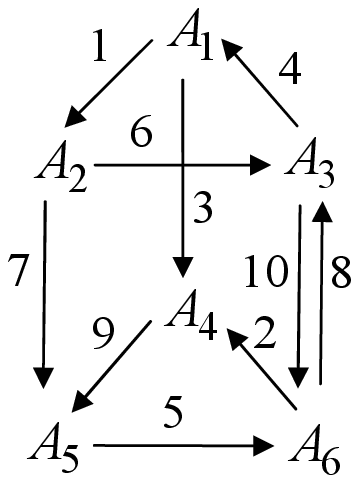} \;
b)\includegraphics[height=23mm]{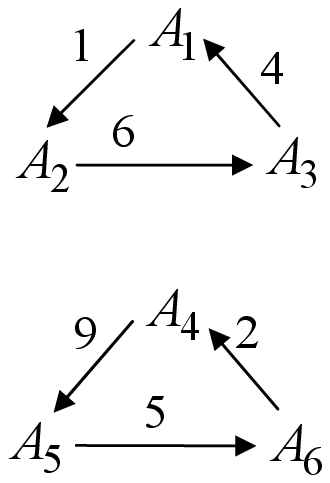}
\;c)\includegraphics[height=23mm]{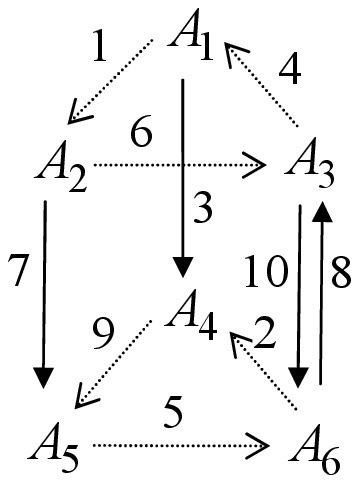} \\
\vspace{3mm} d)\includegraphics[height=23mm]{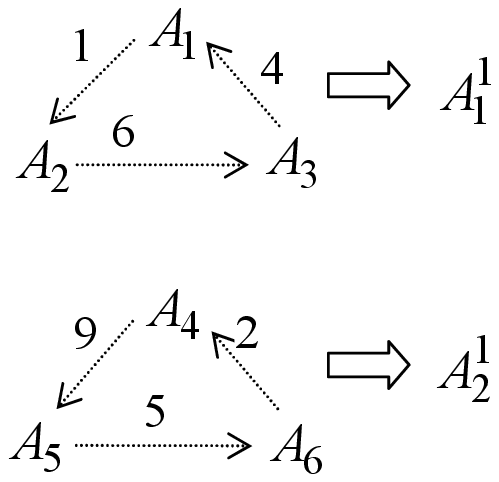}\;
e)\includegraphics[height=23mm]{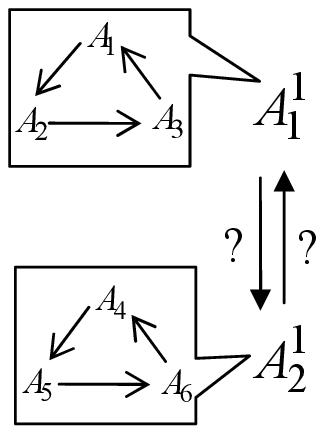} \;
f)\includegraphics[height=23mm]{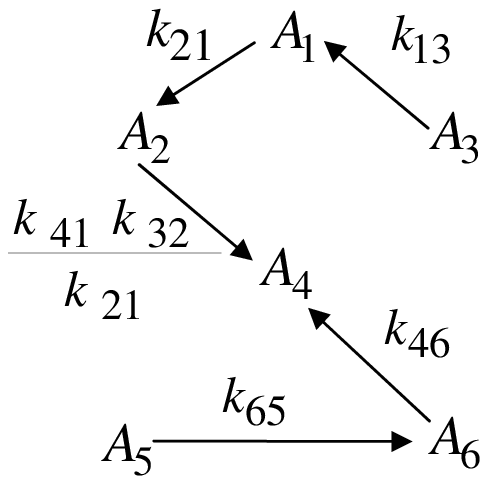} } \vspace{1mm}
\caption{\label{Prism1} Gluing of cycles for the prism of reactions
with a given ordering of rate constants in the case of two
attractors in the auxiliary dynamical system: (a) initial reaction
network, (b) auxiliary dynamical system that consists of two cycles,
(c) connection between cycles, (d) gluing cycles into new
components, (e) network $\mathcal{W}^1$ with glued vertices, (f) an
example of dominant system in the case when $k^1 _{21}= k_{41}
k_{32}/k_{21}$ and $k^1 _{21} > k^1 _{12}$ (by definition, $k^1
_{21}= \max\{k_{41} k_{32}/k_{21},\, k_{52} , \, k_{63} k_{32}
/k_{13} \}$ and $k^1_{12}= k_{36} k_{54}/ k_{46}$), the order of
constants in the dominant system is: $k_{21} >  k_{46} > k_{13}
> k_{65} > k_{41} k_{32}/k_{21} $. }
\end{figure}

\begin{figure}[t]
\centering{ a)\includegraphics[height=23mm]{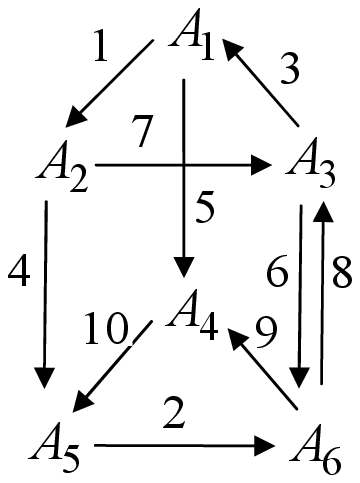} \;
b)\includegraphics[height=23mm]{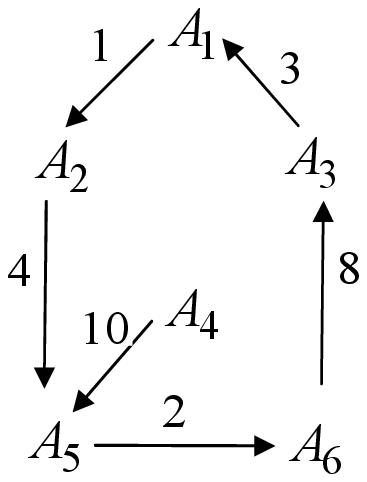}
\;c)\includegraphics[height=23mm]{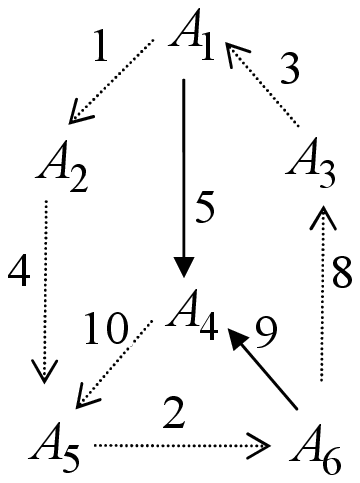} \\ \vspace{3mm}
d)\includegraphics[height=23mm]{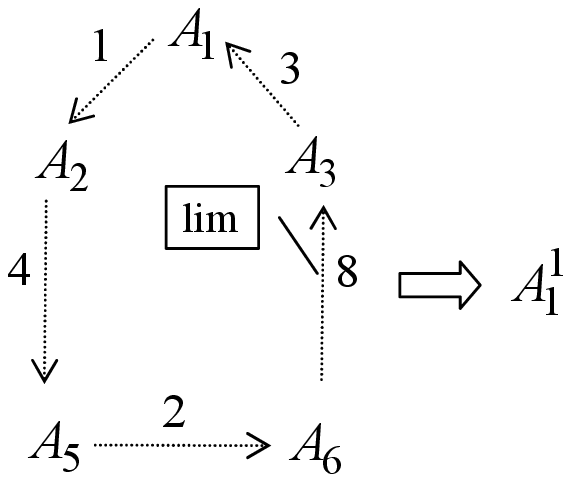}\;
e)\includegraphics[height=23mm]{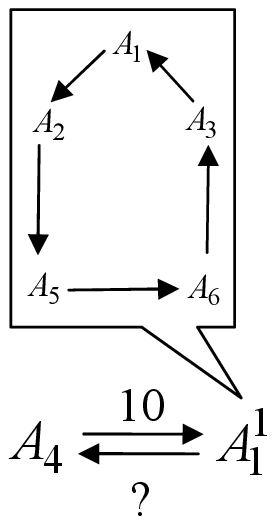}\;
f)\includegraphics[height=23mm]{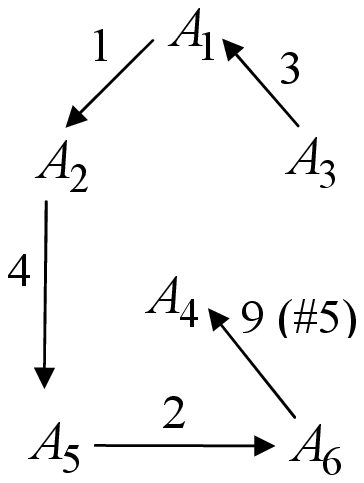} }
\caption{\label{Prism2}Gluing of a cycle for the prism of reactions
with a given ordering of rate constants in the case of one
attractors in the auxiliary dynamical system: (a) initial reaction
network, (b) auxiliary dynamical system that has one attractor, (c)
outgoing reactions from a cycle, (d) gluing of a cycle into new
component, (e) network $\mathcal{W}^1$ with glued vertices,  (f) an
example of dominant system in the case when $k^1= k_{46}$, and,
therefore $k^1 > k_{54}$ (by definition, $k^1= \max\{k_{41} k_{36}/
k_{21}, k_{46} \}$); this dominant system is a linear chain that
consists  of some reactions from the initial system (no nontrivial
monomials among constants). Only one reaction rate constant has in
the dominant system new number (number 5 instead of 9).}
\end{figure}

Let us demonstrate work of the algorithm on a typical example, a
prism of reaction that consists of two connected cycles
(Fig.~\ref{Prism1},\ref{Prism2}). Such systems appear in many
areas of biophysics and biochemistry (see, for example, the paper
of \cite{Kurz}).

For the first example we use the reaction rate constants ordering
presented in Fig.~\ref{Prism1}a. For this ordering, the auxiliary
dynamical system consists of two cycles (Fig.~\ref{Prism1}b) with
the limiting constants $k_{54}$ and $k_{32}$, correspondingly. These
cycles are connected by four reaction (Fig.~\ref{Prism1}c). We glue
the cycles into new components $A^1_1$ and $A^1_2$
(Fig.~\ref{Prism1}d), and the reaction network is transformed into
$A^1_1 \leftrightarrow A^1_2$. Following the general rule ($k^1=k
k_{\lim }/ k_j$), we determine the rate constants: for reaction
$A^1_1 \to A^1_2$ $$k^1 _{21}= \max\{k_{41} k_{32}/k_{21},\, k_{52}
, \, k_{63} k_{32} /k_{13} \},$$ and for reaction $A^1_2 \to A^1_1$
$$k^1_{12}= k_{36} k_{54}/ k_{46}.$$

There are six possible orderings of the constant combinations:
three possibilities for the choice of $k^1 _{21}$ and for each
such a choice there exist two possibilities: $k^1 _{21} > k^1
_{12}$ or $k^1 _{21} < k^1 _{12}$.

The zero order approximation of the steady state depends only on
the sign of inequality between $k^1 _{21}$ and  $k^1 _{12}$. If
$k^1 _{21} \gg k^1 _{12}$ then almost all concentration in the
steady state is accumulated inside $A^1_2$. After restoring the
cycle $A_4 \to A_5 \to A_6 \to A_4$ we find that in the  steady
state almost all concentration is accumulated in $A_4$ (the
component at the beginning of the limiting step of this cycle,
$A_4 \to A_5$). Finally, the eigenvector for zero eigenvalue is
estimated as the vector column with coordinates $(0,0,0,1,0,0)$.

If, inverse, $k^1 _{21} \ll k^1 _{12}$ then almost all
concentration in the steady state is accumulated inside $A^1_1$.
After restoring the cycle $A_1 \to A_2 \to A_3 \to A_1$ we find
that in the steady state almost all concentration is accumulated
in $A_2$ (the component at the beginning of the limiting step of
this cycle, $A_2 \to A_3$). Finally, the eigenvector for zero
eigenvalue is estimated as the vector column with coordinates
$(0,1,0,0,0,0)$.

Let us find the first order (in rate limiting constants)
approximation to the steady states. If $k^1 _{21} \gg k^1 _{12}$
then $k^1 _{12}$ is the rate limiting constant for the cycle
$A^1_1 \leftrightarrow A^1_2$ and almost all concentration in the
steady state is accumulated inside $A^1_2$: $c^1_2 \approx 1- k^1
_{12}/k^1 _{21}$ and $c^1_1 \approx k^1 _{12}/k^1 _{21}$. Let us
restore the glued cycles (Fig.~\ref{Prism1}). In the upper cycle
the rate limiting constant is $k_{32}$, hence, in steady state
almost all concentration of the upper cycle, $c^1_1$, is
accumulated in $A_2$: $c_2\approx c^1_1
(1-k_{32}/k_{13}-k_{32}/k_{21})$, $c_3\approx c^1_1
k_{32}/k_{13}$, and $c_1\approx c^1_1 k_{32}/k_{21}$. In the
bottom cycle the rate limiting constant is $k_{54}$, hence, $c_4
\approx c^1_2 (1- k_{54}/k_{65} - k_{54}/k_{46})$, $c_5 \approx
c^1_2  k_{54}/k_{65}$ and $c_6 \approx c^1_2 k_{54}/k_{46}$.

If, inverse, $k^1 _{21} \ll k^1 _{12}$ then $k^1 _{21}$ is the
rate limiting constant for the cycle $A^1_1 \leftrightarrow A^1_2$
and almost all concentration in the steady state is accumulated
inside $A^1_1$:  $c^1_1 \approx 1- k^1 _{21}/k^1 _{12}$ and $c^1_2
\approx k^1 _{21}/k^1 _{12}$. For distributions of concentrations
in the upper and lover cycles only the prefactors $c^1_1$, $c^1_2$
change their values.

For analysis of relaxation, let us analyze one of the six
particular cases separately.

1. $k^1 _{21}= k_{41} k_{32}/k_{21}$ and $k^1 _{21} > k^1 _{12}$.

In this case, the finite acyclic auxiliary dynamical system,
$\Phi^m=\Phi^1$, is $A^1_1 \to A^1_2$ with reaction rate constant
$k^1 _{21}= k_{41} k_{32}/k_{21}$, and $\mathcal{W}^1$ is $A^1_1
\leftrightarrow A^1_2$. We restore both cycles and delete the
limiting reactions $A_2 \to A_3$ and $A_4 \to A_5$. This is the
common step for all cases. Following the general procedure, we
substitute the reaction $A^1_1 \to A^1_2$ by $A_2 \to A_4$ with
the rate constant $k^1 _{21}= k_{41} k_{32}/k_{21}$ (because $A_2$
is the head of the limiting step for the cycle $A_1 \to A_2 \to
A_3 \to A_1$, and the prototype of the reaction $A^1_1 \to A^1_2$
is in that case $A_1 \to A_4$.

We find the dominant system for relaxation description: reactions
$A_3 \to A_1 \to A_2$ and $A_5 \to A_6 \to A_4$ with original
constants, and reaction $A_2 \to A_4$ with the rate constant $k^1
_{21}= k_{41} k_{32}/k_{21}$.

This dominant system graph is acyclic and, moreover, represents a
discrete dynamical system, as it should be (not more than one
outgoing reaction for any component). Therefore, we  can estimate
the eigenvalues and eigenvectors on the base of formulas
(\ref{right01}), (\ref{left01}). It is easy to determine the order
of constants because $k^1 _{21}= k_{41} k_{32}/k_{21}$: this
constant is the smallest nonzero constant in the obtained acyclic
system. Finally, we have the following ordering of constants: $A_3
{\rightarrow^{\!\!\!\!\!\!3}}\,\, A_1
{\rightarrow^{\!\!\!\!\!\!1}}\,\, A_2
{\rightarrow^{\!\!\!\!\!\!5}}\,\, A_4$, $A_5
{\rightarrow^{\!\!\!\!\!\!4}}\,\, A_6
{\rightarrow^{\!\!\!\!\!\!2}}\,\, A_4$.

So, the eigenvalues of the prism of reaction for the given
ordering are (with high accuracy, with probability close to one)
$- k_{21} < -k_{46} < - k_{13} < -k_{65} < - k_{41}
k_{32}/k_{21}$. The relaxation time is $\tau \approx
k_{21}/(k_{41} k_{32})$.

We use the same notations as in previous sections: eigenvectors
$l^i$ and $r^i$ correspond to the eigenvalue $-\kappa_i$, where
$\kappa_i$ is the reaction rate constant for the reaction $A_i \to
...\,$. The left eigenvectors $l^i$ are:
\begin{equation}\label{leftPrism1}
\begin{split}
&l^1\approx (1,0,0,0,0,0), \; l^2\approx (1,1,1,0,0,0), \\
&l^3\approx (0,0,1,0,0,0), \;  l^4 \approx (1,1,1,1,1,1), \\
&l^5\approx (0,0,0,0,1,0), \; l^6\approx (0,0,0,0,0,1).
\end{split}
\end{equation}
The right eigenvectors $r^i$ are (we represent vector columns as
rows):
\begin{equation}
\begin{split}
&r^1 \approx (1,-1,0,0,0,0),\; r^2 \approx (0,1,0,-1,0,0), \\
&r^3\approx(0,-1,1,0,0,0),\;  r^4 \approx (0,0,0,1,0,0), \\
&r^5\approx (0,0,0,-1,1,0),\; r^6 \approx (0,0,0,-1,0,1)
\end{split}
\end{equation}
The vertex $A_4$ is the fixed point for the discrete dynamical
system. There is no reaction $A_4 \to ... \,$.   For convenience,
we include the eigenvectors $l^4$ and $r^4$ for zero eigenvalue,
$\kappa_4=0$. These vectors correspond to the steady state: $r^4$
is the steady state vector, and the functional $l^4$ is the
conservation law.

The correspondent approximation to the general solution of the
kinetic equation for the prism of reaction (Fig.~\ref{Prism1}a)
is:
\begin{equation}\label{AppRel}
c(t)= \sum_{i=1}^6 r^i (l^i, c(0)) \exp(-\kappa_i t).
\end{equation}
Analysis of other five particular cases is similar. Of course,
some of the eigenvectors and eigenvalues can differ.

Of course, different ordering can lead to very different
approximations. For example, let us consider the same prism of
reactions, but with the ordering of constants presented in
Fig.~\ref{Prism2}a. The auxiliary dynamical system has one cycle
(Fig.~\ref{Prism2}b) with the limiting constant $k_{36}$. This
cycle is not a sink to the initial network, there are outgoing
reactions from its vertices (Fig.~\ref{Prism2}c). After gluing,
this cycles transforms into a vertex $A^1_1$ (Fig.~\ref{Prism2}d).
The glued network, $\mathcal{W}^1$ (Fig.~\ref{Prism2}e), has two
vertices, $A_4$ and $A^1_1$ the rate constant for the reaction
$A_4 \to A^1_1$ is $k_{54}$, and the rate constant for the
reaction $A^1_1 \to A_4$ is $k^1= \max\{k_{41} k_{36}/ k_{21},
k_{46}  \}$. Hence, there are not more than four possible
versions: two possibilities for the choice of $k^1$ and for each
such a  choice there exist two possibilities: $k^1 > k_{54}$ or
$k^1 < k_{54}$ (one of these four possibilities cannot be
realized, because $k_{46}
> k_{54}$)).

Exactly as it was in the previous example, the zero order
approximation of the steady state depends only on the sign of
inequality between $k^1$ and  $k_{54}$. If $k^1 \ll k_{54}$ then
almost all concentration in the steady state is accumulated inside
$A^1$. After restoring the cycle $A_3 \to A_1 \to A_2 \to A_5 \to
A_6 \to A_3$ we find that in the  steady state almost all
concentration is accumulated in $A_6$ (the component at the
beginning of the limiting step of this cycle, $A_6 \to A_3$). The
eigenvector for zero eigenvalue is estimated as the vector column
with coordinates $(0,0,0,0,0,1)$.

If $k^1 \gg k_{54}$ then almost all concentration in the steady
state is accumulated inside $A^4$. This vertex is not a glued
cycle, and immediately we find the approximate eigenvector for
zero eigenvalue, the vector column with coordinates
$(0,0,0,1,0,0)$.

Let us find the first order (in rate limiting constants)
approximation to the steady states. If $k^1 \ll k_{54}$ then $k^1$
is the rate limiting constant for the cycle $A^1_1 \leftrightarrow
A_4$ and almost all concentration in the steady state is
accumulated inside $A^1_1$: $c^1_1 \approx 1- k^1/k_{54}$ and $c_4
 \approx k^1/k_{54}$. Let us restore the glued cycle
(Fig.~\ref{Prism2}). The limiting constant for that cycle is
$k_{36}$, $c_6 \approx c^1_1(1- k_{36}/k_{13} -k_{36}/k_{21} -
k_{36}/k_{52} - k_{36}/k_{65})$, $c_3 \approx c^1_1
k_{36}/k_{13}$, $c_1 \approx c^1_1 k_{36}/k_{21}$, $c_2 \approx
c^1_1 k_{36}/k_{52}$, and $c_5 \approx c^1_1 k_{36}/k_{65}$.

If $k^1 \gg k_{54}$ then $k_{54}$ is the rate limiting constant
for the cycle $A^1_1 \leftrightarrow A_4$ and almost all
concentration in the steady state is accumulated inside $A_4$:
$c_4 \approx 1- k_{54}/k^1$ and $c^1_1  \approx k_{54}/k^1$. In
distribution of concentration inside the cycle only the prefactor
$c^1_1$ changes.

Let us analyze the relaxation process for one of the
possibilities: $k^1= k_{46}$, and, therefore $k^1 > k_{54}$. We
restore the cycle, delete the limiting step, transform the
reaction $A^1_1 \to A_4$ into reaction $A_6 \to A_4$ with the same
constant $k^1 = k_{46}$ and get the chain with ordered constants:
$A_3 {\rightarrow^{\!\!\!\!\!\!3}}\,\, A_1
{\rightarrow^{\!\!\!\!\!\!1}}\,\, A_2
{\rightarrow^{\!\!\!\!\!\!4}}\,\, A_5
{\rightarrow^{\!\!\!\!\!\!2}}\,\, A_6
{\rightarrow^{\!\!\!\!\!\!5}}\,\, A_4$. Here the nonzero rate
constants $k_{ij}$ have the same value as for the initial system
(Fig.~\ref{Prism2}a). The relaxation time is $\tau \approx
1/k_{46}$. Left eigenvectors are (including $l^4$ for the zero
eigenvalue):
\begin{equation}\label{leftPrism2}
\begin{split}
&l^1\approx(1,0,0,0,0,0), \; l^2\approx(1,1,1,0,0,0), \\
&l^3\approx(0,0,1,0,0,0); \; l^4\approx(1,1,1,1,1,1), \\
&l^5\approx(0,0,0,0,1,0), \; l^6\approx(1,1,1,0,1,1).
\end{split}
\end{equation}
Right eigenvectors are (including $r^4$ for the zero eigenvalue):
\begin{equation}
\begin{split}
&r^1\approx(1,-1,0,0,0,0), \; r^2\approx(0,1,0,0,0,-1),\\
&r^3\approx(0,-1,1,0,0,0), \; r^4\approx(0,0,0,1,0,0), \\
&r^5\approx(0,0,0,0,1,-1), \; r^6\approx(0,0,0,-1,0,1).
\end{split}
\end{equation}
Here we represent vector columns as rows.

For the approximation of relaxation in that order we can use
(\ref{AppRel}).

\section{The reversible triangle of reactions: the simple example case study \label{sec:triangle}}

In this section, we illustrate the analysis of dominant systems on
a simple example, the reversible triangle of reactions.
\begin{equation}\label{revtri}
A_1 \leftrightarrow A_2 \leftrightarrow A_3 \leftrightarrow A_1 \,
.
\end{equation}
This triangle appeared in many works as an ideal object for a case
study. Our favorite example is the work of \cite{Wei62}. Now in
our study the triangle (\ref{revtri}) is  not obligatory a closed
system. We can assume that it is a subsystem of a larger system,
and any reaction $A_i \to A_j$ represents a reaction of the form
$\ldots +A_i \to A_j + \ldots$, where unknown but slow components
are substituted by dots. This means that there are no obligatory
relations between reaction rate constants, first of all, no
detailed balance relations, and six reaction rate constants are
arbitrary nonnegative numbers.

There exist $6!=720$ orderings of six reaction rate constants for
this triangle, but, of course, it is not necessary to consider all
these orderings. First of all, because of the permutation
symmetry, we can select an arbitrary reaction as the fastest one.
Let the reaction rate constant $k_{21}$ for the reaction $A_1 \to
A_2$ is the largest. (If it is not, we just have to change the
enumeration of reagents.)

\begin{figure}
\centering{
\includegraphics[width=75mm]{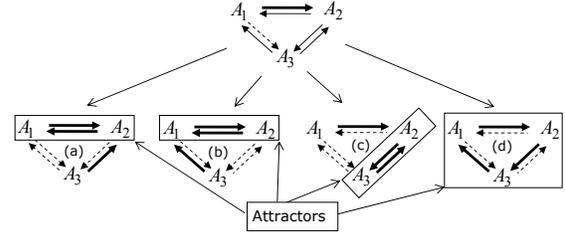}
\caption{\label{TriangleAux} Four possible auxiliary dynamical
systems for the reversible triangle of reactions with $k_{21} >
k_{ij}$ for $(i,j) \neq (2,1)$: (a) $k_{12} > k_{32}$, $k_{23} >
k_{13}$; (b) $k_{12}
> k_{32}$, $k_{13} > k_{23}$; (c) $k_{32} > k_{12}$, $k_{23} >
k_{13}$; (d) $k_{32} > k_{12}$, $k_{13} > k_{23}$. For each vertex
the outgoing reaction with the largest rate constant is
represented by the solid bold arrow, and other reactions are
represented by the dashed arrows. The digraphs formed by solid
bold arrows are the auxiliary discrete dynamical systems.
Attractors of these systems are isolated in frames. }}
\end{figure}

First of all, let us describe all possible auxiliary dynamical
systems for the triangle (\ref{revtri}). For each vertex, we have
to select the fastest outgoing reaction. For $A_1$, it is always
$A_1 \to A_2$, because of our choice of enumeration (the higher
scheme in Fig.~\ref{TriangleAux}). There exist two choices of the
fastest outgoing reaction for two other vertices and, therefore,
only four versions of auxiliary dynamical systems for
(\ref{revtri}) (Fig.~\ref{TriangleAux}).

Because of the choice of enumeration, the vectors of logarithms of
reaction rate constants form a convex cone in $R^6$ which is
described by the system of inequalities $\ln k_{21} > \ln k_{ij}$,
$(i,j) \neq (2,1)$. For each of the possible auxiliary systems
(Fig~\ref{TriangleAux}) additional inequalities between constants
should be valid, and we get four correspondent cones in $R^6$.
This cones form a partitions of the initial one (we neglect
intersections of faces which have zero measure). Let us discuss
the typical behaviour of systems from this cones separately. (Let
us remind that if in a cone for some values of coefficients
$\theta_{ij}$, $\zeta_{ij}$  $\sum_{ij} \theta_{ij} \ln k_{ij} <
\sum_{ij} \zeta_{ij} \ln k_{ij}$, then, typically in this cone
$\sum_{ij} \theta_{ij} \ln k_{ij} < K + \sum_{ij} \zeta_{ij} \ln
k_{ij}$ for any positive $K$. This means that typically
$\prod_{ij} k_{ij}^{\theta_{ij}} \ll \prod_{ij}
k_{ij}^{\zeta_{ij}}$.)

\subsection{Auxiliary system (a): $A_1 \leftrightarrow A_2
\leftarrow A_3$; $k_{12} > k_{32}$, $k_{23} > k_{13}$}

\subsubsection{Gluing cycles}

The attractor is a cycle (with only two vertices) $A_1
\leftrightarrow A_2$. This is not a sink, because two outgoing
reactions exist: $A_1 \to A_3$ and $A_2 \to A_3$. They are
relatively slow: $k_{31} \ll k_{21}$ and $k_{32} \ll k_{12}$. The
limiting step in this cycle is $A_2 \to A_1$ with the rate
constant $k_{12}$. We have to glue the cycle $A_1 \leftrightarrow
A_2$ into one new component $A_1^1$ and to add a new reaction
$A_1^1 \rightarrow A_3$ with the rate constant
\begin{equation}\label{gluedcona}
k_{31}^1=\max\{k_{32}, \, k_{31}k_{12}/k_{21}\}\, .
\end{equation}
This is a particular case of (\ref{renormrecharge}),
(\ref{kappamax}).

As a result, we get a new system, $A_1^1 \leftrightarrow A_3$ with
reaction rate constants $k_{31}^1$ (for $A_1^1 \rightarrow A_3$)
and initial $k_{23}$ (for $A_1^1 \leftarrow A_3$). This cycle is a
sink, because it has no outgoing reactions (the whole system is a
trivial example of a sink).

\subsubsection{Steady states}

To find the steady state, we have to compute the stationary
concentrations for the cycle $A_1^1 \leftrightarrow A_3$, $c^1_1$
and $c_3$. We use the standard normalization condition  $c^1_1 +
c_3= 1$. On the base of  the general formula for a simple cycle
(\ref{CycleRate}) we obtain:
\begin{equation}\label{fstcyca}
w=\frac{1}{\frac{1}{k_{31}^1} + \frac{1}{k_{23}}}\, , \;
c^1_1=\frac{w}{k_{31}^1}\, , \; c_3=\frac{w}{k_{23}}\, .
\end{equation}

After that, we can calculate the concentrations of $A_1$ and $A_2$
with normalization $c_1 + c_2 = c^1_1$. Formula (\ref{CycleRate})
gives:
\begin{equation}\label{scncyca}
w'=\frac{c^1_1}{\frac{1}{k_{21}} +\frac{1}{k_{12}}}\, ,
c_1=\frac{w'}{k_{21}}\, , c_2 =\frac{w'}{k_{12}}\, .
\end{equation}

We can simplify the answer using inequalities between constants,
as it was done in formulas (\ref{CycleLimRate}),
(\ref{CycleLimRateLin}). For example, $\frac{1}{k_{21}}
+\frac{1}{k_{12}} \approx \frac{1}{k_{12}}$, because $k_{21} \gg
k_{12}$. It is necessary to stress that we have used the
inequalities between constants $k_{21} > k_{ij}$ for $(i,j) \neq
(2,1)$, $k_{12} > k_{32}$, and $k_{23}
> k_{13}$ to obtain the simple answer (\ref{fstcyca}),
(\ref{scncyca}),  hence if  we even do not use these inequalities
for the further simplification, this does not guarantee the higher
accuracy of formulas.

\subsubsection{Eigenvalues and eigenvectors}

At the next step, we have to restore and cut the cycles. First
cycle to cut is the result of cycle gluing, $A_1^1 \leftrightarrow
A_3$. It is necessary to delete the limiting step, i.e. the
reaction with the smallest rate constant. If $k_{31}^1 > k_{23}$,
then we get $A_1^1 \rightarrow A_3$. If, inverse, $k_{23} >
k_{31}^1$, then we obtain $A_1^1 \leftarrow A_3$.

\begin{figure}
\centering{
\includegraphics[width=75mm]{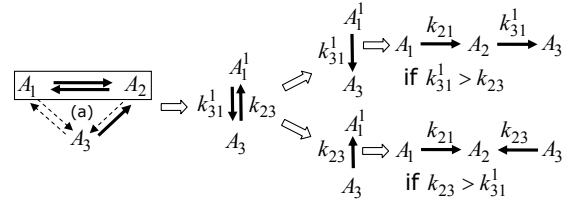}
\caption{\label{Dom1} Dominant systems for case (a) (defined in
Fig.~\ref{TriangleAux})}}
\end{figure}

After that, we have to restore and cut the cycle which was glued
into the vertex $A_1^1$. This is the two-vertices cycle $A_1
\leftrightarrow A_2$. The limiting step for this cycle is $A_1
\leftarrow A_2$, because $k_{21} \gg k_{12}$. If $k_{31}^1 >
k_{23}$, then following the rule visualized by
Fig.~\ref{CycleSurg}, we get the dominant system $A_1 \to A_2 \to
A_3$ with reaction rate constants  $k_{21}$ for $A_1 \to A_2$ and
$k_{31}^1$ for  $A_2 \to A_3$. If $k_{23} > k_{31}^1$ then we
obtain $A_1 \to A_2 \leftarrow A_3$ with reaction rate constants
$k_{21}$ for $A_1 \to A_2$ and $k_{23}$ for $A_2 \leftarrow A_3$.
All the procedure is illustrated by Fig.~\ref{Dom1}.

The eigenvalues and the correspondent eigenvectors for dominant
systems in case (a) are represented below in zero-one asymptotic.
\begin{enumerate}
\item{$k_{31}^1 > k_{23}$, the dominant system $A_1 \to A_2 \to
A_3$,
\begin{equation}
\begin{array}{lll}
  \lambda_0 = 0\, , \;  &r^0\approx(0,0,1)\, , \;
 & l^0=(1,1,1)\, ; \\
  \lambda_1 \approx -k_{21}\, , \; & r^1\approx(1,-1,0)\, , \;
 &l^1\approx(1,0,0)\, ; \\
  \lambda_2 \approx -k_{31}^1\, , \;  & r^2\approx(0,1,-1)\, , \;
 &l^2\approx(1,1,0)\,;
\end{array}
\end{equation}}
\item{$k_{23} > k_{31}^1 $, the dominant system $A_1 \to A_2 \leftarrow
A_3$,
\begin{equation}
\begin{array}{lll}
  \lambda_0 = 0\, , \; & r^0\approx(0,1,0)\, , \;
 &l^0=(1,1,1)\, ; \\
  \lambda_1 \approx -k_{21}\, , \; & r^1\approx(1,-1,0)\, , \;
 &l^1\approx(1,0,0)\, ; \\
  \lambda_2 \approx -k_{23}\, , \; &  r^2\approx(0,-1,1)\, , \;
 &l^2\approx(0,0,1)\,.
\end{array}
\end{equation}}
\end{enumerate}
Here, the value of $k_{31}^1$ is given by formula
(\ref{gluedcona}).

With higher accuracy, in case (a)
\begin{equation}
r^0\approx\left(\frac{w'}{k_{21}}\, , \frac{w'}{k_{12}}\,
,\frac{w}{k_{23}}\right)\, ,
\end{equation}
where $$w=\frac{1}{\frac{1}{k_{31}^1} + \frac{1}{k_{23}}}\, , \;
w'=\frac{c^1_1}{\frac{1}{k_{21}} +\frac{1}{k_{12}}}\, , \;
c^1_1=\frac{w}{k_{31}^1}\, ,$$ in according to (\ref{fstcyca}),
(\ref{scncyca}).

\subsection{Auxiliary system (b): $A_3 \rightarrow A_1 \leftrightarrow A_2$;
$k_{12} > k_{32}$, $k_{13} > k_{23}$}

\subsubsection{Gluing cycles}

The attractor is a cycle $A_1 \leftrightarrow A_2$ again, and this
is not a sink. We have to glue the cycle $A_1 \leftrightarrow A_2$
into one new component $A_1^1$ and to add a new reaction $A_1^1
\rightarrow A_3$ with the rate constant $k_{31}^1$ given by
formula (\ref{gluedcona}). As a result, we get an new system,
$A_1^1 \leftrightarrow A_3$ with reaction rate constants
$k_{31}^1$ (for $A_1^1 \rightarrow A_3$) and initial $k_{13}$ (for
$A_1^1 \leftarrow A_3$). At this stage, the only difference from
the case (a) is the reaction $A_1^1 \leftarrow A_3$ rate constant
$k_{13}$ instead of $k_{23}$.

\subsubsection{Steady states}

For the steady states we have to repeat formulas (\ref{fstcyca})
(\ref{scncyca}) with minor changes (just use $k_{13}$ instead of
$k_{23}$):

\begin{equation}\label{fstcycb}
\begin{split}
w=\frac{1}{\frac{1}{k_{31}^1} + \frac{1}{k_{13}}}\, , \;
c^1_1=\frac{w}{k_{31}^1}\, , \; c_3=\frac{w}{k_{13}}\, ; \\
w'=\frac{c^1_1}{\frac{1}{k_{21}} +\frac{1}{k_{12}}}\, , \;
c_1=\frac{w'}{k_{21}}\, , \; c_2 =\frac{w'}{k_{12}}\, .
\end{split}
\end{equation}

\subsubsection{Eigenvalues and eigenvectors}

\begin{figure}
\centering{
\includegraphics[width=75mm]{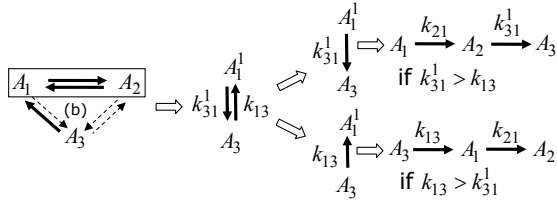}
\caption{\label{Dom2} Dominant systems for case (b) (defined in
Fig.~\ref{TriangleAux})}}
\end{figure}

The structure of the dominant system depends on the limiting step
of the cycle $A^1_1 \leftrightarrow A_3$ (Fig.~\ref{Dom2}). If
$k^1_{31} > k_{13}$, then in the dominant system remains the
reaction $A_1^1 \to A_3$ from this cycle. After restoring the
glued cycle $A_1 \leftrightarrow A_2$ it is necessary to delete
the slowest reaction from this cycle too. This is always $A_1
\leftarrow A_2$, because $A_1 \to A_2$ is the fastest reaction.
The reaction $A_1^1 \to A_3$ transforms into $A_2 \to A_3$,
because $A_2$ is the head of the limiting step $A_1 \leftarrow
A_2$ (see Fig.~\ref{CycleSurg}). Finally, we get $A_1 \to A_2 \to
A_3$.

If $k_{13} > k^1_{31} $, then then in the dominant system remains
the reaction $A_3 \to A_1$, and the dominant system is $A_3 \to
A_1 \to A_2$ (Fig.~\ref{Dom2}).

The eigenvalues and the correspondent eigenvectors for dominant
systems in case (b) are represented below in zero-one asymptotic.
\begin{enumerate}
\item{$k_{31}^1 > k_{13}$, the dominant system $A_1 \to A_2 \to
A_3$,
\begin{equation}
\begin{array}{lll}
  \lambda_0 = 0\, , \;  &r^0\approx(0,0,1)\, , \;
 & l^0=(1,1,1)\, ; \\
  \lambda_1 \approx -k_{21}\, , \; & r^1\approx(1,-1,0)\, , \;
 &l^1\approx(1,0,0)\, ; \\
  \lambda_2 \approx -k_{31}^1\, , \;  & r^2\approx(0,1,-1)\, , \;
 &l^2\approx(1,1,0)\,;
\end{array}
\end{equation}}
\item{$k_{13} > k_{31}^1 $, the dominant system $A_3 \to A_1 \to A_2$,
\begin{equation}
\begin{array}{lll}
  \lambda_0 = 0\, , \; & r^0\approx(0,1,0)\, , \;
 &l^0=(1,1,1)\, ; \\
  \lambda_1 \approx -k_{21}\, , \; & r^1\approx(1,-1,0)\, , \;
 &l^1\approx(1,0,0)\, ; \\
  \lambda_2 \approx -k_{13}\, , \; &  r^2\approx(0,-1,1)\, , \;
 &l^2\approx(0,0,1)\,.
\end{array}
\end{equation}}
\end{enumerate}
Here, the value of $k_{31}^1$ is given by formula
(\ref{gluedcona}). The only difference from case (a) is the rate
constant $k_{23}$ instead of $k_{13}$.

With higher accuracy, in case (b)
\begin{equation}
r^0\approx\left(\frac{w'}{k_{21}}\, , \frac{w'}{k_{12}}\,
,\frac{w}{k_{13}}\right)\, ,
\end{equation}
where $w$ and $w'$ are given by formula (\ref{fstcycb}).

\subsection{Auxiliary system (c): $A_1 \rightarrow A_2 \leftrightarrow A_3$;
$k_{32} > k_{12}$, $k_{23} > k_{13}$}

\subsubsection{Gluing cycles}

The attractor is a cycle $A_2 \leftrightarrow A_3$. This is not a
sink, because two outgoing reactions exist: $A_2 \to A_1$ and $A_3
\to A_1$.  We have to glue the cycle $A_2 \leftrightarrow A_3$
into one new component $A_2^1$ and to add a new reaction $A_2^1
\rightarrow A_1$ with the rate constant $k^1_{12}$. The definition
of this new constant depends on normalized  the steady-state
distribution in this cycle. If $c_2^*$, $c_3^*$ are the
steady-state  concentrations (with normalization $c_2^* + c_3^*
=1$), then $$k^1_{12}\approx \max\{k_{12}c_2^*, \, k_{13} c_3^*
\}.$$ If we use limitation in the glued cycle explicitly, then we
get the direct analogue of (\ref{gluedcona}) in two versions: one
for  $k_{32} > k_{23}$, another for $k_{23} > k_{32}$. But we can
skip this simplification and write
\begin{equation}
k^1_{12}\approx \max\{k_{12}w^*/k_{32} , \, k_{13} w^*/k_{23} \},
\end{equation}
where $$w^*=\frac{1}{\frac{1}{k_{32}}+\frac{1}{k_{23}}}.$$

\subsubsection{Steady states}

Exactly as in the cases (a) and (b) we can find approximation of
steady state using steady states in cycles $ A_1 \leftrightarrow
A_2^1$ and $A_2 \leftrightarrow A_3$:

\begin{equation}\label{fstcycc}
\begin{split}
w=\frac{1}{\frac{1}{k_{12}^1} + \frac{1}{k_{21}}}\, , \;
c^1_2=\frac{w}{k_{12}^1}\, , \; c_1=\frac{w}{k_{21}}\, ; \\
w'=\frac{c^1_2}{\frac{1}{k_{32}} +\frac{1}{k_{23}}}\, , \;
c_2=\frac{w'}{k_{32}}\, , \; c_3 =\frac{w'}{k_{23}}\, .
\end{split}
\end{equation}

\subsubsection{Eigenvalues and eigenvectors}
\begin{figure}
\centering{
\includegraphics[width=75mm]{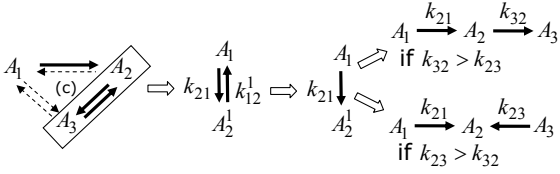}
\caption{\label{Dom3} Dominant systems for case (c) (defined in
Fig.~\ref{TriangleAux})}}
\end{figure}

The limiting step in the cycle $ A_1 \leftrightarrow A_2^1$ in
known, this is $ A_1 \leftarrow A_2^1$. There are two
possibilities for the choice on limiting step in the cycle $A_2
\leftrightarrow A_3$. If $k_{32}> k_{23}$, then this limiting step
is $A_2 \leftarrow A_3$, and the dominant system is $A_1 \to A_2
\to A_3$. If $k_{23}> k_{32}$, then the dominant system is $A_1
\to A_2 \leftarrow A_3$ (Fig.~\ref{Dom3}).

The eigenvalues and the correspondent eigenvectors for dominant
systems in case (b) are represented below in zero-one asymptotic.
\begin{enumerate}
\item{$k_{32}> k_{23}$, the dominant system $A_1 \to A_2 \to
A_3$,
\begin{equation}
\begin{array}{lll}
  \lambda_0 = 0\, , \;  &r^0\approx(0,0,1)\, , \;
 & l^0=(1,1,1)\, ; \\
  \lambda_1 \approx -k_{21}\, , \; & r^1\approx(1,-1,0)\, , \;
 &l^1\approx(1,0,0)\, ; \\
  \lambda_2 \approx -k_{32}\, , \;  & r^2\approx(0,1,-1)\, , \;
 &l^2\approx(1,1,0)\,;
\end{array}
\end{equation}}
\item{$k_{23} > k_{32} $, the dominant system $A_1 \to A_2 \leftarrow A_3 $,
\begin{equation}
\begin{array}{lll}
  \lambda_0 = 0\, , \; & r^0\approx(0,1,0)\, , \;
 &l^0=(1,1,1)\, ; \\
  \lambda_1 \approx -k_{21}\, , \; & r^1\approx(1,-1,0)\, , \;
 &l^1\approx(1,0,0)\, ; \\
  \lambda_2 \approx -k_{23}\, , \; &  r^2\approx(0,-1,1)\, , \;
 &l^2\approx(0,0,1)\,.
\end{array}
\end{equation}}
\end{enumerate}

With higher accuracy the value of $r^0$ is given by formula  of
the steady state concentrations (\ref{fstcycc}).

\subsection{Auxiliary system (d): $A_1 \rightarrow A_2 \rightarrow A_3 \rightarrow A_1$;
$k_{32} > k_{12}$, $k_{13} > k_{23}$}

This is a simple cycle. We discussed this case in details several
times. To get the dominant system it is sufficient just to delete
the limiting step. Everything is determined by the choice of the
minimal constant in the couple $\{k_{32},\, k_{13}\}$. Formulas
for steady state are well known too: (\ref{CycleRate}),
(\ref{CycleLimRate}), (\ref{CycleLimRateLin}).

This is not necessary to discuss all orderings of constants,
because some of them are irrelevant to the final answer. For
example, in this case (d) interrelations between constants
$k_{31}$, $k_{23}$, and $k_{12}$ are not important.

\subsection{Resume: zero-one multiscale asymptotic for the reversible reaction
triangle}

We found only three topologically different version of dominant
systems for the reversible reaction triangle: (i) $A_1 \to A_2 \to
A_3$, (ii) $A_1 \to A_2 \leftarrow A_3$, and (iii) $A_3 \to A_1
\to A_2$. Moreover, there exist only two versions of zero-one
asymptotic for eigenvectors: the fastest eigenvalue is always
$-k_{21}$ (because our choice of enumeration), the correspondent
right and left eigenvectors (fast mode) are: $r^1\approx(1,-1,0)$,
$l^1=(1,0,0)$. (The difference between systems (ii) and (iii)
appears in the first order of the slow/fast constants ratio.)

If in the steady state (almost) all mass is concentrated in $A_2$
(this means that  $r^0\approx (0,1,0)$, dominant systems (ii) or
(iii)), then $r^2\approx (0,-1,1)$ and $l^2 \approx (0,0,1)$. If in
the steady state (almost) all mass is concentrated in $A_3$ (this
means that  $r^0\approx (0,0,1)$, dominant system (i)), then
$r^2\approx (0,1,-1)$ and $l^2 \approx (0,1,0)$. We can see that the
dominant systems of the forms (ii) and (iii) produce the same
zero-one asymptotic of eigenvectors. Moreover, the right
eigenvectors $r^2\approx (0,1,-1)$ coincide for all cases (there is
no difference between $r^2$ and $-r^2$), and the difference appears
in the left eigenvector $l^2$. Of course, this peculiarity
(everything is regulated by the steady state asymptotic) results
from the simplicity of this example.

In the zero-one asymptotic, the reversible reaction triangle is
represented by one of the reaction mechanisms, (i) or (iii). The
rate constant of the first reaction $A_1 \to A_2$ is always
$k_{12}$. The direction of the second reaction is determined by a
system of linear uniform inequalities between {\it logarithms} of
rate constants. The logarithm of effective constant of this
reaction is the piecewise linear function of the logarithms of
reaction rate constants, and the switching between different
pieces is regulated by linear inequalities. These inequalities are
described in this section, and most of them are represented in
Figs.~\ref{TriangleAux}-\ref{Dom3}. One can obtain the first-order
approximation of eigenvectors in the slow/fast constants ratio
from the Appendix 1 formulas.

\section{Three zero-one laws and nonequilibrium phase transitions in multiscale systems \label{sec:01}}

\subsection{Zero-one law for steady states of weakly ergodic reaction networks}

Let us take a weakly ergodic network $\mathcal{W}$ and apply the
algorithms of auxiliary systems construction and cycles gluing. As
a result we obtain an auxiliary dynamic system with one fixed
point (there may be only one minimal sink). In the algorithm of
steady state reconstruction (Subsec.~\ref{subsecSurgery}) we
always operate with one cycle (and with small auxiliary cycles
inside that one, as in a simple example in
Subsec.~\ref{subsecCycle+1}). In a cycle with limitation almost
all concentration is accumulated at the start of the limiting step
(\ref{CycleLimRateLin}), (\ref{CycleLimZero}). Hence, in the whole
network almost all concentration will be accumulated in one
component. The dominant system for a weekly ergodic network is an
acyclic network with minimal element. The minimal element is such
a component $A_{\min}$ that there exists a oriented path in the
dominant system from any element to $A_{\min}$. Almost all
concentration in the steady state of the network $\mathcal{W}$
will be concentrated in the component $A_{\min}$.

\subsection{Zero-one law for nonergodic multiscale networks}

The simplest example of nonergodic but connected reaction network
is $A_1 \leftarrow A_2 \to A_3$ with reaction rate constants $k_1,
k_2$. For this network, in addition to $b^0(c)=c_1+c_2+c_3$ a
kinetic conservation law exist, $b^{\rm k}(c) =\frac{c_1}{k_1}-
\frac{c_3}{k_2}$. The result of time evolution, $\lim_{t \to
\infty} \exp(Kt) c$ (\ref{TimeEv}),  is described by simple
formula (\ref{massDistri}): $$\lim_{t \to \infty} \exp(Kt) c
=b^1(c)(1,0,0)+b^2(c)(0,1,1),$$ where  $b^1(c)+b^2(c)=b^0(c)$ and
$\frac{k_1+k_2}{k_1}b^1(c)- \frac{k_1+k_2}{k_2}b^2(c)=b^{\rm
k}(c)$. If $k_1 \gg k_2$ then $b^1(c)\approx c_1+c_2$ and
$b^2(c)\approx c_3$. If $k_1 \ll k_2$ then $b^1(c)\approx c_1$ and
$b^2(c)\approx c_2+c_3$. This simple zero-one law (either almost
all amount of $A_2$ transforms into $A_1$, or almost all amount of
$A_2$ transforms into $A_3$) can be generalized onto all
nonergodic multiscale systems.

Let us take a multiscale network and perform the iterative process
of auxiliary dynamic systems construction and cycle gluing, as it
is prescribed  in Subsec.~\ref{subsecSurgery}. After the final
step the algorithm gives the discrete dynamical system $\Phi^m$
with fixed points $A^m_{fi}$.

The fixed points $A^m_{fi}$ of the discrete dynamical system
$\Phi^m$ are the glued ergodic components $G_i \subset
\mathcal{A}$ of the initial network $\mathcal{W}$. At the same
time, these points are attractors of $\Phi^m$. Let us consider the
correspondent decomposition of this system with partition
$\mathcal{A}^m = \cup_i Att(A^m_{fi})$. In the cycle restoration
during construction of dominant system ${\rm dom \,
mod}(\mathcal{W})$ this partition transforms into partition of
$\mathcal{A}$: $\mathcal{A}= \cup_i U_i$, $Att(A^m_{fi})$
transforms into $U_i$ and $G_i \subset U_i$ (and $U_i$ transforms
into $Att(A^m_{fi})$ in hierarchical gluing of cycles).

It is straightforward to see that during construction of dominant
systems for $\mathcal{W}$ from the network $\mathcal{V}^m$  no
connection between $U_i$ are created. Therefore, the reaction
network ${\rm dom \, mod}(\mathcal{W})$ is a union of networks on
sets $U_i$ without any link between sets.

If $G_1, \ldots G_m$ are all ergodic components of the system,
then there exist $m$ independent positive linear functionals
$b^1(c)$, ...  $b^m(c)$ that describe asymptotical behaviour of
kinetic system when $t \to \infty$ (\ref{TimeEv}). For ${\rm dom
\, mod}(\mathcal{W})$ these functionals are: $b^l(c)=\sum_{A \in
U_l} c_A$ where $c_A$ is concentration of $A$. Hence, for the
initial reaction network $\mathcal{W}$ with well separated
constants
\begin{equation}\label{zero-one}
b^l(c) \approx \sum_{A \in U_l} c_A.
\end{equation}
This is the zero--one law for multiscale networks: for any $l, i$,
the value of functional $b^l$ (\ref{TimeEv}) on basis vector
$e^i$, $b^l (e^i)$, is either close to one or close to zero (with
probability close to 1).  We already mentioned this law in
discussion of a simple example (\ref{massDistri}). The approximate
equality (\ref{zero-one}) means that for each reagent $A \in
\mathcal{A}$ there exists such an ergodic component $G$ of
$\mathcal{W}$ that $A$ transforms when $t \to \infty$ preferably
into elements of $G$ even if there exist paths from $A$  to other
ergodic components of $\mathcal{W}$.

\subsection{Dynamic limitation and ergodicity boundary}

Dominant systems are acyclic. All the stationary rates in the
first order are limited by limiting steps of some cycles. Those
cycles are glued in the hierarchical cycle gluing procedure, and
their limiting steps are deleted in the cycles surgery procedures
(see Subsec.~\ref{subsecSurgery} and Fig.~\ref{CycleSurg}).

Relaxation to steady state of the network is multi-exponential,
and now we are interested in estimate of the longest relaxation
time $\tau$:
\begin{equation}
\tau = 1 / \min \{ - Re \lambda_i |  \lambda_i \neq 0 \}
\end{equation}
Is there a constant that limits the relaxation time? The general
answer for multiscale system is: $1/\tau$ is equal to the minimal
reaction rate constant of the dominant system. It is impossible to
guess a priori, before construction of the dominant system, which
constant it is. Moreover, this may be not a rate constant for a
reaction from the initial network, but a monomial of such
constants.

Nevertheless, sometimes it is possible to point the reaction rate
constant that is limiting for the relaxation in the following
sense. For known topology of reaction network and given ordering
of reaction rate constants we find such a constant (ergodicity
boundary) $k_{\tau}$ that
\begin{equation}
\tau \approx  \frac{1}{a k_{\tau}} , \label{eq11}
\end{equation}
with $ a \lesssim 1$ is a function of constants $k_{j}> k_{\tau}$.
This means that ${1}/{k_{\tau}}$ gives the lower estimate of the
relaxation time, but $\tau$ could be larger. In addition, we show
that there is a zero-one alternative too: if the constants are
well separated then either $a\approx 1$ or $a\ll 1$.

We study a multiscale system with a given reaction rate constants
ordering, $k_{j_1}>k_{j_2}> \ldots > k_{j_n}$. Let us suppose that
the network is weakly ergodic (when there are several ergodic
components, each one has its longest relaxation time that can be
found independently). We say that $k_{j_r}$, $1 \leq r \leq n$ is
the
  {\em ergodicity boundary} $k_{\tau}$
 if the network of reactions with parameters $k_{j_1},k_{j_2}, \ldots ,
k_{j_r}$ (when $k_{j_{r+1}}=...k_{j_n}=0$) is weakly ergodic, but
the network with parameters $k_{j_1},k_{j_2}, \ldots , k_{j_r-1}$
(when $k_{j_{r}}=k_{j_{r+1}}=...k_{j_n}=0$) it is not. In other
words, when eliminating reactions in decreasing order of their
characteristic times, starting with the slowest one, the
ergodicity boundary is the constant of the first reaction whose
elimination breaks the ergodicity of the reaction digraph. This
reaction we also call the ``ergodicity boundary".

Let us describe the possible location of the ergodicity boundary
in the general multiscale reaction network $(\mathcal{W})$. After
deletion of reactions with constants
$k_{j_{r}},k_{j_{r+1}},...k_{j_n}$ from the network two ergodic
components (minimal sinks) appear, $G_1$ and $G_2$. The ergodicity
boundary starts in one of the ergodic components, say $G_1$, and
ends at the such a reagent $B$ that another ergodic component,
$G_2$, is reachable by $B$ (there exists an oriented path from $B$
to some element of $G_2$).

An estimate of the longest relaxation time can be obtained by
applying the perturbation theory for linear operators to the
degenerated case of the zero eigenvalue of the matrix ${K}$. We
have ${K}={K}_{<r}(k_{j_1},k_{j_2}, \ldots , k_{j_r-1})+k_{j_r}Q
+o(k_r)$, where ${K}_{<r}$ is obtained from ${K}$ by letting
$k_r=k_{r+1}=\ldots k_n=0$, $Q$ is a constant matrix of rank 1,
and $o(k_r)$ includes terms that are negligible relative to $k_r$.
The zero eigenvalue is twice degenerate in ${K}_{<r}$ and not
degenerate in ${K}_{<r}+k_rQ$. One gets the following estimate:
\begin{equation}
\overline{a}\frac{1}{k_{\tau}} \geq \tau \geq
\underline{a}\frac{1}{k_{\tau}}, \label{eq10}
\end{equation}
where $\overline{a}, \underline{a} >0$ are some positive functions
of $k_1,k_2, \ldots , k_{r-1}$ (and of the reaction graph
topology).

\begin{figure}
\begin{centering}
a)\includegraphics[width=20mm]{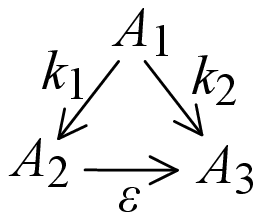}
b)\includegraphics[width=20mm]{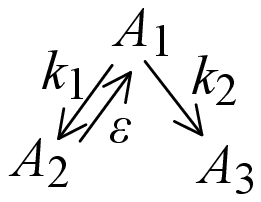} \caption
{\label{Fig:Triangl} Two basic examples of ergodicity boundary
reaction: (a) Connection between ergodic components; (b)
Connection from one ergodic component to element that is connected
to the both ergodic components by oriented paths. In both cases,
for $\varepsilon=0$, the ergodic components are $\{A_2\}$ and
$\{A_3\}$.}
\end{centering}
\end{figure}

Two simplest examples demonstrate two types of dependencies of
$\tau$ on $k_{\tau}$:
\begin{enumerate}
\item{For the reaction mechanism
Fig.~\ref{Fig:Triangl}a $$\min_{\lambda \neq 0} \{- Re \lambda
\}=\varepsilon,$$ if $\varepsilon < k_1+k_2$.}
\item{For the reaction mechanism
Fig.~\ref{Fig:Triangl}b $$\min_{\lambda \neq 0} \{- Re \lambda
\}=\varepsilon k_2/(k_1+k_2) + o({ \varepsilon}),$$ if $\varepsilon
< k_1+k_2$. For well separated parameters there exists a zero-one
(trigger) alternative: if $k_1 \ll k_2$ then $\min_{\lambda \neq 0}
\{- Re \lambda \}\approx \varepsilon$; if, inverse, $k_1 \gg k_2$
then $\min_{\lambda \neq 0} \{- Re \lambda \}= o({ \varepsilon})$.}
\end{enumerate}

In general multiscale network, two type of obstacles can violate
approximate equality $\tau \approx 1/k_{\tau}$. Following the
zero-one law for  nonergodic multiscale networks (previous
subsection) we can split the set of all vertices into two subsets,
$U_1$ and $U_2$. The dominant  reaction network ${\rm dom \,
mod}(\mathcal{W})$ is a union of networks on sets $U_{1,2}$
without any link between sets.

If the ergodicity boundary reaction starts in the ergodic
component $G_1$ and ends at $B$ which does not belong to the
``opposite" basin of attraction $U_2$, then $\tau \gg 1/k_{\tau}$.
This is the first possible obstacle.

Let the ergodicity boundary reaction start at $A \in G_1$  and end
at $B \in U_2$. We define the maximal linear chain of reactions in
dominant system with start at $B$: $B \to ...$~. This chain
belongs to $U_2$. Let us extend this chain from the left by the
ergodicity boundary: $A \to B \to ...$~. Relaxation time for the
network of $r$ reactions (with the kinetic matrix $K_{\leq
r}={K}_{<r}(k_{j_1},k_{j_2}, \ldots , k_{j_r-1})+k_{j_r}Q$) is,
approximately,  the relaxation time of this chain, i.e. $1/k$,
where $k$ is the minimal constant in the chain. There may appear a
monomial constant $k \ll k_{\tau}$. In that case, $\tau \gg
1/k_{\tau}$, and relaxation is limited by this minimal $k$ or by
some of constants $k_{j_p}$, $p>r$ or by some of their
combinations. This existence of a monomial constant $k \ll
k_{\tau}$ in the maximal chain $A \to B \to ...$ from the dominant
system is the second possible obstacle for approximate equality
$\tau \approx 1/k_{\tau}$.

If there is neither the first obstacle, nor the second one, then
$\tau \approx 1/k_{\tau}$. The possibility of these obstacles
depends on the definition of multiscale ensembles we use. For
example for the log-uniform distribution of rate constants in the
ordering cone $k_{j_1}>k_{j_2}> \ldots > k_{j_n}$
(Sec.~\ref{sec3}) the both obstacles have nonzero probability, if
they are topologically possible. On the other hand, if we study
asymptotic of relaxation time at $\epsilon \to 0$ for $k_{i_r}=
\epsilon k_{j_r-1}$ for  given values of $k_{j_1},k_{j_2}, \ldots
, k_{j_r-1}$, then for sufficiently small $\epsilon >0$ the second
obstacle is impossible.

Thus, the well known concept of stationary reaction rates {\it
limitation} by ``narrow places" or ``limiting steps" (slowest
reaction) should be complemented by the {\it ergodicity boundary}
limitation of relaxation time. It should be stressed that the
relaxation process is limited  not by the classical limiting steps
(narrow places), but by  reactions that may be absolutely
different. The simplest example of this kind is an irreversible
catalytic cycle: the stationary rate is limited by the slowest
reaction (the smallest constant), but the relaxation time is
limited by the reaction constant with the second lowest value (in
order to break the weak ergodicity of a cycle two reactions must
be eliminated).

\subsection{Zero-one law for relaxation modes (eigenvectors)
and lumping analysis}

For kinetic systems with well-separated constants the left and
right eigenvectors can be explicitly estimated. Their coordinates
are close to $\pm 1$ or 0.  We analyzed these estimates first for
linear chains and cycles (\ref{01cyrcleAsy}) and then for general
acyclic auxiliary dynamical systems (\ref{rightAcyc}),
(\ref{leftAcyc}) (\ref{right01}), (\ref{left01}). The distribution
of zeros and $\pm 1$ in the eigenvectors components depends on the
rate constant ordering and may be rather surprising. Perhaps, the
simplest example gives the asymptotic equivalence (for $k_i^- \gg
k_i, \, k_{i+1}$) of the reaction network $A_i \leftrightarrow
A_{i+1} \to A_{i+2}$ with rate constants $k_i$, $k_i^-$ and
$k_{i+1}$ to the reaction network $A_{i+1} \to A_i \to A_{i+2}$
with rate constants $k_i^-$ (for the reaction $A_{i+1} \to A_i$)
and $k_{i+1} k_i /k_i^-$ (for the reaction $A_i \to A_{i+2}$)
presented in in Subsec.~\ref{subsecCycle+1}.

For reaction networks with well-separated constants coordinates of
left eigenvectors $l^i$ are close to 0 or 1. We can use the left
eigenvectors for coordinate change. For the new coordinates $z_i =
l^i c$ (eigenmodes) the simplest equations hold:
$\dot{z}_i=\lambda_i z_i$. The zero-one law for left eigenvectors
means that the eigenmodes are (almost) sums of some components:
$z_i = \sum_{i \in V_i} c_i$ for some sets of numbers $V_i$. Many
examples, (\ref{LeftApproxLin}), (\ref{leftAcycBran}),
(\ref{leftPrism1}), (\ref{leftPrism2}), demonstrate that some of
$z_i$ can include the same concentrations: it may be that $V_i
\cap V_j \neq \varnothing$ for some $i\neq j$. Aggregation of some
components (possibly with some coefficients) into new group
components for simplification of kinetics is the major task of
lumping analysis.

Wei and Kuo  studied conditions for exact (\cite{LumpWei1}) and
approximate (\cite{LumpWei2}) linear lumping. More recently,
sensitivity analysis and Lie group approach were applied to lumping
analysis (\cite{LumpLiRab1,LumpLiRab2}), and more general nonlinear
forms of lumped concentrations are used (for example, $z_i$ could be
rational function of $c$). The power of lumping using a time-scale
based approach was demonstrated by
\cite{LumpAthmTime,LumpLiaoBiochem}. This computationally cheap
approach combines ideas of sensitivity analysis with simple and
useful grouping of species with similar lifetimes and similar
topological properties caused by connections of the species in the
reaction networks. The lumped concentrations in this approach are
simply sums of concentrations in groups.

Kinetics of multiscale systems studied in this paper and developed
theory of dynamic limitation demonstrates that in multiscale limit
lumping analysis can work (almost) exactly. Lumped concentrations
are sums in groups, but these groups can intersect and usually
there exist several intersections.

\subsection{Nonequilibrium phase transitions in multiscale systems}

For each zero-one law specific sharp transitions exist: if two
systems in a  one-parametric family have different zero-one steady
states or relaxation modes, then somewhere between a point of jump
exists. Of course, for given finite values of parameters this will
be not a point of discontinuity, but rather a thin zone of fast
change. At such a point the dominant system changes. We can call
this change a {\it nonequilibrium phase transition}. Here we
identify a ``multiscale nonequilibrium phase" with a dominant
system.

A point of phase transition can be a point where the order of
parameters changes. But not every change of order causes the
change of dominant systems. On the other hand, change of order of
some monomials can change the dominant system even if the order of
parameters persists (examples are presented in previous section).
Evolution of a parameter--dependent multiscale reaction network
can be represented as a sequence of sharp change of dominant
system. Between such sharp changes there are periods of evolution
of dominant system parameters without qualitative changes.

\section{Limitation in modular structure and solvable
modules}\label{sec5}

\subsection{Modular limitation} The simplest one-constant
limitation concept cannot be applied to all systems. There is
another very simple case based on exclusion of ``fast equilibria"
$A_i \rightleftharpoons A_j$. In this limit, the ratio of reaction
constants $K_{ij}=k_{ij}/k_{ji}$ is bounded, $0 < a < K_{ij} <b
<\infty$, but for different pairs $(i,j), \, (l,s)$ one of the
inequalities $k_{ij} \ll k_{ls}$ or $k_{ij} \gg k_{ls}$ holds. (One
usually calls these $K$ ``equilibrium constant", even if there is no
relevant thermodynamics.)  \cite{Ray} discussed that case
systematically for some real examples. Of course, it is possible to
create the theory for that case very similarly to the theory
presented above. This should be done, but it is worth to mention now
that the limitation concept can be applied to {\it any} modular
structure of reaction network. Let for the reaction network
$\mathcal{W}$ the set of elementary reactions $\mathcal{R}$ is
partitioned on some modules: $\mathcal{R}= \cup_i \mathcal{R}_i$. We
can consider the related multiscale ensemble of reaction constants:
let the ratio of any two rate constants inside each module be
bounded (and separated from zero, of course), but the ratios between
modules form a well separated ensemble. This can be formalized by
multiplication of rate constants of each module $\mathcal{R}_i$ on a
time scale coefficient $k_i$. If we assume that $\ln k_i$ are
uniformly and independently distributed on a real line (or $k_i$ are
independently and log-uniformly distributed on a sufficiently large
interval) then we come to the problem of modular limitation. The
problem is quite general: describe the typical behavior of
multiscale ensembles for systems with given modular structure: each
module has its own time scale and these time scales are well
separated.

Development of such a general theory is outside the scope of our
paper, and here we just find building blocks for the future
theory, {\it solvable reaction modules}. There may be many various
criteria of selection the reaction modules. Here are several
possible choices: individual reactions (we developed the theory of
multiscale ensembles of individual reactions in this paper),
couples of mutually inverse reactions, as we mentioned above,
acyclic reaction networks, ... \ .

Among the possible reasons for selection the class of reaction
mechanisms for this purpose, there is one formal, but important:
the possibility to solve the kinetic equation for every module in
explicit analytical (algebraic) form with quadratures. We call
these systems ``solvable".

\subsection{Solvable reaction mechanisms}

Let us describe all solvable reaction systems (with mass action
law), linear and nonlinear.

Formally, we call the set of reaction solvable, if there exists a
linear transformation of coordinates $c \mapsto a$  such that
kinetic equation in new coordinates for all values of reaction
constants has the triangle form:
\begin{equation}\label{triangle}
\frac{\D a_i}{\D t}=f_i(a_1, a_2,... \, a_i).
\end{equation}
This system has the lower triangle Jacobian matrix $\partial
\dot{a}_i / \partial a_j$.

To construct the general mass action law system we need: the list
of components, $\mathcal{A} =\{ A_1, ...\, A_n\}$ and the list of
reactions (the reaction mechanism):
\begin{equation}\label{stoi}
\sum_i \alpha_{ri}A_i \to \sum_{k} \beta_{rk} A_k,
\end{equation}
where $r$ is the reaction number, $\alpha_{ri}, \, \beta_{rk}$ are
nonnegative integers (stoichiometric coefficients). Formally, it
is possible that all $\beta_k=0$ or all $\alpha_i=0$. We allow
such reactions. They can appear in reduced models or in auxiliary
systems.

A real variable $c_i$ is assigned to every component $A_i$, $c_i$
is the concentration of $A_i$, $c$ is the concentration vector
with coordinates $c_i$. The reaction kinetic equations are
\begin{equation}\label{MALkinur}
\frac{\D c}{\D t}= \sum_r \gamma_r w_r(c),
\end{equation}
where $\gamma_r$ is the reaction stoichiometric vector with
coordinates $\gamma_{ri}=\beta_{ri}- \alpha_{ri}$, $w_r(c)$ is the
reaction rate. For mass action law,
\begin{equation}\label{MAL}
w_r(c)= k_r \prod_i c_i^{\alpha_{ri}},
\end{equation}
where $k_r$ is the reaction constant.

Physically, equations (\ref{MALkinur}) correspond to reactions in
fixed volume, and in more general case  a multiplier $V$ (volume)
is necessary: $$\frac{\D (Vc)}{\D t}= V \sum_r \gamma_r w_r(c).$$
Here we study the systems (\ref{MALkinur}) and postpone any
further generalization.

The first example of solvable systems give the sets of reactions
of the form
\begin{equation}\label{acycle}
\alpha_{ri}A_i \to \sum_{k, \, k>i} \beta_{rk} A_k
\end{equation}
(components $A_k$ on the right hand side have higher numbers $k$
than the component $A_i$ on the left hand side, $i < k$). For
these systems, kinetic equations (\ref{MALkinur}) have the
triangle form from the very beginning.

The second standard example gives the couple of mutually inverse
reactions:
\begin{equation}\label{inverse}
\sum_i \alpha_{i}A_i \rightleftharpoons \sum_{k} \beta_{k} A_k,
\end{equation}
these reactions have stoichiometric vectors $\pm \gamma$,
$\gamma_i =\beta_i - \alpha_i$. The kinetic equation
$\dot{c}=(w^+-w^-)\gamma$ has the triangle form (\ref{triangle})
in any orthogonal coordinate system with the last coordinate $a_n
= (\gamma,c)=\sum_i \gamma_i c_i$. Of course, if there are several
reactions with proportional stoichiometric vectors, the kinetic
equations have the triangle form in the same coordinate systems.

The general case of solvable systems is essentially a combination
of that two (\ref{acycle}), (\ref{inverse}), with some
generalization. Here we follow the book by \cite{Ocherki} and
present an algorithm for analysis of reaction network solvability.
First, we introduce a relation between reactions ``$r$th reaction
directly affects the rate of $s$th reaction" with notation $r
\rightarrow s$:  $r \rightarrow s$ if there exists such $A_i$ that
$\gamma_{ri}\alpha_{si} \neq 0$. This means that concentration of
$A_i$ changes in the $r$th reaction ($\gamma_{ri} \neq 0$) and the
rate of the $s$th reaction depends on $A_i$ concentration
($\alpha_{si} \neq 0$). For that relation we use $r \rightarrow
s$. For transitive closure of this relation we use notation $r
\succeq s$ (``$r$th reaction affects the rate of $s$th reaction"):
$r \succeq s$ if there exists such a sequence $s_1, s_2, ...\,
s_q$ that $r \to s_1 \to s_2 \to ...\, s_q \to s$.

The {\it hanging component} of the reaction network $\mathcal{W}$
is such $A_i \in \mathcal{A}$ that for all reactions
$\alpha_{ri}=0$. This means that all reaction rates do not depend
on concentration of $A_i$. The {\it hanging reaction} is such
element of $\mathcal{R}$ with number $r$ that $r \succeq s$ only
if $\gamma_s = \lambda \gamma_r$ for some number $\lambda$. An
example of hanging components gives the last component $A_n$ for
the triangle network (\ref{acycle}). An example of hanging
reactions gives a couple of reactions (\ref{inverse}) if they do
not affect any other reaction.

In order to check solvability of the reaction network
$\mathcal{W}$ we should find all hanging components and reactions
and delete them from $\mathcal{A}$ and $\mathcal{R}$,
correspondingly. After that, we get a new system, $\mathcal{W}_1$
with the component set $\mathcal{A}_1$ and the reaction set
$\mathcal{R}_1$. Next, we should find all hanging components and
reactions for $\mathcal{W}_1$ and delete them from $\mathcal{A}_1$
and $\mathcal{R}_1$. Iterate until no hanging components or
hanging reactions could be found. If the final set of components
is empty, then the reaction network $\mathcal{W}$ is solvable. If
it is not empty, then $\mathcal{W}$ is not solvable.

For example, let us consider the reaction mechanism with
$\mathcal{A}=\{A_1,A_2,A_3,A_4\}$ and reactions $A_1+A_2 \to 2
A_3$, $A_1+A_2 \to A_3 + A_4$, $A_3 \to A_4$, $A_4 \to A_3$. There
are no hanging components, but two hanging reactions, $A_3 \to
A_4$ and $A_4 \to A_3$. After deletion of these two reactions, two
hanging components appear, $A_3$ and $A_4$. After deletion these
two components, we get two hanging reactions, $A_1+A_2 \to 0$,
$A_1+A_2 \to 0$ (they coincide). We delete these reactions and get
two components $A_1$, $A_2$ without reactions. After deletion
these hanging components we obtain the empty system. The reaction
network is solvable.

An oriented cycle of the length more than two is not solvable. For
each number of vertices one can calculate the set of all maximal
solvable mechanisms. For example, for five components there are
two maximal solvable mechanisms of monomolecular reactions:
\begin{enumerate}
\item{$A_1 \to A_2 \to A_4$, $A_1 \to A_4$, $A_2\to A_3$, $A_1 \to A_3 \to A_5$,
$A_1 \to A_5$, $A_4 \leftrightarrow A_5$;}
\item{$A_1 \to A_2$, $A_1 \to A_3$, $A_1 \to A_4$, $A_1 \to A_5$,
$A_2 \leftrightarrow A_3$, $A_4 \leftrightarrow A_5$.}
\end{enumerate}
It is straightforward to check solvability of these mechanism. The
first mechanism has a couple of hanging reactions, $A_4
\leftrightarrow A_5$. After deletion of these reactions, the
system becomes acyclic, of the form (\ref{acycle}). The second
mechanism has two couples of hanging reactions, $A_2
\leftrightarrow A_3$ and $A_4 \leftrightarrow A_5$. After deletion
of these reactions, the system also transforms into the triangle
form (\ref{acycle}). It is impossible to add any new monomolecular
reactions between $\{A_1,A_2,A_3,A_4,A_5\}$ to these mechanisms
with preservation of solvability, and any solvable monomolecular
reaction network with five reagents is a subset of one of these
mechanisms.

Finally, we should mention connections between solvable reaction
networks and solvable Lie algebras (\cite{Lie,LieGraa1}). Let us
remind that matrices $M_1, ... \, M_q$ generate a solvable Lie
algebra if and only if they could be transformed simultaneously
into a triangle form by a change of basis.

The Jacobian matrix for the mass action law kinetic equation
(\ref{MALkinur}) is:

\begin{equation}\label{JacKin}
J = \left(\frac{\partial c_i }{\partial c_j} \right)= \sum_r w_r
J_r = \sum_{rj} \frac{w_r}{c_j} M_{rj} ,
\end{equation}
where
\begin{equation}
\begin{split}
 J_r&=\gamma_r \alpha_r^{\top} {\rm diag}\left\{\frac{1}{c_1},
\frac{1}{c_2}, ... \, \frac{1}{c_n} \right\}=\sum_j \frac{1}{c_j}
M_{rj} ,\\ M_{rj}&=\alpha_{rj} \gamma_r e^{j\top},
\end{split}
\end{equation}
$\alpha_r^{\top}$ is the vector row $(\alpha_{r1}, ... \,
\alpha_{rn})$, $e^{j\top}$ is the $j$th basis vector row with
coordinates $e^{j\top}_k= \delta_{jk}$.

The Jacobian matrix (\ref{JacKin}) should have the lower triangle
form in coordinates $a_i$ (\ref{triangle}) for all nonnegative
values of rate constants and concentrations. This is equivalent to
the lower triangle form of all matrices $M_{rj}$ in these
coordinates. Because usually there are many zero matrices among
$M_{rj}$, it is convenient to describe the set of nonzero
matrices.

For the $r$th reaction $I_r=\{i \, | \, \alpha_{ri} \neq 0 \} $.
The reaction rate $w_r$ depends on $c_i$ if and only if $i \in
I_r$. For each $i=1,...\, n$ we define a matrix
$$m_{ri}=\left[0,0, ... \, \underbrace{\gamma_r}_i \, ...\,
0\right].$$  The $i$th column of this matrix coincides with the
vector column $\gamma_r$. Other columns are equal to zero. For
each $r$ we define a set of matrices $\mathcal{M}_r=\{m_{ri}\, |
\, i \in I_r \}$, and $\mathcal{M} = \cup_r \mathcal{M}_r$. The
reaction network $\mathcal{W}$ is solvable if and only if the
finite set of matrices $\mathcal{M}$ generates a solvable Lie
algebra.

Classification of finite dimensional solvable Lie algebras remains a
difficult problem (\cite{LieGraa1,LieGraa2}). It seems plausible
that the classification of solvable algebras associated with
reaction networks can bring new ideas into this field of algebra.

\section{Conclusion: Concept of limit simplification in
multiscale systems \label{sec:conclud} }

In this paper, we study networks of linear reactions. For any
ordering of reaction rate constants we look for the dominant
kinetic system. The dominant system is, by definition, the system
that gives us the main asymptotic terms of the stationary state
and relaxation in the limit for well separated rate constants. In
this limit any two constants are connected by the relation $\gg$
or $\ll$.

The topology of dominant systems is rather simple; they are those
networks which are  graphs of discrete dynamical systems on the
set of vertices. In such graphs each vertex has no more than one
outgoing reaction. This allows us to construct the explicit
asymptotics of eigenvectors and eigenvalues. In the limit of well
separated constants, the coordinates of eigenvectors for dominant
systems can take only three values: $\pm 1$ or 0. All algorithms
are represented topologically by transformation of the graph of
reaction (labeled by reaction rate constants). We call these
transformations ``cycles surgery", because the main operations are
gluing cycles and cutting cycles in graphs of auxiliary discrete
dynamical systems.

In the simplest case, the dominant system is determined by the
ordering of constants. But for sufficiently complex systems we
need to introduce auxiliary elementary reactions. They appear
after cycle gluing and have monomial rate constants of the form
$k_{\varsigma}=\prod_i k_i^{\varsigma_i}$. The dominant system
depends on the place of these monomial values among the ordered
constants.

Construction of the dominant system clarifies the notion of
limiting steps for relaxation. There is an exponential relaxation
process that lasts much longer than the others in
(\ref{ApproxRelAux}), (\ref{AppRel}). This is the slowest
relaxation and it is controlled by one reaction in the dominant
system, the limiting step. The limiting step for relaxation is not
the slowest reaction, or the second slowest reaction of the whole
network, but the slowest reaction of the dominant system. That
limiting step constant is not necessarily  a reaction rate
constant for the initial system, but can be represented by a
monomial of such constants as well.

The idea of dominant subsystems in asymptotic analysis was
proposed by Newton and developed by \cite{Kruskal}. A modern
introduction with some historical review is presented by
\cite{White}. In our analysis we do not use the powers of small
parameters (as it was done by
\cite{ViLju,Lid,Akian,Kruskal,White}), but operate directly with
the rate constants ordering.

To develop the idea of systems with well separated constants to
the state of a mathematical notion, we introduce multiscale
ensembles of constant tuples. This notion allows us to discuss
rigorously uniform distributions on infinite space and gives the
answers to a question: what does it mean ``to pick a multiscale
system at random".

Some of results obtained are rather surprising and unexpected.
First of all is the zero-one asymptotic of eigenvectors. Then, the
good approximation to eigenvectors does not give approximate
eigenvectors (the inverse situation is more common: an approximate
eigenvector could be far from the eigenvector). The almost exact
lumping analysis provided by the zero-one approximation of
eigenvectors has an unexpected property: the lumped groups for
different eigenvalues can intersect. Rather unexpected seems the
change of reaction sequence when we construct the dominant
systems. For example, asymptotic equivalence (for $k_i^- \gg k_i,
\, k_{i+1}$) of the reaction network $A_i \leftrightarrow A_{i+1}
\to A_{i+2}$ with rate constants $k_i$, $k_i^-$ and $k_{i+1}$ to
the reaction network $A_{i+1} \to A_i \to A_{i+2}$ with rate
constants $k_i^-$ (for the reaction $A_{i+1} \to A_i$) and
$k_{i+1} k_i /k_i^-$ (for the reaction $A_i \to A_{i+2}$) is
simple, but surprising (Subsec.~\ref{subsecCycle+1}). And, of
course, it was surprising to observe how the dynamics of linear
multiscale networks transforms into the dynamics on finite sets of
reagent names.

Now we have the complete theory and the exhaustive construction of
algorithms for linear reaction networks with well separated rate
constants. There are several ways of using the developed theory
and algorithms:
\begin{enumerate}
 \item{For direct computation of steady states and relaxation dynamics;
 this may be useful for complex systems because of the simplicity of the
 algorithm and
 resulting formulas and because often we do not know the  rate
 constants for complex networks, and kinetics that is ruled by orderings
 rather than by exact values of rate constants may be very useful;}
 \item{For planning of experiments and mining the experimental data
 -- the observable kinetics is more sensitive to reactions from
 the dominant network, and much less sensitive to other reactions,
 the relaxation spectrum of the dominant network is  explicitly connected
 with the correspondent reaction rate constants,
 and the  eigenvectors  (``modes") are sensitive to the constant ordering,
 but not to exact values;}
 \item{The steady states and dynamics of the dominant system could
 serve as a robust first approximation in perturbation theory or as a
 preconditioning in numerical methods.}
\end{enumerate}

The developed methods are computationally cheap, for example, the
algorithm for construction of dominant system has linear
complexity ($\sim$ number of reactions). From a practical point of
view, it is attractive to use exact rational expressions for the
dominant system modes (\ref{ChainEigen}), (\ref{rightAcyc}),
(\ref{leftAcyc}) instead of the zero-one approximation. Also, we
can use exact formula (\ref{CycleRate}) for irreversible cycle
steady state instead of linear approximation
(\ref{CycleLimRateLin}). These improvements are computationally
cheap and may enhance accuracy of computations.

From a theoretical point of view the outlook is more important.
Let us answer the question: what has to be done, but is not done
yet? Three directions for further development are clear now:
\begin{enumerate}
 \item{Construction of dominant systems for the reaction network that
 has a group of constants with comparable values (without
 relations $\gg$ between them). We considered cycles with several
 comparable constants in Sec.~\ref{sec2}, but the general theory still
 has to be developed.}
 \item{Construction of dominant systems for reaction networks with
 modular structure. We can assume that the ratio of any two rate
 constants inside each module be bounded and separated from zero, but
 the ratios between modules form a well separated ensemble. A reaction network that
 has a group of constants with comparable values gives us an
 example of the simplest modular structure: one module includes
 several reactions and other modules arise from one reaction.
 In Sec.~\ref{sec5} we describe all solvable modules such that it is possible
 to solve the kinetic equation for every module in explicit analytical
 (algebraic) form with quadratures (even for nonconstant in time reaction rate constants).}
 \item{Construction of dominant systems for nonlinear reaction
 networks. The first idea here is the representation of a nonlinear
 reaction as a pseudomonomolecular reaction: if for reaction $A+B \to ...$
 concentrations $c_A$ and $c_B$ are well separated, say, $c_A \gg c_B$, then we
 can consider this reaction as $B\to ... $ with rate constant dependent on
 $c_A$. The relative change of $c_A$ is slow, and we can consider this reaction
 as pseudomonomolecular until the relation $c_A \gg c_B$ changes
 to $c_A \sim c_B$. We can assume that in the general case only for small
 fraction of nonlinear reactions the pseudomonomolecular approach
 is not applicable, and this set of genuinely nonlinear reactions
 changes in time, but remains small. For nonlinear systems, even
 the realization of the limiting step idea for steady states
 of a one-route mechanism of a catalytic reaction is nontrivial
 and was developed through the concept of kinetic polynomial (\cite{YabLazLim})}.
\end{enumerate}

Finally, the concept of ``limit simplification" will be developed.
For multiscale nonlinear reaction networks the expected dynamical
behaviour is to be approximated by the system of dominant
networks. These networks may change in time but remain small
enough.

This hypothetical picture should give an answer to a very practical
question: how to describe kinetics beyond the standard
quasi-steady-state and quasiequilibrium approximations
(\cite{nonLumpMaini}). We guess that the answer has the following
form: during almost all time almost everything could be simplified
and the whole system behaves as a small one. But this picture is
also nonstationary: this small system change in time. Almost always
``something is very small and something is very big", but due to
nonlinearity this ordering can change in time. The whole system
walks along small subsystems, and constants of these small
subsystems change in time under control of the whole system state.
The dynamics of this walk supplements the dynamics of individual
small subsystems.

The corresponding structure of fast--slow time separation in phase
space is not necessarily  a smooth slow invariant manifold, but
may be similar to a ``crazy quilt" and may consist of fragments of
various dimensions that do not join smoothly or even continuously.

\section*{Appendix~1: Estimates of eigenvectors
for diagonally dominant matrices with diagonal gap condition}

The famous Gershgorin theorem gives estimates of eigenvalues. The
estimates of correspondent eigenvectors are not so well known. In
the paper we use some estimates of eigenvectors of kinetic
matrices. Here we formulate and prove these estimates for general
matrices. Below $A=(a_{ij})$ is a complex $n\times n$ matrix, $P_i
= \sum_{j, j\neq i} |a_{ij}|$ (sums of non-diagonal elements in
rows), $Q_i = \sum_{j, j\neq i} |a_{ji}|$ (sums of non-diagonal
elements in columns).

Gershgorin theorem (\cite{MM}, p.~146): The characteristic roots
of $A$ lie in the closed region $G^P$ of the $z$-plane
\begin{equation}
G^P=\bigcup_i G^P_i \; \; (G^P_i=\{z \, \bigl| \, |z-a_{ii}|\leq
P_i\}.
\end{equation}
Analogously, the characteristic roots of $A$ lie in the closed
region $G^Q$ of the $z$-plane
\begin{equation}
G^Q=\bigcup_i G^Q_i \; \; (G^Q_i=\{z \, \bigl| \, |z-a_{ii}|\leq
Q_i\}.
\end{equation}
Areas $G^P_i$ and $G^Q_i$ are the Gershgorin discs.

Gershgorin discs $G^P_i$ ($i=1, \ldots n$) are isolated, if $G^P_i
\cap G^P_j= \varnothing$ for $i\neq j$. If discs $G^P_i$ ($i=1,
\ldots n$) are isolated, then the spectrum of $A$ is simple, and
each Gershgorin disc $G^P_i$ contains one and only one eigenvalue of
$A$ (\cite{MM}, p.~147). The same is true for discs $G^Q_i$.

Below we assume that Gershgorin discs $G^Q_i$ ($i=1, \ldots n$)
are isolated, this means that for all $i,j$
\begin{equation}
|a_{ii}-a_{jj}| > Q_i + Q_j.
\end{equation}
Let us introduce the following notations:
\begin{equation}
\begin{split}
&\frac{Q_i}{|a_{ii}|}=\varepsilon_i, \;\; \frac{|a_{ij}|}{|a_{jj}|}
= \chi_{ij} \; \; \left(\varepsilon_i =\sum_l \delta_{li}\right), \\
& \min_j \frac{|a_{ii}-a_{jj}|}{|a_{ii}|}=g_i.
\end{split}
\end{equation}
Usually, we consider $\varepsilon_i$ and $\chi_{ij}$ as sufficiently
small numbers. In contrary, $g_i$ should  not be small, (this is the
{\it gap condition}). For example, if for any two diagonal elements
$a_{ii}$, $a_{jj}$ either $a_{ii} \gg a_{jj}$ or $a_{ii} \ll
a_{jj}$, then $g_i \gtrsim 1$ for all $i$.

Let $\lambda_1 \in G^Q_1$ be the eigenvalue of $A$ ($|\lambda_1  -
a_{11}| < Q_1$). Let us estimate the correspondent right eigenvector
$x^{(1)} = (x_i) $: $A x^{(1)} = \lambda_1 x^{(1)}$. We take $x_1=1$
and write equations for $x_i$ ($i\neq 1$):
\begin{equation}
(a_{ii}-a_{11}-\theta_1)x_i + \sum_{j,\, j\neq 1,i}
a_{ij}x_j=-a_{i1},
\end{equation}
where $\theta_1=\lambda_1  - a_{11}$, $|\theta_1|< Q_1$.

Let us introduce new variables $$\tilde{x}=(\tilde{x}_i), \; \;
\tilde{x}_i = {x_i}({a_{ii}-a_{11}}) \;\; (i=2, \ldots n).$$ In
these variables,
\begin{equation}\label{MatrixB}
\left(1-\frac{\theta_1}{a_{ii}-a_{11}}\right)\tilde{x}_i +
\sum_{j,\, j\neq 1,i}
\frac{a_{ij}}{a_{jj}-a_{11}}\tilde{x}_j=-a_{i1},
\end{equation}
or in matrix notations: $(1-B)\tilde{x}=-\tilde{a}_1$, where
$\tilde{a}_1$ is a vector column with coordinates $a_{i1}$. because
of gap condition and smallness of $\varepsilon_i$ and $\chi_{ij}$ we
$\chi_{ij}$ we can consider matrix $B$ as a small matrix, for
assume that $\|B\| <1$ and $(1-B)$ is reversible (for detailed
estimate of $\|B\|$ see below).

For $\tilde{x}$ we obtain:
\begin{equation}\label{tildasol}
\tilde{x} = -\tilde{a}_1 - B(1-B)^{-1}\tilde{a}_1,
\end{equation}
and for residual estimate
\begin{equation}
\|B(1-B)^{-1}\tilde{a}_1\| \leq \frac{\|B\|}{1-\|B\|} \|
\tilde{a}_1\|.
\end{equation}
For eigenvector coordinates we get from (\ref{tildasol}):
\begin{equation}
x_i=-
\frac{a_{i1}}{a_{ii}-a_{11}}-\frac{(B(1-B)^{-1}\tilde{a}_1)_i}{a_{ii}-a_{11}}
\end{equation}
and for residual estimate
\begin{equation}
 \frac{|(B(1-B)^{-1}\tilde{a}_1)_i|}{|a_{ii}-a_{11}|}
 \leq \frac{\|B\|}{1-\|B\|} \frac{\| \tilde{a}_1\|}{|a_{ii}-a_{11}|}.
\end{equation}

Let us give more detailed estimate of residual. For vectors we use
$l_1$ norm: $\|x\|=\sum |x_i|$. The correspondent operator norm of
 matrix $B$ is $$\|B\|=\max_{\|x\|=1} \|Bx\| \leq \sum_i \max_j |b_{ij}|.$$

With the last estimate for matrix $B$ (\ref{MatrixB}) we find:
\begin{equation}\label{syrye}
\begin{split}
|b_{ii}| \leq \frac{Q_1}{|a_{ii}-a_{11}|} \leq
\frac{\varepsilon_1}{g_1} \leq\frac{\varepsilon}{g}, \\
|b_{ij}|=\frac{|a_{ij}|}{|a_{jj}-a_{11}|} \leq
\frac{\chi_{ij}}{g_j}\leq\frac{\chi}{g} \;\; (i\neq j),
\end{split}
\end{equation}
where $\varepsilon=\max_i \varepsilon_i$, $\chi=\max_{i,j}
\chi_{ij}$, $g=\min_i g_i$. By definition, $\varepsilon \geq \chi$,
and for all $i,j$ the simple estimate holds: $|b_{ij}| \leq
\varepsilon /g$. Therefore, $\|Bx\| \leq n \varepsilon /g$ and,
${\|B\|}/({1-\|B\|}) \leq n \varepsilon /(g-n \varepsilon )$ (under
condition $g > n \varepsilon$). Finally,  $\| \tilde{a}_1\|=Q_1$ and
for residual estimate we get:
\begin{equation}\label{secOrEs}
\left|x_i+\frac{a_{i1}}{a_{ii}-a_{11}}\right|\leq \frac{n
\varepsilon^2}{g(g-n \varepsilon )} \; \; (i\neq 1).
\end{equation}
More accurate estimate can be produced from inequalities
(\ref{syrye}), if it is necessary. For our goals it is sufficient
to use the following consequence of (\ref{secOrEs}):
\begin{equation}\label{firOrEs}
|x_i| \leq \frac{\chi}{g}+\frac{n \varepsilon^2}{g(g-n \varepsilon
)}  \; \; (i\neq 1).
\end{equation}
With this accuracy, eigenvectors of $A$ coincide with standard
basis vectors, i.e. with eigenvectors of diagonal part of $A$,
${\rm diag}\{a_{11}, \ldots a_{nn}\}$.

\section*{Appendix~2: Time separation and averaging in cycles}

In Sec.~\ref{sec2} we analyzed relaxation of a simple cycle with
limitation as a perturbation of the linear chain relaxation by one
more step that closes the chain into the cycle. The reaction rate
constant for this perturbation is the smallest one. For this
analysis we used explicit estimates (\ref{01cyrcleAsy}) of the
chain eginvectors for reactions with well separated constants.

Of course, one can use estimates (\ref{rightAcyc}), (\ref{right01})
(\ref{leftAcyc}) and (\ref{left01}) to obtain a similar perturbation
analysis for more general acyclic systems (instead of a linear
chain). If we add a reaction to an acyclic system (after that a
cycle may appear) and assume that the reaction rate constant for
additional reaction is smaller than all other reaction constants,
then the generalization is easy.

This smallness with respect to all constants is required only in a
very special case when the additional reaction has a form $A_i \to
A_j$ (with the rate constant $k_{ji}$) and there is no reaction of
the form $A_i \to ...$ in the non-perturbed system. In
Sec.~\ref{sec5} and Appendix~1 we demonstrated that if in a
non-perturbed acyclic system there exists another reaction of the
form $A_i \to ...$ with rate constant $\kappa_i$, then we need
inequality $k_{ji} \ll \kappa_i$ only. This inequality allows us to
get the uniform estimates of eigenvectors for all possible values of
other rate constants (under the diagonally gap condition in the
non-perturbed system).

For substantiation of cycles surgery we need additional perturbation
analysis for zero eigenvalues. Let us consider a simple cycle $A_1
\to A_2 \to ... \to A_n \to A_1$ with reaction $A_i \to ...$ rate
constants $\kappa_i$. We add a perturbation $A_1 \to 0$ (from $A_1$
to nothing) with rate constant $\epsilon \kappa_1$. Our goal is to
demonstrate that the zero eigenvalue moves under this perturbation
to $\lambda_0= - \epsilon w^*(1 + \chi_w)$, the correspondent left
and right eigenvectors $r^0$ and $l^0$ are $r^0_i= c_i^*(1+
\chi_{ri})$  and $l^0_i=1+ \chi_{li}$, and $\chi_w$, $\chi_{ri}$ and
$\chi_{li}$ are uniformly small for a given sufficiently small
$\epsilon$ under all variations of rate constants. Here, $w^*$ is
the stationary cycle reaction rate and $c^*_i$ are stationary
concentrations for a cycle (\ref{CycleRate}) normalized by condition
$\sum_i c^*_i =1$. The estimate $\epsilon w^*$ for $-\lambda_0$ is
$\epsilon$-small with respect to any reaction of the cycle:
$w^*=\kappa_i c^*_i <\kappa_i $ for all $i$ (because $c^*_i <1$),
and $\epsilon w^* \ll \kappa_i$ for all $i$.

The kinetic equation for the perturbed system is:
\begin{equation}\label{kinPerCyc}
\begin{split}
&\dot{c}_1=-(1+\epsilon)\kappa_1 c_1 + \kappa_n c_n, \\ &\dot{c}_i
=-\kappa_i c_i + \kappa_{i-1} c_{i-1} \; \; (\mbox{for $i\neq 1$}).
\end{split}
\end{equation}
In the matrix form we can write
\begin{equation}
\dot{c} =Kc= (K_0 - \epsilon k_1 e^1 e^{1\top})c ,
\end{equation}
where $K_0$ is the kinetic matrix for non-perturbed cycle. To
estimate the right perturbed eigenvector $r^0$ and eigenvalue
$\lambda_0$ we are looking for transformation of matrix $K$ into
the form $K=K_r - \theta r e^{1\top}$, where $K$ is a kinetic
matrix for extended reaction system with components $A_1,... A_n$,
$K_r r=0$ and $\sum_i r_i=1$. In that case, $r$ is the
eigenvector, and $\lambda = -\theta r_1$ is the correspondent
eigenvalue.

To find vector $r$, we add to the cycle new reactions $A_1 \to
A_i$ with rate constants $\epsilon \kappa_1 r_i$ and subtract the
correspondent kinetic terms from the perturbation term $\epsilon
e^1 e^{1\top}c$. After that, we get $K=K_r - \theta r e^{1\top}$
with $\theta= \epsilon k_1$ and
\begin{equation}
\begin{split}
(K_r c)_1&= -k_1 c_1 - \epsilon k_1 (1-r_1) c_1 +k_n c_n, \\ (K_r
c)_i&=-k_i c_i + \epsilon k_1 r_i c_1 + k_{i-1}c_{i-1} \;\;
\mbox{for}\; \; i>1
\end{split}
\end{equation}
We have to find a positive normalized solution $r_i > 0$, $\sum_i
r_i =1$ to equation $K_r r =0$. This is the fixed point equation:
for every positive normalized $r$ there exists unique positive
normalized steady state $c^*(r)$: $K_r c^*(r)=0$, $c^*_i>0$,
$\sum_i c^*_i(r) =1$. We have to solve the equation $r=c^*(r)$.
The solution exists because the Brauer fixed point theorem.

If $r=c^*(r)$ then $k_i r_i - \epsilon k_1 r_i r_1 =
k_{i-1}r_{i-1}$. We use notation $w^*_i(r)$ for the correspondent
stationary reaction rate along the ``non-perturbed route":
$w^*_i(r)=k_i r_i$. In this notation, $w^*_i(r)- \epsilon r_i
w^*_1(r)= w^*_{i-1}(r)$. Hence, $|w^*_i(r)-w^*_1(r)| < \epsilon
w^*_1(r)$ (or $|k_i r_i - k_1 r_1| < \epsilon k_1 r_1$). Assume
$\epsilon< 1/4$ (to provide $1- 2\epsilon < 1/(1\pm\epsilon) <
1+2\epsilon$). Finally,
\begin{equation}\label{rightZero}
r_i= \frac{1}{k_i}\frac{1+\chi_i}{\sum_j \frac{1}{k_j}}=(1+\chi_i)
c^*_i
\end{equation}
where the relative errors $|\chi_i|<3 \epsilon$ and $c^*_i=c^*_i(0)$
is the normalized steady state for the non-perturbed  system. For
cycles with limitation, $r_i \approx (1+\chi_i)k_{\rm lim}/k_i$ with
$|\chi_i| < 3 \epsilon$. For the eigenvalue we obtain
\begin{equation}\label{lambdaZero}
\begin{split}
\lambda_0 &=- \epsilon w_1^* (r) = - \epsilon w_i^* (r)(1 +
\varsigma_i) \\ &= - \epsilon w^* (1+\chi)=- \epsilon k_i c^*_i(0)
(1+\chi)
\end{split}
\end{equation}
for all $i$, with $|\varsigma_i| < \epsilon$ and $|\chi|<3\epsilon$.
$|\chi|<3\epsilon$. Therefore, $\lambda_0$ is $\epsilon$-small
rate constant $k_i$ of the non-perturbed cycle. This implies that
$\lambda_0$ is $\epsilon$-small with respect to the real part of
every non-zero eigenvalue of the non-perturbed kinetic matrix $K_0$
(for given number of components $n$). For the cycles from multiscale
ensembles these eigenvalues are typically real and close to $-k_i$
for non-limiting rate constants, hence we proved for $\lambda_0$
even more than we need.

Let us estimate the  correspondent left eigenvector $l^0$ (a
vector row). The eigenvalue is known, hence it is easy to do just
by solution of linear equations. This system of $n-1$ equations
is:
\begin{equation}
\begin{split}
&-l_1 (1+\epsilon)k_1+l_2 k_1=\lambda_0 l_1 \\
& -l_i k_i+l_{i+1}k_i=\lambda_0 l_i, \; i=2,... n-1.
\end{split}
\end{equation}
For normalization, we take $l_1=1$ and find:
\begin{equation}\label{leftZero}
l_2=\left(\frac{\lambda_0}{k_1}+1+\epsilon\right)l_1, \;\;
l_{i+1}=\left(\frac{\lambda_0}{k_i}+1\right)l_i\;\; i>2.
\end{equation}
Formulas (\ref{rightZero}), (\ref{lambdaZero}) and (\ref{leftZero})
give the backgrounds for  surgery of cycles with outgoing reactions.
The left eigenvector gives the slow variable: if there are some
incomes to the cycle, then
\begin{equation}\label{kinPerCycExt}
\begin{split}
&\dot{c}_1=-(1+\epsilon)\kappa_1 c_1 + \kappa_n c_n + \phi_1(t), \\
&\dot{c}_i =-\kappa_i c_i + \kappa_{i-1} c_{i-1}+\phi_i(t) \; \;
(\mbox{for $i\neq 1$})
\end{split}
\end{equation}
and for slow variable $\widetilde{c}=\sum l_i c_i$ we get
\begin{equation}
\frac{\D \widetilde{c}}{\D t}= \lambda_0 \widetilde{c} + \sum_i l_i
\phi_i(t).
\end{equation}
This is the kinetic equation for a glued cycle. In the leading term,
all the outgoing reactions $A_i \to 0$ with rate constants $k =
\epsilon k_i$ give the same eigenvalue $- \epsilon w^*$
(\ref{lambdaZero}).

Of course, similar results for perturbations of zero eigenvalue
are valid for more general ergodic chemical reaction network with
positive steady state, and not only for simple cycles, but for
cycles we get simple explicit estimates, and this is enough for
our goals.

{\bf Acknowledgements.} This work was supported by British Council
Alliance Franco-British Research Partnership Programme.

\end{document}